\documentclass[aps,pra,10pt,twocolumn,notitlepage,superscriptaddress,dvipsnames,longbibliography,tightenlines,floatfix]{revtex4-2}
\pdfoutput=1
\usepackage{xcolor}
\usepackage{bm}
\usepackage{bbold}
\usepackage{mathtools}
\usepackage{siunitx}
\usepackage{graphicx}
\usepackage{amsmath,amsfonts,amssymb}
\usepackage[caption=false, subrefformat=parens, labelformat=parens]{subfig}
\usepackage{enumitem}
\usepackage{diagbox}
\usepackage{booktabs}
\usepackage[normalem]{ulem}
\usepackage{bibunits}
\usepackage[toc,page,title]{appendix}

\makeatletter
\let\start@align@nopar\start@align
\let\start@gather@nopar\start@gather
\let\start@multline@nopar\start@multline
\long\def\start@align{\par\start@align@nopar}
\long\def\start@gather{\par\start@gather@nopar}
\long\def\start@multline{\par\start@multline@nopar}
\makeatother
\usepackage{xcolor}
\definecolor{dark-red}{rgb}{0.4,0.15,0.15}
\definecolor{dark-blue}{rgb}{0.15,0.15,0.4}
\definecolor{medium-blue}{rgb}{0,0,0.5}
\usepackage{hyperref}
\usepackage[all]{hypcap}%fix problems with figures and tables of the 'hyperref' package
\usepackage[capitalize]{cleveref}% Package hperref works wrongly within the 'subequations' environment without this.
\hypersetup{
    colorlinks, linkcolor={dark-red},
    citecolor={dark-blue}, urlcolor={medium-blue}
} %These setups replace the ugly and distracting red and green boxes of the 'hyperref' package with colored texts.
\usepackage{cleveref}
\crefname{equation}{Eq.}{Eqs.}
\Crefname{equation}{Equation}{Equations}
\crefname{algorithm}{Alg.}{Algs.}
\Crefname{algorithm}{Algorithm}{Algorithms}
\crefname{figure}{Fig.}{Figs.}
\Crefname{figure}{Figure}{Figures}
\crefname{section}{Sec.}{Secs.}
\Crefname{section}{Section}{Sections}
\crefname{appendix}{Appendix}{Apps.}
\Crefname{appendix}{Appendix}{Apps.}
\crefname{paragraph}{Sec.}{Secs.}
\crefname{table}{Table}{Tables}

\usepackage[braket,qm]{qcircuit}

\usepackage{scalerel,stackengine}
\newcommand\pig[1]{\scalerel*[5pt]{\big#1}{%
\ensurestackMath{\addstackgap[1.5pt]{\big#1}}}}
\newcommand\pigl[1]{\mathopen{\pig{#1}}}
\newcommand\pigr[1]{\mathclose{\pig{#1}}}
\stackMath
\newcommand\tsup[2][2]{%
 \def\useanchorwidth{T}%
  \ifnum#1>1%
    \stackon[-1.2ex]{\tsup[\numexpr#1-1\relax]{#2}}{\mathchar"307E}%
  \else%
    \stackon[-1ex]{#2}{\mathchar"307E}%
  \fi%
}

\newcommand{\nsp}{\hspace{-0.6pt}}
\newcommand{\xssp}{\hspace{0.4pt}}
\newcommand{\xmsp}{\hspace{0.6pt}}
\newcommand{\norm}[1]{\lvert #1 \rvert}
\newcommand{\normb}[1]{\big\lvert #1 \big\rvert}
\newcommand{\proj}[1]{\ket{#1}\! \bra{#1}}

\newcommand{\braket}[2]{\langle\xssp #1\xssp\vert\xssp #2 \xssp\rangle}

\newcommand{\anticommute}[2]{{\{\xssp #1,\, #2 \xssp\}}}
\newcommand{\half}{\frac{1}{2}}
\newcommand{\dif}{d}

\newcommand{\hc}{\mathrm{h.c.}}

\DeclareMathOperator{\sgn}{sgn}

\newcommand{\FF}{F}

\newcommand{\LL}{L}

\newcommand{\RR}{R}
\renewcommand{\SS}{\RR}
\newcommand{\uodd}{\WW_\mathrm{odd}}
\newcommand{\ueven}{\WW_\mathrm{even}}
\newcommand{\VV}{V}
\newcommand{\UU}{U}

\newcommand{\CZ}{{\scalebox{0.96}{CZ}}}

\newcommand{\iSWAP}{\mbox{\small $\mathrm{iSWAP}$}}
\newcommand{\SWAP}{\mbox{\small $\mathrm{SWAP}$}}

\newcommand{\nn}{n}
\newcommand{\mm}{m}

\newcommand{\ff}{C}

\newcommand{\omegap}{\nu}

\newcommand{\polar}{{\varrho}}

\newcommand{\CC}{C}

\newcommand{\WW}{W}
\newcommand{\overbar}[1]{\mkern 1.5mu\overline{\mkern-1.5mu#1 \rule{0pt}{2.5mm}\mkern-1.5mu}\mkern 1.5mu}

\newcommand{\RX}{X}
\newcommand{\RY}{Y}
\newcommand{\RZ}{Z}
\newcommand{\RXY}{\mathrm{PhX}}

\newcommand{\sigmaZ}{Z}

\newcommand{\epsilonX}{\epsilon_{\scalebox{0.65}{$X$}}}
\newcommand{\epsilonY}{\epsilon_{\scalebox{0.65}{$Y$}}}
\newcommand{\epsilonZ}{\epsilon_{\scalebox{0.65}{$Z$}}}

\DeclareMathOperator{\arccot}{arccot}

\graphicspath{{./circuits/}, {./figures/}}

\mathchardef\mhyphen="2D % Define a "math hyphen"

\begin{document}

\title{
Characterizing Coherent Errors using Matrix-Element Amplification
}
\date{\today}

\author{Jonathan A. Gross}
\email[Corresponding author e-mail: ]{jarthurgross@google.com }
\affiliation{Google Quantum AI, Venice, CA 90291, USA}

\author{\'{E}lie Genois}
\affiliation{Google Quantum AI, Venice, CA 90291, USA}
\affiliation{Institut quantique \& D\'epartement de Physique, Universit\'e de Sherbrooke, Quebec J1K2R1, Canada}

\author{Dripto M. Debroy}
\affiliation{Google Quantum AI, Venice, CA 90291, USA}

\author{Yaxing Zhang}
\affiliation{Google Quantum AI, Venice, CA 90291, USA}

\author{Wojciech Mruczkiewicz}
\affiliation{Google Quantum AI, Venice, CA 90291, USA}

\author{Ze-Pei Cian}
\affiliation{Google Quantum AI, Venice, CA 90291, USA}
\affiliation{Department of Physics \& Joint Quantum Institute, University of Maryland, MD 20742, USA}

\author{Zhang Jiang}
\email[Corresponding author e-mail: ]{qzj@google.com }
\affiliation{Google Quantum AI, Venice, CA 90291, USA}

\begin{abstract}
Repeating a gate sequence multiple times amplifies systematic errors coherently, making it a useful tool for characterizing quantum gates.
However, the precision of such an approach is limited by low-frequency noise, while its efficiency is hindered by time-consuming scans required to match up the phases of the off-diagonal matrix elements being amplified.
Here, we overcome both challenges by interleaving the gate of interest with dynamical decoupling sequences in a protocol we call Matrix-Element Amplification using Dynamical Decoupling (MEADD).
Using frequency-tunable superconducting qubits from a Google Sycamore quantum processor, we experimentally demonstrate that MEADD surpasses the accuracy and precision of existing characterization protocols for estimating systematic errors in single- and two-qubit gates.
We use MEADD to estimate coherent parameters of $\CZ$ gates with 5 to 10 times the precision of existing methods and to characterize previously undetectable coherent crosstalk, reaching a precision below one milliradian.
\end{abstract}

\maketitle

\begin{bibunit}[apsrev4-1_with_title]

\section{Introduction}
\label{sec:intro}

\begin{figure*}[t]
    \centering
    \includegraphics[width=1.9\columnwidth]{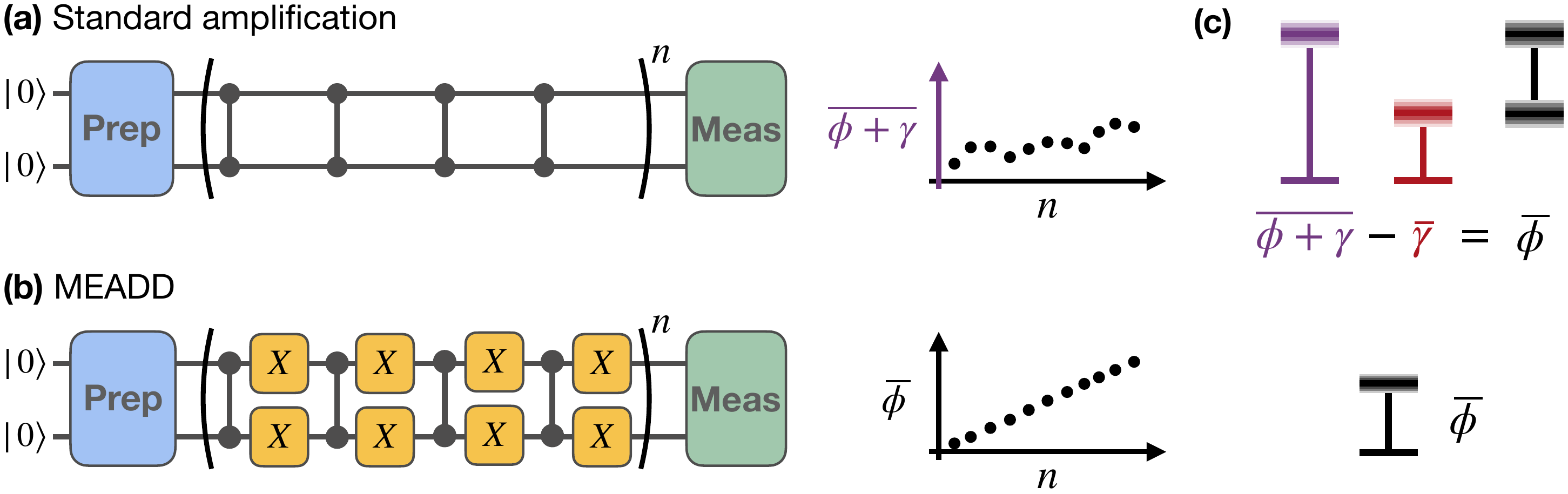}
    \caption{
    Schematic of the working principle of MEADD.
    \textbf{(a)} Standard gate repetition amplifies the matrix elements to estimate, but also amplifies other coherent and incoherent (noise) processes.
    \textbf{(b)} Making use of single-qubit dynamical decoupling sequences, MEADD filters out specific matrix elements while additionally echoing out the noisy time-varying single-qubit phases. The parameter we are estimating---together with other matrix element commuting with the DD gates---then builds up linearly with depth and can be fitted from the slope, with Heisenberg scaling in precision with circuit depth.
    All MEADD circuits possess this simple structure, with a preparation and pre-measurement layer containing at most a single entangling operation together with at most four single-qubit gates, and terminated by standard single-qubit Z-basis measurement.
    \textbf{(c)} Fluctuations in a noisy parameter $\gamma$ (red), representing single-qubit phases for example, can overshadow the estimation of a stable parameter $\phi$ (black) in a simple gate repetition scheme that provides estimates depending on both of these parameters (purple).
    Note that a separate, and necessarily noisy, estimation of $\gamma$ is therefore required to find $\phi$.
    In contrast, MEADD provides a direct estimate of the stable parameter $\phi$ as illustrate on the bottom, thus yielding more precise and accurate characterization information.
    }
    \label{fig:schematic}
\end{figure*}

Scalable quantum computation suffers from systematic errors, such as over-rotation and crosstalk.
Characterizing these errors often requires coherently amplifying them by repeating the same gate sequence $\nn$ times in a quantum circuit, a core principle of many schemes~\cite{kimmel_robust_2015,rudinger_experimental_2017,arute_observation_2020,neill_accurately_2021,nielsen_gate_2021}.
This approach also allows one to distinguish errors in the gate of interest from state preparation and measurement (SPAM) errors; while gate errors accumulate linearly in $\nn$, SPAM errors only introduce a constant offset that is independent of $\nn$.
However, in practice, noise in certain nuisance parameters often overshadows small, systematic coherent errors of interest, limiting the extent to which they can be amplified and characterized precisely.
In particular, frequency-tunable superconducting qubits suffer from fluctuations and drift in qubit frequencies~\cite{chiorescu_coherent_2003, quintana_observation_2017}, imposing a noise floor that limits the precision of standard characterization approaches.

In this work, we introduce Matrix-Element Amplification using Dynamical Decoupling (MEADD) – a protocol to amplify coherent errors for their characterization, even in the presence of low-frequency noise.
This is achieved by interleaving the gate of interest with carefully chosen single-qubit gates that implement a dynamical decoupling (DD) sequence~\cite{carr_effects_1954,meiboom_modified_1958,viola_dynamical_1999,khodjasteh_fault_2005,uhrig_keeping_2007}, selecting a single target parameter by echoing away the noisy and irrelevant parameters.
We experimentally demonstrate its accuracy, reliability, and efficiency for characterizing single- and two-qubit gates in superconducting qubits.

When only a few parameters need to be characterized, MEADD saves time by focusing on the specified parameters. Learning one parameter at a time makes data processing much easier and more reliable, especially when there are unmodeled errors.
When measuring small off-diagonal transitions, MEADD also eliminates costly scans needed for identifying the resonance conditions, i.e., phase matching the transition matrix elements so that they add up constructively in a gate sequence.
This gives MEADD greater efficiency in the number of experiments to run, which need only scale logarithmically in the maximum circuit depth, compared to similar noise-robust methods based on signal processing~\cite{dong_beyond_2022}, which scale linearly in circuit depth.
In contrast to a hardware-agnostic method like Gate Set Tomography (GST)~\cite{nielsen_gate_2021}, MEADD is a special-purpose method that leverages knowledge of the noise and systematic errors in the hardware, which is often important to achieving high-quality results and efficient implementations.
We expect that insights making MEADD useful in characterizing superconducting qubits can be applied to other qubit platforms, including leveraging the most precise control parameters for estimating less precise ones, isolating stable parameters from less stable and noisy ones, and targeting only a few parameters at a time. That being said, it will likely require domain expertise.

The MEADD protocol combines ideas from several established gate-characterization techniques.
Like Robust Phase Estimation (RPE)~\cite{kimmel_robust_2015, rudinger_experimental_2017} and Floquet Characterization (FC)~\cite{arute_observation_2020, neill_accurately_2021}, MEADD measures eigenvalue differences, which can be amplified by repeating the same gate sequence $\nn$ times.
If the procedure works, the measured total phase fits a linear function of $\nn$ well. The eigenvalue difference can be extracted from the slope of the linear function, while SPAM errors only contribute to the offset.
The additional ingredient of MEADD over FC is to use DD sequences to ensure the eigenvalue difference only depends on one gate parameter at a time, much like what is done in Hamiltonian learning~\cite{wang_hamiltonian_2015} and Hamiltonian error amplifying tomography~\cite{sundaresan_reducing_2020}.
This isolation simplifies data processing, making it easy to debug and relatively robust to processes such as stray couplings and leakage.

An essential requirement of MEADD is robustness to systematic errors in the DD implementation, such as over rotations.
These errors can also be amplified coherently, ultimately spoiling the estimation of the intended parameter. In~\cref{sec:robustness}, we present a systematic approach to constructing DD sequences such that these types of errors cancel each other to the first order, ensuring accuracy and reliability in practice.
Such constructions depend on the gate of interest, and MEADD is most suitable for characterizing gates whose errors can be thought of as perturbations of the ideal gate.
Therefore, a typical use of MEADD is to refine gates that are already calibrated with less precise but more versatile methods, such as Cross-Entropy Benchmarking (XEB)~\cite{boixo_characterizing_2018,arute2019quantum}.
In practice MEADD can tolerate gates with angle errors at least as large as several hundred milliradians, as one can see by the distributions of gate angles successfully characterized in Supplementary Information~\ref{sec:cz-parameters}.

We illustrate the MEADD protocol using the \CZ\ gate as an example in~\cref{fig:schematic}. To describe systematic errors in implemented \CZ\ gates, we consider an arbitrary excitation-preserving two-qubit unitary $\WW$ that is native to our hardware~\cite{foxen2020}. Up to an overall phase, such a unitary can be parameterized as
\begin{multline}
\label{eq:nc_2q_u}
\WW = 
\begin{pmatrix}
e^{i\xssp\gamma}    & 0    & 0   & 0\\[3pt]
0 & e^{-i\xssp\zeta} \cos\theta &  -i\, e^{i\xssp\chi}\sin \theta & 0\\[3pt]
0 & -i\, e^{-i\xssp\chi}\sin \theta &e^{i\xssp\zeta} \cos\theta& 0\\[3pt]
0 & 0 & 0 & e^{-i(\gamma + \phi)}
\end{pmatrix}\,,
\end{multline}
where $0\leq\theta\leq \pi/2$ is the swap angle, $\chi$ the swap azimuthal phase, $\phi$ the controlled phase, $\zeta$ the 10-01 Z phase, and $\gamma$ the 00-11 Z phase. Here, we use the lexicographically ordered basis $\{\ket{00},\ket{01},\ket{10},\ket{11}\}$, where the left bit corresponds to the left qubit in tensor products and the top wire in circuit diagrams. The ideal \CZ\ gate takes the parameters
\begin{align}\label{eq:CZ}
\theta = \zeta = \gamma = 0\,,\quad \phi = \pi\,,
\end{align}
and $\chi$ is indefinite for $\theta = 0$. The swap angle $\theta$ and controlled phase $\phi$ determine the entangling properties of the two-qubit gate. The swap azimuthal phase $\chi$ can be regarded as a basis change of the two-qubit unitary, corresponding to a relative Z rotation of the two qubits; therefore, it must be considered in reference to those of the surrounding gates. The Z phases $\zeta$ and $\gamma$ are especially susceptible to flux noise and drift in the control electronics present in our system. This ultimately hinders one's ability to characterize precisely the entangling gate parameters, as shown schematically in~\cref{fig:schematic}(c).

\begin{figure*}[t]
    \centering
    \includegraphics[width=2\columnwidth]{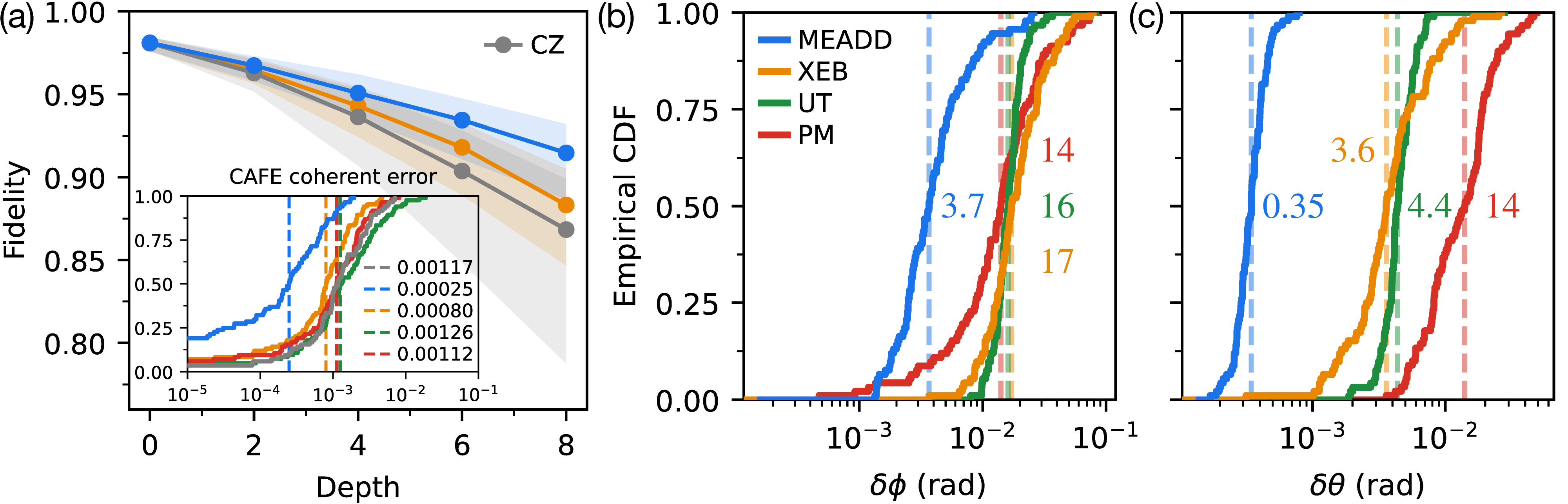}
    \caption{
    Experimental results of MEADD on $\CZ$ gates (excitation-conserving two-qubit gates close to the ideal \CZ). All data are collected from a 69-qubit Google Sycamore processor, see Supplementary Information in~\cite{arute2019quantum} for details of the processor.
    \textbf{(a)} Fidelities obtained using the Context-Aware Fidelity Estimation (CAFE) protocol~\cite{cafe_paper_2023}.
    These fidelities indicate how well a particular unitary characterization corresponds to the implementation of the $\CZ$ gate, and are presented for
    unitaries characterized by MEADD (blue), Cross-Entropy Benchmarking (XEB, orange), and the perfect $\CZ$ unitary (gray).
    Solid lines with markers are the median over 85 different $\CZ$ gates of a single Sycamore device, and the shaded regions are the inner quartile regions.
    A higher CAFE fidelity corresponds to a more accurate unitary characterization of the gate, see text for details.
    Inset shows the empirical cumulative distribution functions (CDF) of the coherent errors inferred from the CAFE curves of every $\CZ$ gate, with median values reported in the legend.
    Also shown are the results obtained using the unitary tomography (UT, green) and the phase method (PM, red) characterization protocols.
    \textbf{(b)} Empirical CDFs of run-to-run standard deviations in the characterized controlled phase $\phi$, and \textbf{(c)} swap angle $\theta$, measured over 12 sequential characterizations of the 85 $\CZ$ gates.
    The different protocols were interleaved with one another over several hours to average out device inconsistencies.
    The median standard deviations of the different methods are reported in milliradian in the plots.
    }
    \label{fig:experimental-results-cz}
\end{figure*}

In~\cref{sec:characterize_two_qubit_gates}, we discuss how to use MEADD for characterizing general two-qubit excitation-preserving gates, including \CZ, $\iSWAP$, and $\sqrt{\mathrm{iSWAP}}$.

\section{Results and discussion}

In this section, we discuss how to characterize both single- and two-qubit gate parameters using MEADD. Remarkably, MEADD can estimate the entangling parameters of \CZ\ gates with 5x and 10x higher precision than previous methods for the controlled-phase and swap angle, respectively.
This characterization identifies systematic errors in the implemented quantum operations which can then be corrected for by adjusting the control parameters.
In addition, we present experimental results of MEADD on characterizing unwanted crosstalk couplings that were too small to measure previously.

\subsection{Characterizing CZ}
\label{sec:cz}

% \subsubsection{Experimental results}

Before describing the MEADD protocol, we first benchmark it against other methods with the \CZ\ gates calibrated using our existing procedure, which can be regard as excitation-preserving two-qubit gates~\eqref{eq:nc_2q_u} close to the ideal \CZ. In~\cref{fig:experimental-results-cz}, we compare MEADD to cross-entropy benchmarking (XEB)~\cite{arute2019quantum}, unitary tomography (UT)~\cite{foxen2020} and the phase method (PM)~\cite{neill_accurately_2021}.
XEB uses random circuits to scramble errors into an effective depolarizing channel and infers unitary parameters by fitting the observed outcome probabilities to those predicted by the unitary in \cref{eq:nc_2q_u} combined with depolarizing noise.
Unitary tomography measures observables whose expectation values correspond to the real and imaginary parts of the matrix elements of an excitation-preserving two-qubit gate.
This is achieved by using the eigenstate $\ket{00}$ as a phase reference.
The phase method uses the same initial states and observables as unitary tomography, while repeating the gate to be characterized along with additional Z phases for various cycle numbers to amplify coherent errors.
By varying these Z phases, one can determine all the matrix elements of the gate of interest by merely estimating the eigenvalue differences of the cycle unitary.

In~\cref{fig:experimental-results-cz}(a), we use a recently introduced protocol named Context-Aware Fidelity Estimation (CAFE)~\cite{cafe_paper_2023} to assess the accuracy of the different characterization methods, where the average gate fidelity between the learned unitary and the physical gate implemented is estimated.
CAFE estimates this fidelity from an ensemble of circuits with an average expectation value dependent only on the fidelity.
Much like direct fidelity estimation~\cite{flammia_direct_2011} and randomized benchmarking~\cite{knill_randomized_2008}, this provides an estimate of fidelity that is simultaneously independent of any characterization scheme and more efficient than process tomography.
Using different numbers of gate repetitions $n$ gives the protocol robustness to SPAM errors and allows for error budgeting in terms of incoherent and coherent contributions, as these errors build up linearly and quadratically with $n$ to lowest order~\cite{cafe_paper_2023}.

The unitaries extracted with MEADD outperform all other unitary characterization techniques as measured by CAFE. 
Specifically, looking at the inset of~\cref{fig:experimental-results-cz}(a), we see that MEADD captures significantly more of the coherent error budget (80 percent of the difference to the ideal gate), resulting in a median of $2.5\times10^{-4}$ uncharacterized coherent error across the 85 $\CZ$ gates characterized, 3-5x smaller than the other methods.
The remaining 20 percent unaccounted coherent error derives from statistical noise and qubit-frequency fluctuations that CAFE does not echo away.
We note that a variant of CAFE with DD gates, called DECAF~\cite{cafe_paper_2023}, can provide a more fine-grained benchmark of errors that excludes qubit-frequency fluctuations, though we exclusively cite CAFE numbers in this work to demonstrate the holistic utility of using MEADD.

\begin{figure}[t]
    \centering
    \includegraphics[width=1\columnwidth]{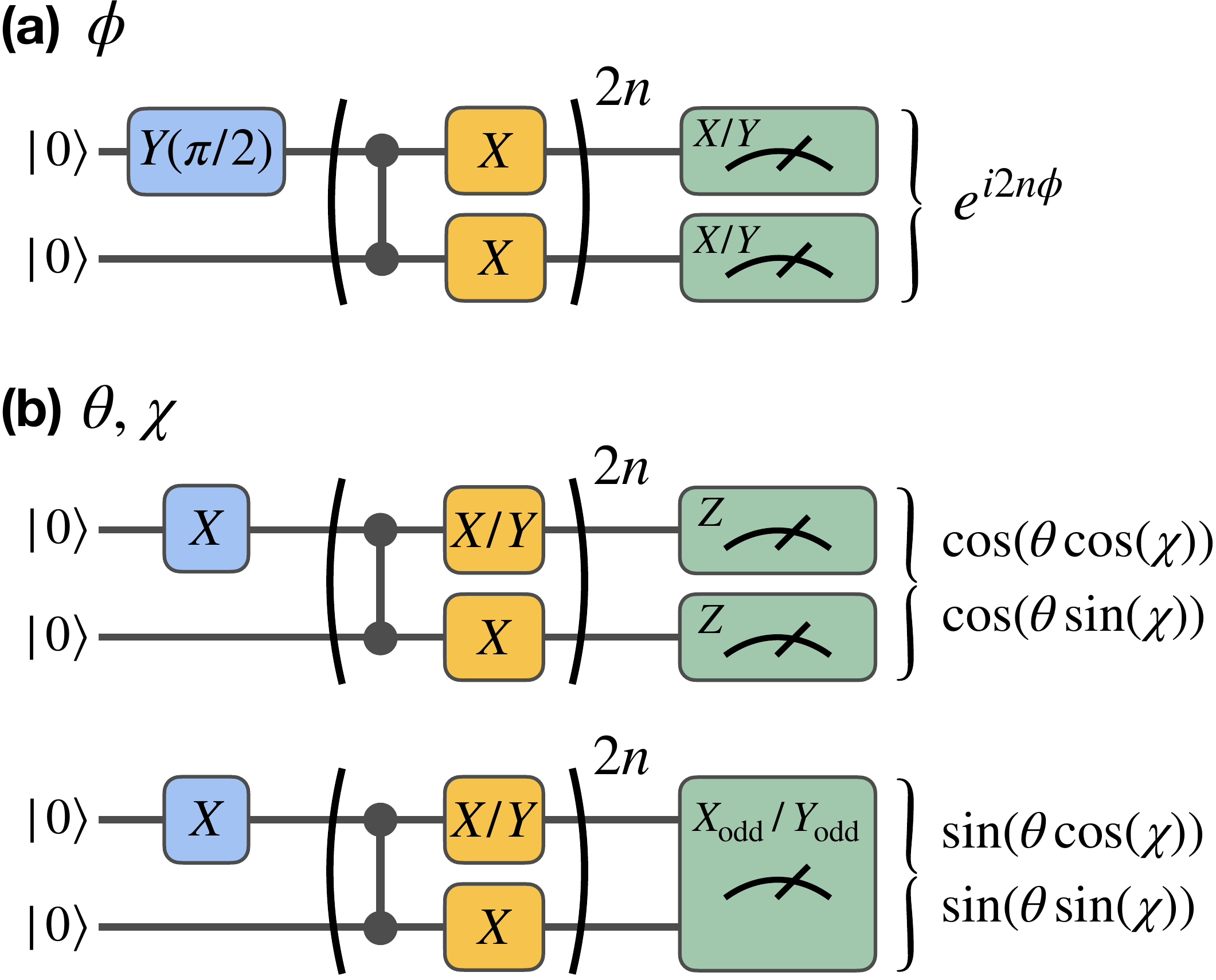}
    \caption{MEADD circuits used to characterize the entangling parameters of $\CZ$ gates.
    \textbf{(a)} The circuits are composed of three steps. Prepare $\ket{+0}$ with a microwave gate (blue), or similarly $\ket{0+}$ (not shown), interleave $X\otimes X$ dynamical decoupling gates (yellow) with the repeated $\CZ$ gates for even repetitions $2n$, and finally measure both qubits in the $X$ and $Y$ bases (green). As explained in the main text, combining the resulting measurement bits for different depths $n$ yields an estimate of the controlled phase $\phi$ that is robust to both low-frequency noise and DD gate imperfections. 
    \textbf{(b)} Similarly for the swapping angle $\theta$ and phase $\chi$, we prepare $\ket{10}$ (or $\ket{01}$, not shown), use $X\otimes X$ and $Y\otimes X$ as DD gates, and measure either in the computational basis of both qubits (top circuit) or in the Bell basis of the odd parity subspace (bottom circuit), where $X_\mathrm{odd}=\ket{10}\!\bra{01} + \ket{01}\!\bra{10}$ and $Y_\mathrm{odd}=i\ket{10}\!\bra{01} -i \ket{01}\!\bra{10}$.
    }
    \label{fig:meadd-circuits}
\end{figure}

\begin{figure}[t]
    \centering
    \includegraphics[width=0.84\columnwidth]{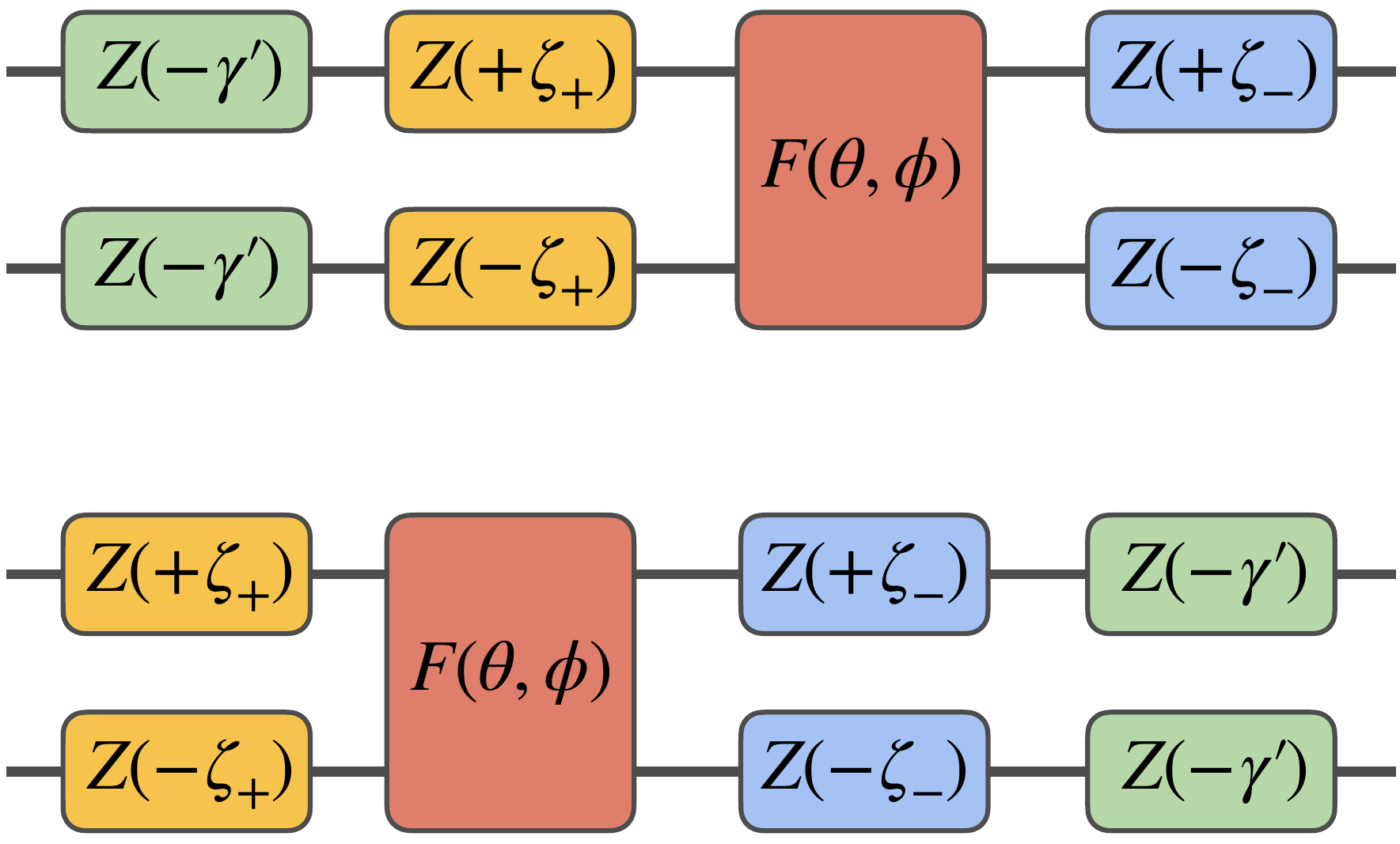}
    \caption{
        Two equivalent ways to decompose the arbitrary excitation-preserving two-qubit gate defined in Eq.~\eqref{eq:nc_2q_u} into single-qubit Z rotations and the fundamental entangler $\FF(\theta,\phi) = e^{-i\theta(X\otimes X + Y\otimes Y)/2-i\phi Z\otimes Z/4}$.
        While the differential phases $\zeta_\pm = (\zeta\pm\chi)/2$ do not commute with $\FF(\theta,\phi)$, the symmetric phase $\gamma'= \gamma+\phi/2$ does and can be placed either before or after the entangling operation.
    }
    \label{fig:excite-preserve-kak}
\end{figure}

In~\cref{fig:experimental-results-cz}(b-c), we compare the precision of MEADD to the other methods by repeating the different unitary characterization techniques and reporting the run-to-run standard deviation of the estimated controlled phase $\phi$ and swap angle $\theta$ for all 85 $\CZ$ gates of the device.
MEADD provides significantly more consistent characterizations than any other method, with 5x and 10x improvements over XEB for the controlled phase $\phi$ and swap angle $\theta$, respectively.
Remarkably, MEADD achieves a $3.4\times10^{-4}$~rad standard deviation for estimating the residual swap angle $\theta$ of our $\CZ$ gates.
Such precision, below one milliradian, promises to be a key asset in the calibration of stable high fidelity two-qubit gates as required for scalable quantum error correction applications.
The actual values of $\theta$ and $\phi$ characterized by the various protocols are presented in Supplementary Information~\ref{sec:cz-parameters}.

% \subsubsection{Steps in characterizing CZ with MEADD} 

Having presented the benchmarking results for our \CZ\ gates, we now overview the four main steps of the MEADD protocol:
\begin{enumerate}
\item Precalibrate the \CZ\ gates to be close to the ideal \CZ, with a more generic method.
\item Characterize the controlled phase $\phi$ using the circuits illustrated in~\cref{fig:meadd-circuits}(a).
\item  Characterize the swap angle $\theta$ and azimuthal phase $\chi$ using the circuits illustrated in~\cref{fig:meadd-circuits}(b).
\item Characterize the Z phases $\zeta$ and $\gamma$ using the method introduced in our previous work~\cite{arute_observation_2020,neill_accurately_2021}, also see Supplementary Information~\ref{sec:floquet}.
\end{enumerate}

\Cref{sec:characterize_two_qubit_gates} covers the details behind characterizing $\phi$ and $\theta$.
Measuring the expectation values $\langle X\rangle+i\langle Y\rangle$ on each qubit for the different initial states gives one the matrix elements for the evolution matrix (non-unitary due to decoherence) in 
the single-excitation (odd-parity) subspace.
Taking the determinant of these matrices and fitting the complex angle as a function of $\nn$ to a line allows one to extract $\phi$ from the slope.

Estimation of $\theta$ can be broken down into two independent measurements of the real and imaginary part of the swap matrix element $e^{i\xssp\chi}\sin \theta$.
The real and imaginary parts are selected by choosing the relative phase of the DD gates that echo away the unwanted part. 
By preparing a single-excitation state and measuring both population transfer between the two qubits and coherence in different Bell bases, one extracts both Bloch-vector components in the plane of rotation.
Fitting the angle of this Bloch vector as a function of $\nn$ allows one to extract the swap around each axis from the slopes.

Supplementary Information~\ref{sec:floquet} details how to extract the Z phases $\zeta$ and $\gamma$ by getting matrix elements from the expectation values $\langle X\rangle+i\langle Y\rangle$ like is done for characterizing $\phi$, though using circuits without DD.
The Z phase $\gamma$ is extracted via fitting the complex angle of the determinant of the evolution matrix in the odd-parity subspace, while $\zeta$ is extracted by fitting phases of the singular-value decomposition of the same matrix.

For the analysis to carry through in the DD circuits for measuring $\theta$ and $\phi$, the $\CZ$ gate must be repeated an even number of times.
The robustness analysis of \cref{sec:robustness} reveals that microwave errors are cancelled to first order when the $\CZ$ repetitions are a multiple of 4, though the fact that these errors do not coherently build up means that simply using multiples of 2 works in practice.
For the circuits characterizing $\zeta$ and $\gamma$, one can use arbitrary repetitions of the $\CZ$ gate.

Choosing the maximum number of repetitions in practice depends on the coherence of the device and precision of the microwave gates, see~Supplementary Information~\ref{sec:parameter_estimation_decoherence}.
One can evaluate this by running a single experiment with many depths and observing when the data starts deviating significantly from linearity.
We find that a maximum depth of 14 repetitions on our device yields reliable data and produces the precision we cite in this work.
The same arguments advanced for RPE~\cite{kimmel_robust_2015, rudinger_experimental_2017} for using logarithmically-spaced depths apply to our method, though for modest maximum depths taking linearly spaced data adds little overhead and produces datasets that are more informative for diagnosing failure modes.

\begin{figure*}[t]
    \centering
    \includegraphics[width=1.0\textwidth]{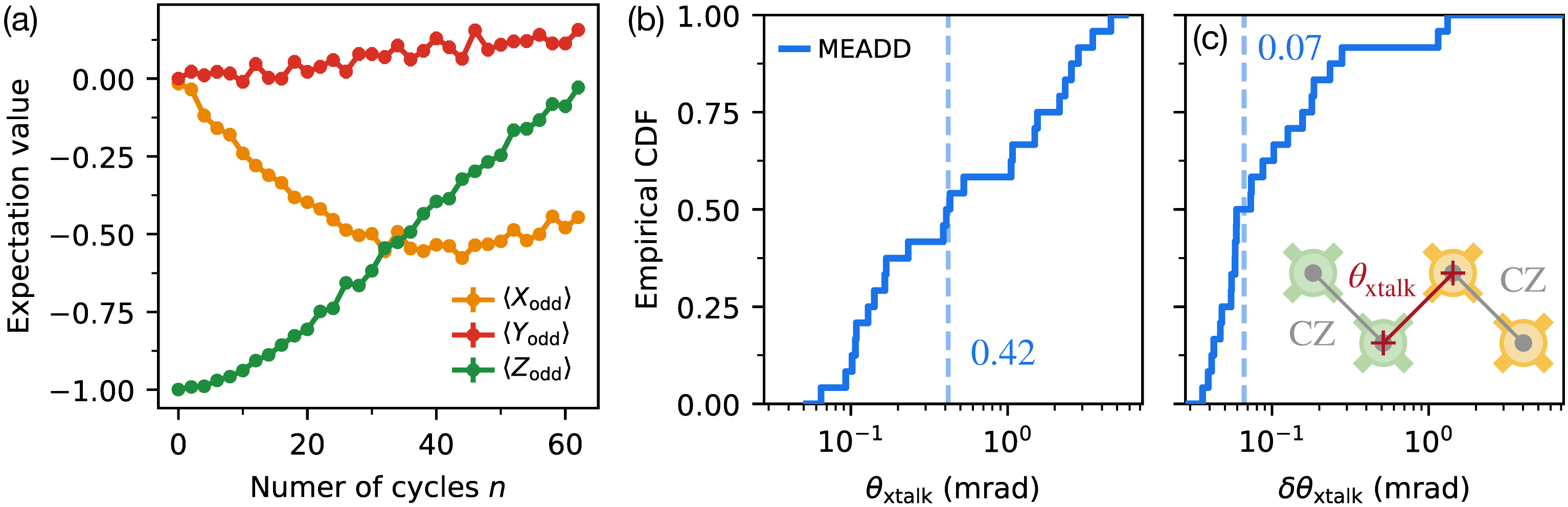}
    \caption{
    Experimental results of MEADD for characterizing crosstalk.
    \textbf{(a)} Evolution of the components of the Bloch vector in the odd-parity sector, $X_\mathrm{odd}=\ket{10}\!\bra{01} + \ket{01}\!\bra{10}$, $Y_\mathrm{odd}=i\ket{10}\!\bra{01} -i \ket{01}\!\bra{10}$, and $Z_\mathrm{odd}=\ket{01}\!\bra{01} - \ket{10}\!\bra{10}$, for two neighboring qubits involved in $2n$ parallel $\CZ$ gates. The corresponding crosstalk angle is about \SI{4}{\milli\radian} per cycle.
    Error bars correspond to the standard deviation over 2000 shots and are smaller than the data markers.
    Distributions of \textbf{(b)} the median crosstalk swap angle $\theta_\mathrm{xtalk}$ and \textbf{(c)} the run-to-run standard deviation in this estimate after repeating the measurement 10 times over one hour for 24 neighboring $\CZ$ gates on the Sycamore processor, revealing a median signal 6 times larger than the median noise.
    As shown in the inset, the \SI{34}{\nano\second} $\CZ$ gates (gray) are performed simultaneously on neighboring qubit pairs (green and yellow), and we use MEADD to characterize the residual swap angle produced by these operations, yielding $\theta_\mathrm{xtalk}=(0.42\pm0.07)~\SI{}{\milli\radian}$ on median.
    }
    \label{fig:crosstalk}
\end{figure*}

% \subsubsection{Calibrating the CZ pulses} \label{sec:pulses}

After learning the deviations of the \CZ\ gate parameters, we can correct them by fine tuning the control pulses in our quantum processors. Here, we briefly discuss this procedure. The exact details are beyond the scope of this work, as they are necessarily implementation dependent and often involve making trade offs among multiple parameters~\cite{barends_diabatic_2019,foxen2020}.  

One can separate the entangling parameters $\theta$ and $\phi$ from the rest using a special case of the KAK decomposition~\cite{drury_constructive_2008}, as shown in \cref{fig:excite-preserve-kak}.
Therefore, we can zero out the Z phases $\zeta$ and $\gamma$ by applying compensating single-qubit Z rotations before and after the \CZ\ gate. This can be implemented by simply changing the phases of the microwave pulses, without introducing any additional flux pulses, see virtual single-qubit Z rotations in Sec.~\ref{sec:single-qubit_gates}. We can also zero out the swap azimuthal angle $\chi$ using the same method. However, it is inconsequential for $\theta\ll 1$. 

Correcting $\theta$ and $\phi$ involves adjusting the pulses used for generating entanglement between the qubits.
Our diabatic \CZ\ gates are performed by a triplet of control pulses~\cite{barends_diabatic_2019}.
Two flux pulses on both qubits bring the two interacting levels 02 and 11 into resonance.
At the same time, a third pulse on the coupler activates a tunable excitation-preserving interaction between the two qubits.
The controlled phase $\phi$ is sensitive to the detuning between the two interacting levels controlled by the first two pulses, while the swap angle $\theta$ can be fine tuned by adjusting the rise time of the third pulse.
Synchronized minimization of the leakage and swap errors can be achieved by tuning the hold time (duration of the plateau) and height of the third pulse.

\subsection{Crosstalk}
\label{sec:crosstalk}

In our hardware, crosstalk between qubits can arise due to unwanted charge-charge (capacitive) couplings, which results in a qubit-subspace Hamiltonian of the form
\begin{align}\label{eq:crosstalk_h}
  H_\mathrm{xtalk} &\propto  e^{i\chi}\,\sigma_+\otimes\sigma_- +\hc,
\end{align}
where $\sigma_+ = \sigma_-^\dag = \ket{0}\!\bra{1}$ and we neglect the counter-rotating terms. This coupling produces an excitation swapping between the two qubits with angle $\theta_\mathrm{xtalk}$. These terms are typically on the order of $\SI{100}{kHz}$, much smaller than the detuning of the two qubits on the order of $\SI{100}{MHz}$. Therefore, they are suppressed and often hard to measure. The errors this incurs for typical gate times (10s of nanoseconds) yields average gate infidelities between $\SI{1e-5}{}$ and $\SI{1e-4}{}$, which while small can be particularly damaging due to their nonlocal nature within error-correcting circuits~\cite{google2023suppressing}.

A typical strategy for measuring $\theta_\mathrm{xtalk}$ is to tune the qubits to resonance.
Since the qubit frequency uncertainty is typically on the order of $\SI{100}{kHz}$, such a strategy requires scanning over the qubit-frequency control to find the resonance, which is time-consuming to perform in practice.
In contrast, we can use MEADD to directly estimate the unwanted crosstalk without any qubit frequency tuning.

Experimental results are shown in~\cref{fig:crosstalk}. 
In (a), we plot data with an unambiguous signal for a crosstalk swapping angle on the order of \SI{1}{\milli\radian}. 
We can fit this data to reliably characterize the angle with a precision below \SI{0.1}{\milli\radian}, as shown in panel (c), which corresponds to an uncertainty on the effective swapping coupling strength of less than \SI{0.5}{\mega\hertz}.
Note that this protocol can readily be used to characterize long-range crosstalk across distant qubits on the processor.

This kind of crosstalk cannot be calibrated away by simply tuning the control pulses, and has to be tackled at the hardware level.
That being said, we can minimize their effect by putting the involved qubits at frequencies that are far apart from each other during idling and gates.

\subsection{Single-qubit gates}
\label{sec:single-qubit_gates}

Leveraging the hierarchy of precision in the different controls available on a quantum device is central to designing highly accurate, reliable, and efficient characterization protocols. To that end, we present the considerations for calibrating single-qubit gates in tunable-frequency superconducting qubits, highlighting the principal noise sources therein~\cite{quintana_observation_2017, vool_introduction_2017, krantz_quantum_2019, blais_circuit_2021}. 

We first describe the way that single-qubit gates are implemented in our hardware. Understanding these details is crucial to designing efficient characterization protocols for our device. Consider a qubit driven by a microwave pulse, where we ignore leakage errors into energy levels lying outside of the qubit subspace.
Under the rotating wave approximation (RWA), the Hamiltonian for this two-level system is ($\hbar=1$)
\begin{align}\label{eq:rwa_Hamiltonian}
 H_\text{\scriptsize RWA}(t) = -\frac{1}{2}\,\omega_{01}(t)\,\sigmaZ + \pigl(\xmsp\Omega(t)\, e^{-i \omegap t}\xssp \sigma_- + \hc\pigr)\xssp,
\end{align}
where $\sigmaZ=\proj{0}-\proj{1}$, $\ket{0}$ and $\ket{1}$ representing the ground and first excited states of the qubit, respectively.
The terms with $\sigma_- = \ket{1}\!\bra{0}$ and its Hermitian conjugate $\sigma_+= \ket{0}\!\bra{1}$ capture the microwave control field, where $\Omega(t)$ is an envelope function with bandwidth \SI{\sim 100}{\mega\hertz} and $\omegap/2\pi$ is the nominal microwave frequency, sitting at several \SI{}{\giga\hertz}.
The parameter $\omega_{01}(t)$ is the angular frequency of the qubit, adjustable via the SQUID loop flux.
Ideally, $\omega_{01}(t)$ matches $\omegap$ during idle periods and rotations about axes in the XY plane.
However, noise in the flux control line and surrounding electromagnetic environment can cause $\omega_{01}(t)$ to drift, leading to various errors when performing qubit operations.
This can be seen directly by moving to the interaction frame of the microwave frequency $\ket{\psi_I(t)} = e^{-i\xssp \omegap t\xssp \sigmaZ/2}\ket{\psi(t)}$, which eliminates the rapidly oscillating phase $e^{-i \omegap t}$ in \cref{eq:rwa_Hamiltonian} as follows
\begin{align}
  H_\text{\scriptsize I}(t) &= e^{-i\xssp \omegap t\xssp\sigmaZ/2} H_\text{\scriptsize RWA}(t)\, e^{i\xssp \omegap t\xssp \sigmaZ/2}  + \half\, \omegap\xssp\sigmaZ\\[3pt]
  &= -\half\,\delta\omega_{01}(t)\,\sigmaZ + \pigl(\xmsp\Omega(t)\,\sigma_- + \hc\pigr)\xssp.
  \label{eq:int_hamiltonian}
\end{align}
Here, one can directly see how any non-zero detuning $\delta\omega_{01}(t) = \omega_{01}(t) - \omegap \ll \omegap$ can lead to unwanted $\sigmaZ$ rotations.
The resulting unitary produced by this drive can be formally written as
\begin{align}\label{eq:propagator}
 \UU_I = \mathcal T  \exp\biggl(- i \int_{t_i}^{t_f} H_I(t)\, \dif t\biggr) \,,
\end{align}
where $\mathcal T$ is the time ordering operator and $t_i$ ($t_f$) the initial (final) time of the pulse.

We now discuss how to implement quantum logical circuits in the above interaction picture, i.e. control $H_I(t)$ such that $\UU_I$ corresponds to the desired quantum gate.
Note that as we always initialize and measure the qubits in the Z basis, there is no need to implement the unitary $e^{-i\xssp \omegap t\xssp \sigmaZ/ 2}$ for the frame change before the final measurement.
However, any gate operation that is not diagonal in the Z basis needs to account for the frame change, a process known as phase tracking or matching.

\subsubsection{Single-qubit XY rotations (microwave gates)} 
An arbitrary $\mathrm{SU}(2)$ matrix can be parameterized using three angles as
\begin{align}\label{eq:single-qubit_gate}
\RR(\mu, \zeta, \chi) &= 
\begin{pmatrix}
e^{-i\zeta} \cos\frac{\mu}{2} &  -i\, e^{i\chi}\sin\frac{\mu}{2}  \\[5pt]
-i\, e^{-i\chi}\sin\frac{\mu}{2} & e^{i \zeta} \cos\frac{\mu}{2}
\end{pmatrix}\,,
\end{align}
where we have recycled some Greek letters used in the single-excitation subspace of the two-qubit gate~\eqref{eq:nc_2q_u}, which shares the same $\mathrm{SU}(2)$ structure.
This parametrization directly leads to an Euler decomposition of arbitrary single-qubit gates as an X rotation sandwiched by two Z rotations:
\begin{align}\label{eq:euler_decomposition}
\RR(\mu, \zeta, \chi) &= 
\RZ(\zeta-\chi)\, \RX(\mu)\, \RZ(\zeta+\chi)\,,
\end{align}
where
\begin{align}
    X(\varphi)
    &=
    e^{-i\varphi X/2}\;,
    &
    Z(\varphi)
    &=
    e^{-i\varphi Z/2}
    \,.
    \label{eq:single-qubit-gate}
\end{align}
The rotation $\RX(\mu)$ can be achieved by applying a microwave pulse to the qubit, e.g., $\Omega(t) = \mu\, \ff(t)$, where $\ff(t)$ is the shifted cosine function
\begin{align}
\ff(t) =  \biggl\{\begin{array}{cl}
        \frac{1}{2 T} \bigl(1 + \cos\frac{2\pi t}{T}\bigr), & \text{for $-\frac{T}{2}\leq t\leq \frac{T}{2}$}\,,\\[6pt]
        0, & \text{for $\norm{t} > \frac{T}{2}$}\,,
        \end{array}
\end{align}
where the pulse time $T$ is around \SI{15}{\nano\second} in our system.
As such, the parameter $\mu$ of the single-qubit gate is directly determined by the amplitude of the microwave drive, which can be controlled to a high precision using standard arbitrary waveform generators.

More generally, we consider a rotation along an arbitrary axis in the XY plane, which we notate as a \textit{phased-X} rotation
\begin{align}
  \RXY(\mu, \vartheta) \equiv \RZ(\vartheta)\, \RX(\mu)\, \RZ(-\vartheta)\,,
\end{align}
where $\mu$ and $\vartheta$ specify the angle and axis of the rotation, respectively.
It can be implemented by simply changing the phase of the microwave pulse 
\begin{align}
  \Omega(t) = \mu\xmsp e^{i\vartheta} \ff(t)\,.
\end{align}
This microwave phase can be used to control $\chi$ much more precisely than the uncertainties on the qubit phase, making it an ideal control knob to use in designing characterization protocols.
Additionally, the freedom we have in the microwave phase allows us to eliminate one Z rotation from our single-qubit gate
\begin{align}\label{eq:zrot_post}
\RR(\mu, \zeta, \chi) 
&= \RZ(2\zeta)\xmsp \RXY(\mu, -\zeta-\chi) \,,
\end{align}
which can save time and reduce errors.

In practice, we additionally make corrections to the pulse, such as compensating for the AC Stark effect by introducing a detuning $\Delta (\mu)$,
\begin{align}\label{eq:modified_pulses}
 \Omega(t) = \mu\xmsp e^{i\vartheta} e^{-i \Delta (\mu)\xmsp t }\ff(t)\,,
\end{align}
which can also be used to adjust the gate parameter $\zeta$.
Finally, we make use of Derivative Removal by Adiabatic Gate (DRAG)~\cite{motzoi_simple_2009} pulses to mitigate leakage errors, which consist of mixing the original pulse $\Omega(t)$ with its first time derivative;
see Supplementary Information~\ref{sec:drag} for a simple derivation.
Impedance mismatches in the control lines can reflect microwave pulses, which cause delayed tails after the main pulses.
Here we assume that these tails do not bleed into the entangling pulses such as to preserve the independence of consecutive gates.

\begin{figure*}[t]
    \centering
    \includegraphics[width=2\columnwidth]{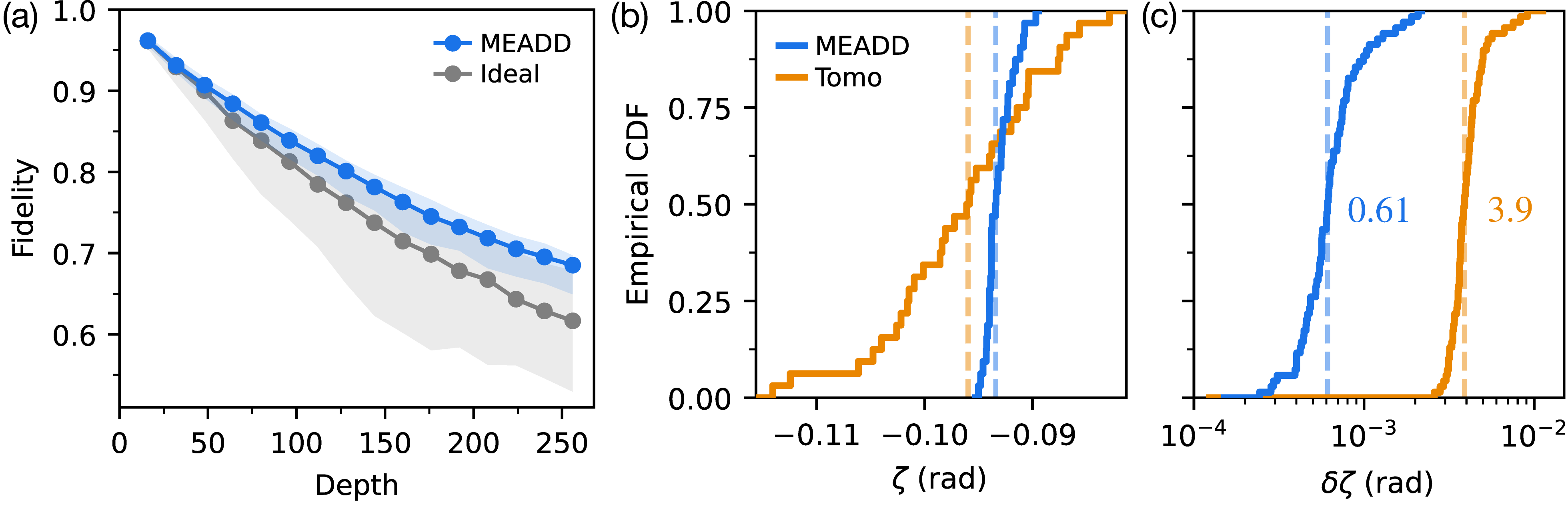}
    \caption{
    Experimental results of MEADD for interleaved single-qubit $X(-\pi/2)$ and $X(\pi)$ gates.
    \textbf{(a)} CAFE results for the single-qubit unitaries characterized by MEADD (blue) compared to assuming a perfect gate (gray) for 69 qubits of a single Sycamore device. The cycle circuit being repeated in CAFE is the same as in the MEADD protocol, namely a $X(-\pi/2)$ rotation followed by a $X(\pi)$ rotation, performed in parallel on all qubits.
    Solid lines with markers are the median over the 69 qubits, and the shaded regions are the inner quartile regions.
    \textbf{(b)} Distributions of the characterized relative rotation axis parameter $\zeta=\chi_{\pi/2}-\chi_\pi$ defined in~\cref{eq:parameters_compositeXY} for a representative single-qubit gate characterized 32 times interleaved between MEADD (blue) and 4-axis tomography (orange), using the same number of measurements for both methods.
    \textbf{(c)} Distributions of run-to-run standard deviations in $\zeta$ estimates for the two methods over the 32 sequential characterizations of the 69 qubits.
    The median standard deviations of the two methods are reported in milliradian.
    }
    \label{fig:experimental-results-sq}
\end{figure*}

\subsubsection{Physical and virtual single-qubit Z rotations}  
Rotations around the Z axis can be implemented simply by moving the qubit frequency away from $\omegap$ for some duration using a flux pulse, while leaving $\Omega(t) = 0$. This can be seen directly by considering the evolved qubit state in the interaction frame under the propagator of~\cref{eq:propagator}
\begin{align}
\ket{\psi_I(t_f)} = e^{i\xssp \varphi(t_f,\,t_i)\xmsp \sigmaZ/2}\ket{\psi_I(t_i)}\,,
\end{align}
where the Z-rotation angle reads
\begin{align}
 \varphi(t_f,\,t_i) = \int_{t_i}^{t_f} \delta\omega_{01}(t)\,\dif t\,. 
\end{align}
This type of Z rotation typically takes \SI{\sim 20}{\nano\second} to implement on our hardware (including a \SI{\sim 10}{\nano\second} pulse and two \SI{5}{\nano\second} paddings before and after the pulse), regardless of the angle, with a corresponding gate infidelity of about $10^{-3}$.
However, these physical Z pulses typically have nonzero tails lasting much longer than the gate time~\cite{rol_fast_2019,negirneac_high_2021,johnson_controlling_2011}, on the order of hundreds of \SI{}{\nano\second} in our devices, which can disturb many subsequent operations.

For quantum circuits consisting of only single-qubit gates and diagonal two-qubit gates, such as the \CZ~gate, the Z rotations can be implemented in software without actually sending flux pulses to the qubits~\cite{krantz_quantum_2019}.
This is because Z rotations commute with diagonal two-qubit gates, and when commuted through a microwave gate, only change its phase $\chi$, which we can control with high precision relative to physical Z rotations.

In more detail, one can push single-qubit Z rotations from the beginning of the circuit down---through any two-qubit gates---and combine them with other Z rotations along the way until they hit single-qubit gates that are not Z rotations. One can then use Eq.~\eqref{eq:zrot_post} to put the Z rotations after the XY rotations, and repeat the same procedure until all the single-qubit Z rotations are pushed to the end of the circuit. Finally, they can be directly eliminated because the qubits are measured in the Z basis. The above process also works for two-qubit gates that are diagonal up to a $\SWAP$, such as $\iSWAP$. However, physical Z rotations are still required for most other two-qubit gates, and they are typically applied before every two-qubit pulses in these cases.

\subsubsection{Relative axes of XY rotations}  
Consider a circuit consisting of only XY rotations and diagonal two-qubit gates, such as \CZ. One can implement the following gauge transformation on all the XY rotations without affecting the measurement results
\begin{align}
\chi_j \rightarrow \chi_j + \text{constant}, 
\end{align}
where $\chi_j$ is the rotation axis of the $j$-th XY gate. 
Therefore, the absolute value of $\chi$ has no physical significance and cannot be measured using such circuits. The relative $\chi$ between two different XY rotations, however, is physical and needs to be calibrated. For example, we consider the relative $\chi$ between a $\pi$ gate and a $\pi/2$ gate, which is nonzero due to effects such as ac-Stark shifts imparting $\chi$ differently for different values of $\mu$. Suppose one has a well calibrated $-\pi/2$ pulse, meaning $\mu=-\pi/2$ and $\zeta$ has been removed through virtual Z phase updates,
\begin{align}
\SS(-\pi/2,0,\chi_{\pi/2}) = \RZ(-\chi_{\pi/2})\, \RX(-\pi/2)\, \RZ(\chi_{\pi/2})\,,
\end{align}
and a $\pi$ pulse with well calibrated rotation angle $\mu$
\begin{align}
\SS(\pi,0,\chi_{\pi}) = \RZ(-\chi_\pi)\, \RX(\pi)\, \RZ(\chi_\pi)\,.
\end{align}
One can characterize the relative phase $\chi_{\pi/2}-\chi_\pi$ by considering the composite gate:
\begin{align}
\SS&(\pi,0,\chi_{\pi})\xssp \SS(-\pi/2,0,\chi_{\pi/2})\\
    &=  \RZ(\chi_{\pi/2}-2\chi_\pi)\,\RX(\pi/2)\,\RZ(\chi_{\pi/2})
    \\[2pt]
    & = \SS(\pi/2,\zeta,\chi) \,,
\end{align}
where
\begin{align}\label{eq:parameters_compositeXY}
    \zeta
    &=
    \chi_{\pi/2}-\chi_\pi\,,
    \qquad
    \chi
    =
    \chi_\pi
    \,.
\end{align}
The relative phase is now $\zeta$ of the composite gate, and can be estimated using our characterization circuits.

In~\cref{fig:experimental-results-sq}, we present experimental results for interleaved single-qubit $X(-\pi/2)$ and $X(\pi)$ gates over the 69 qubits of the device. This can be regarded as an example of MEADD for single-qubit gates, where low-frequency flux noise is echoed away by the $X(\pi)$ gates.
In panel~(a), we compare the CAFE results obtained using the unitaries constructed from the MEADD estimates (blue) of the relative axis $\zeta=\chi_{\pi/2}-\chi_\pi$ in~\cref{eq:parameters_compositeXY} to the ones assuming perfect gates (gray).
The higher fidelities obtained from the MEADD unitaries indicate that these characterizations accurately describe the implemented gates.
In panels (b) and (c), we compare the MEADD characterization results to a direct 4-axis tomography protocol which consists of preparing an equatorial state, applying an $X(\pi)$ pulse, and then measuring expectation values of $X$ and $Y$.
Since the state preparation and measurements use $X(-\pi/2)$ gates, this protocol measures the relative axes of the $X(\pi)$ and $X(-\pi/2)$ gates just like MEADD with interleaved $X(-\pi/2)$ and $X(\pi)$ gates does, as described at the end of \cref{sec:characterize_single_qubit_gates}.
MEADD achieves a median run-to-run standard deviation of $6.11\times10^{-4}$~rad on estimating $\zeta$, a 6x improvement over the tomography protocol, which is the single-qubit equivalent of unitary tomography.
This is illustrated in panel (b) for a representative qubit, showing the tight distribution of repeated MEADD characterizations.
We do not compare with XEB, as the single-qubit implementation only reports the gate fidelity, not a unitary characterization.

\section{Methods}\label{sec:methods}

In this section, we first discuss the theoretical framework for characterizing single-qubit gates using matrix-element amplification.
We then extend this framework to characterizing general two-qubit excitation-preserving gates, including \CZ, $\iSWAP$, and $\sqrt{\mathrm{iSWAP}}$.
Finally, we discuss how to construct MEADD protocols that are robust to implementation errors in the DD gates.

\subsection{Theory on characterizing single-qubit gates}
\label{sec:characterize_single_qubit_gates}

Having presented the experimental results of MEADD, we now describe the theory behind the protocol, starting with the single-qubit version.
The single-qubit protocol bears similarity to RPE~\cite{kimmel_robust_2015, rudinger_experimental_2017}, though we are able to simplify the protocol by taking advantage of the high precision with which microwave phases can be set relative to other errors in the system, as discussed in \cref{sec:single-qubit_gates}.

For the arbitrary single-qubit gate $\SS(\mu,\zeta,\chi)$ in \cref{eq:single-qubit_gate}, we set
$\mu = 2\theta$ to draw an explicit connection to the two-qubit parametrization~\eqref{eq:nc_2q_u}, 
\begin{align}
\SS(\mu,\zeta,\chi) &= I\cos\Omega(\theta, \zeta) -i\, \sigma(\theta, \zeta, \chi) \sin\Omega (\theta, \zeta) \,,
\end{align}
where
\begin{align}
\label{eq:rabi-angle}
\cos\Omega
&=
\cos\theta\cos \zeta\,,\quad \Omega \in [\theta, \pi-\theta]\,,
\end{align}
and the spin-$1/2$ operator 
\begin{align}
\sigma
&=
\frac{\sin\theta}{\sin \Omega}\Bigl(X\cos\chi - Y \sin\chi  + Z \cot\theta\sin\zeta\Bigr)
\,.
\end{align}
In matrix form, we have
\begin{align}
\sigma(\theta, \zeta, \chi) = \begin{pmatrix}
\cos \polar &  e^{i\chi}\xssp \sin \polar  \\[4pt]
e^{-i\chi}\xssp\sin \polar & -\cos \polar 
\end{pmatrix}\,,
\end{align}
where the polar angle $\polar$ is defined by
\begin{align}
\polar = \arccot (\cot \theta \sin\zeta)\,,\quad \polar\in [\theta, \pi-\theta]
\,.
\end{align}
For example, we have $\Omega = \pi/2$, $\polar = \pi/2$,  and $\chi =0$ for the ideal $X(\pi)$ gate.

We can measure the eigenvalues $\pm\Omega(\theta,\zeta)$ of $R$ precisely by preparing and measuring uniform superpositions of the eigenstates of $R$, 
\begin{align}
\ket{\sigma^+}
&=
e^{i\chi/2}\xssp\cos (\varrho/2)\, \ket{0} + e^{-i\chi/2}\xssp \sin (\varrho/2)\, \ket{1}
\\[2pt]
\ket{\sigma^-}
&=
e^{i\chi/2}\xssp\sin (\varrho/2)\, \ket{0} - e^{-i\chi/2}\xssp \cos (\varrho/2)\, \ket{1}\,,
\end{align}
where $\sigma\ket{\sigma^\pm}=
\pm\ket{\sigma^\pm}$.
We parametrize these superpositions by the relative phase $\tau$ between them
\begin{align}
 \ket{\Phi(\tau)} = \frac{1}{\sqrt 2} \pigl(e^{-i\tau/2}\xssp\ket{\sigma^+} +  e^{i\tau/2}\xssp \ket{\sigma^-}\pigr)
 \,.
\end{align}
After applying the cycle unitary $n$ times to the initial state, we have
\begin{align}
 \SS^\nn \ket{\Phi(\tau_0)} = \ket{\Phi(\tau_0 + 2\nn\Omega)}\,.
\end{align}
We then measure the probability of finding $ \ket{\Phi(0)}$ ,
\begin{align}
P_0 &= \normb{\braket{\Phi(0)}{\Phi(\tau_0 + 2n\Omega)}}^2\\[3pt]
&= \frac{1 + \cos (\tau_0 + 2\nn\Omega)}{2}\,,
\end{align}
and the probability to find $ \ket{\Phi(\pi/2)}$ is
\begin{align}
P_{\pi/2} &= \normb{\braket{\Phi(\pi/2)}{\Phi(\tau_0 + 2n\Omega)}}^2\\[3pt]
&= \frac{1 + \sin (\tau_0 + 2\nn\Omega)}{2}\,.
\end{align}
Estimating these two measurement probabilities for a sequence of repetitions will give us the measurement record
\begin{align}
    2P_0-1+i(2P_{\pi/2}-1)
    &=
    \exp\big[i(\tau_0+2\nn\Omega)\big],
\end{align}
from which we can extract $\Omega$ by fitting the complex argument as a linear function of repetitions.
This procedure is robust to SPAM errors.
In particular, the initial states can be prepared assuming ideal microwave gates, since errors in these gates result in non uniform superpositions of the eigenstates $|\sigma^\pm\rangle$ that merely reduce the contrast of the oscillations in $P_0$ and $P_{\pi/2}$.

To extract the gate parameters $\mu$ and $\zeta$, we perform a set of similar experiments, where we add different virtual Z rotations between the gates, giving us unitaries
\begin{align}
    Z(z)\xssp\SS(\mu,\zeta,\chi)
    &=
    Z(\zeta+z-\chi)\,X(\mu)\,Z(\zeta+\chi)
    \\[2pt]
    &=
    \SS\bigl(\mu,\zeta+z/2,\chi-z/2\bigr)\,.
\end{align}
By measuring $\Omega(\theta,\zeta+z/2)$ as a function of $z$, one can fit the curve to the form given in \cref{eq:rabi-angle}, obtaining $\mu$ and $\zeta$.
Note this technique is identical to the technique for extracting two-qubit gate parameters within the single-excitation subspace given in~\cite{arute_observation_2020}.
Like there, the angle $\chi$ does not manifest in the eigenphase $\Omega(\theta,\zeta)$, and so we cannot coherently amplify errors in that parameter without additional interventions. In other words, $\chi$ corresponds to a gauge transformation
\begin{align}
\SS(\mu,\zeta,\chi)^\nn = \RZ(-\chi)\, \SS(\mu,\zeta,0)^\nn\, \RZ(\chi)\,,
\end{align}
which does not add up when the same gate is repeated.
That being said, the relative axis of two different single-qubit gates can be estimated with amplified precision.
For example, we amplify the relative axis between $X(\pi)$ and $X(-\pi/2)$ defined in~\cref{eq:parameters_compositeXY} by interleaving them many times.
Any low-frequency noise in the qubit frequency that would cause phase to accumulate between the gates will be echoed away, since the time padding on either side of the $X(\pi)$ gate is the same, making the measurement of this relative phase robust to these fluctuations.

\subsection{Theory on characterizing two-qubit gates}
\label{sec:characterize_two_qubit_gates}

The entangling parameters $\phi$ and $\theta$ of our two-qubit gates in~\cref{eq:nc_2q_u} are more consistent over time than the single-qubit phases that are susceptible to low-frequency flux noise.
It would therefore be useful to design characterization circuits that are sensitive only to the stable entangling parameters. 
To that end, we introduce microwave gates that are interleaved with the repeated two-qubit gate, as illustrated in \cref{fig:schematic}.
To remove flux noise, MEADD is designed to be insensitive to the Z phases $\zeta$ and $\gamma$ in~\cref{eq:nc_2q_u}; we use an adapted Floquet characterization technique without DD to estimate these parameters, as described in Supplementary Information~\ref{sec:floquet}.

For the simple case of interleaving an excitation-preserving two-qubit unitary $\WW$ in~\cref{eq:nc_2q_u} with X gates, one obtains the cycle unitary
\begin{align}
    \CC = (X\otimes X) \xssp\WW\,.
\end{align}
A useful representation for the following discussion is given by decomposing the two-qubit unitary $\WW$ into the even and odd parity subspaces ${\mathcal H_\mathrm{even} \oplus \mathcal H_\mathrm{odd}}$,
\begin{align}
    \WW
    &=
    \big(e^{-i\phi/2}\,\WW_\mathrm{even}\big)\oplus \WW_\mathrm{odd} \label{eq:direct_sum_structure}
    \\[4pt]
    &=
    \left[
    e^{-i\phi/2}
    \begin{pmatrix}
        e^{i(\gamma+\phi/2)} & 0
        \\[3pt]
        0 & e^{-i(\gamma+\phi/2)}
    \end{pmatrix}
    \right]
    \nonumber
    \\[4pt]
    &\qquad
    \oplus
    \begin{pmatrix}
        e^{-i\xssp\zeta}\cos\theta & -i\,e^{i\xssp\chi}\sin\theta
        \\[3pt]
        -i\,e^{-i\xssp\chi}\sin\theta & e^{i\xssp\zeta}\cos\theta
    \end{pmatrix}
    \,.
\end{align}
This decomposition is a direct sum of two $\mathrm{SU}(2)$ unitaries with a relative phase $e^{-i(\phi/2)}$ between the even and odd parity sectors.
Conveniently, $X\otimes X$ decomposes into the direct sum $X\oplus X$ in this representation, giving the decomposition for the cycle unitary
\begin{align}\label{eq:direct_sum_C}
\CC = \big(e^{-i\xssp\phi/2}\xssp\CC_\mathrm{even}\big)\oplus \CC_\mathrm{odd}\,,
\end{align}
where $\CC_\mathrm{even} = X\xssp\ueven$ and $\CC_\mathrm{odd} = X\xssp\uodd$.
The 2-cycle unitary is trivial in the even-parity subspace
\begin{align}
\CC_\mathrm{even}^2 &= X\xssp\ueven\xssp X\xssp\ueven = I\,,
\end{align}
a feature we use to design our characterization protocol.

\subsubsection{Controlled phase}
For measuring $\phi$, we need to measure a relative phase between the even- and odd-parity subspaces.
Because the 2-cycle unitary is trivial on the even-parity subspace, we can choose any state such as $|00\rangle$ to serve as our even-parity reference.

To measure the relative phase $-\phi/2$ we will therefore prepare states of the form
\begin{align}
\ket{\psi_0}
&=
\frac{1}{\sqrt 2} \pigl(\ket{00} + \ket{\psi_{0,\xssp\mathrm{odd}}}\pigr)
\\
&=
\frac{1}{\sqrt 2} \pigl(\ket{00} + \alpha\ket{01} + \beta\ket{10}\pigr)
\,,
\end{align}
where $\ket{\psi_{0,\xssp\mathrm{odd}}}\in \mathcal H_\mathrm{odd}$.
Applying even powers of the cycle unitary to this initial state puts a phase on $\ket{00}$ and transforms the odd-parity part of the state separately
\begin{align}
\label{eq:even-cycle-state}
    C^{2\nn}\ket{\psi_0}
    &=
    \frac{1}{\sqrt{2}}
    \pigl(
    e^{-i\nn\phi}\ket{00}
    +\alpha'\ket{01}
    +\beta'\ket{10}
    \pigr)
    \,,
\end{align}
where
\begin{gather}
\label{eq:cphase-final-01-amplitude}
\alpha' = \alpha\xssp\langle01|\CC_\mathrm{odd}^{2\nn}|01\rangle
    +\beta\xssp\langle01|\CC_\mathrm{odd}^{2\nn}|10\rangle\,,\\[3pt]
\label{eq:cphase-final-10-amplitude}
\beta'  = \alpha\xssp\langle10|\CC_\mathrm{odd}^{2\nn}|01\rangle
    +\beta\xssp\langle10|\CC_\mathrm{odd}^{2\nn}|10\rangle\,.
\end{gather}

We will make separable measurements on the qubits to infer $\phi$, which one can understand by looking at the final density matrix $C^{2n}\ket{\psi_0}\!\bra{\psi_0}C^{2n\dagger}$ and taking partial traces.
Tracing out the right qubit gives
\begin{align}
    \label{eq:cphase-trace-out-right}
    \begin{pmatrix}
        1+|\alpha'|^2 & e^{-i\nn\phi}\beta^{\prime *}
        \\[3pt]
        e^{i\nn\phi}\beta' & |\beta'|^2
    \end{pmatrix}\,,
\end{align}
and tracing out the left qubit gives
\begin{align}
    \label{eq:cphase-trace-out-left}
    \begin{pmatrix}
        1+|\beta'|^2 & e^{-i\nn\phi}\alpha^{\prime *}
        \\[3pt]
        e^{i\nn\phi}\alpha' & |\alpha'|^2
    \end{pmatrix}
    \,.
\end{align}
Observe that by going between choosing $\alpha$ or $\beta$ to be 1 (and the other 0), one engineers the amplitudes $\alpha'$ and $\beta'$ to take on the values of all the odd-parity cycle matrix elements.
By preparing the states $\ket{0+}$ and $\ket{+0}$ and measuring $X$ and $Y$ on each qubit to reconstruct the off-diagonal reduced-density-matrix elements, one obtains the matrix of values $e^{i\nn\phi}C_\mathrm{odd}^{2\nn}$,
as illustrated in~\cref{fig:meadd-circuits}(a).
As pointed out earlier, $C_\mathrm{odd}$ is a special unitary, with determinant one, and we have
\begin{align}
    \det\pigl(e^{in\phi}C_\mathrm{odd}^{2\nn}\pigr)
    &=
    e^{i2\nn\phi}
    \det\pigl(C_\mathrm{odd}^{2\nn}\pigr)
    =
    e^{i2\nn\phi}
    \,.
    \label{eq:meadd-phi-determinant}
\end{align}
This procedure amplifies $\phi$ coherently $2\nn$ times, suppressing the variance of our estimates of $\phi$ by a factor of $\nn^2$.
The ability to repeat this measurement for multiple depths $\nn$ and fit the slope of the accumulated phase makes us robust to SPAM errors.
Fluctuations in single-qubit phases are also swept away; identically within the even-parity subspace, and to a large extent within the odd-parity subspace as well.
The extent to which they survive will be discussed in more detail when considering the swap angle $\theta$.
Decoherence errors are also mitigated, because they tend to affect only the magnitude of the determinant, not its phase.
Altogether, this simple protocol yields highly accurate and precise estimates of the controlled phase of the \CZ\ gates, as demonstrated experimentally in~\cref{fig:experimental-results-cz}.

\begin{figure*}[t]
    \centering
    \includegraphics[width=1.\textwidth]{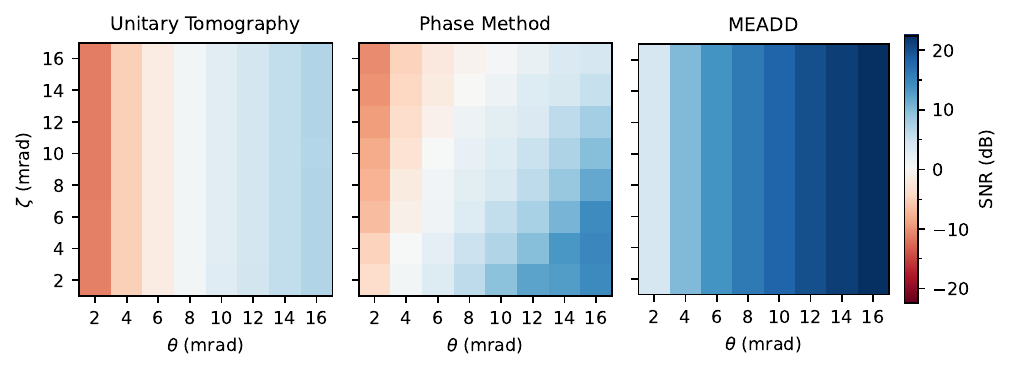}
    \caption{
    Numerical simulation results on signal to noise ratio (SNR) for measuring the swap angle.
    Variance is measured over 100 different realizations for each parameter point.
    The number of simulated measurements is normalized so that the total number of measurements per parameter inferred is the same across the methods (see Supplementary Information~\ref{sec:simulation-parameters} for details).
    The phase method \cite{neill_accurately_2021} has an advantage over unitary tomography in that it coherently amplifies errors. However, this amplification is suppressed as the value of the differential single-qubit phase $\zeta$ becomes comparable to $\theta$.
    In contrast, MEADD exhibits strong SNR through coherent amplification with no visible dependence on the detuning $\zeta$.}
    \label{fig:theta_snr_comparison}
\end{figure*}

\subsubsection{Swap angle and axis}
We now show how the circuits illustrated in~\cref{fig:meadd-circuits}(b) can be used to estimate $\theta$ and $\chi$.
The 2-cycle unitary in the odd-parity subspace $\{\ket{01}, \ket{10}\}$ reads
\begin{align}
\CC_\mathrm{odd}^2
&= X\xssp \uodd\xssp X\xssp\uodd \\[2pt]
&= \RZ(-\zeta)\, e^{-i\theta(\cos\nsp\chi X + \sin\nsp\chi Y)}\xssp \RZ(-\zeta)\nonumber\\
&\quad \times \RZ(\zeta)\, e^{-i\theta(\cos\nsp\chi X - \sin\nsp\chi Y)}\xssp \RZ(\zeta) \\[2pt]
&\approx  \RZ(-\zeta)\xssp \RX(4 \theta\nsp\cos\nsp\chi)\, \RZ(\zeta)
\,,\label{eq:c_odd}
\end{align}
where the approximation applies for $\theta\ll 1$, which is relevant when implementing controlled-phase gates like CZ (we discuss larger $\theta$ to conclude the section).
The dynamical decoupling sequence ensures that the only part of the swap that remains is around the X axis in the odd-parity sector.
The corresponding cycle Rabi angle $\Omega \approx \norm{\theta \cos\nsp\chi}$ can be estimated by observing the population transfer rate between $\ket{01}$ and $\ket{10}$.
To determine the swap around the Y axis, we also introduce the complementary cycle unitary
\begin{align}
\label{eq:complementary-cycle-unitary}
\overbar\CC &= (Y\otimes X) \xssp\WW = \overbar\CC_\mathrm{even}\oplus \overbar\CC_\mathrm{odd}\,,
\end{align}
where the second equation is a result of $Y\otimes X=Y\oplus Y$. For $\theta\ll 1$, we have
\begin{align}
\overbar\CC_\mathrm{odd}^2 &= Y \xssp \uodd\xssp Y\xssp\uodd \\[2pt]
&\approx \RZ(-\zeta)\xssp \RY(-4\theta\sin\nsp\chi)\xssp \RZ(\zeta)\,,
\end{align}
which can be used to estimate $\overbar\Omega \approx \norm{\theta \sin\nsp\chi}$.
Again, the estimation error of $\overbar\Omega$ due to finite sample sizes scales as $O(1/\nn)$, while the small-angle approximation introduces an additional deterministic error on top of that. Measuring only the population transfer between $\ket{01}$ and $\ket{10}$ leaves $\cos\chi$ and $\sin\chi$ with ambiguous signs. By additionally measuring in the Bell-type basis
\begin{align}
    \ket{\pm}_\mathrm{odd}
    &=
    \tfrac{1}{\sqrt{2}}\bigl(\ket{01}\pm \ket{10}\bigr)\,,
    \\[3pt]
    \ket{\pm i}_\mathrm{odd}
    &=
    \tfrac{1}{\sqrt{2}}\bigl(\ket{01}\pm i\ket{10}\bigr)\,,
\end{align}
one resolves these ambiguities and also mitigates the effect of dephasing errors, since one is effectively measuring the complete odd-parity Bloch vector and can distinguish shrinkage (such as comes from dephasing) from rotation.
Since both $\CC$ and $\overbar\CC$ preserve the parity, we can also mitigate single-qubit relaxation errors to first order by postselecting on parity.

For the case of large $\theta$, we can estimate $\theta$ and $\chi$ using the relations
\begin{gather}
\Omega = \arccos\bigl(1 - 2\xssp(\sin\theta \cos\nsp\chi)^2\bigr) \big/ 2\,,\\[2pt]
\overbar\Omega = \arccos\bigl(1 - 2\xssp(\sin\theta \sin\nsp\chi)^2\bigr) \big/ 2\,,
\end{gather}
where the Rabi angles $\Omega$ and $\overbar\Omega$ can be learned similarly as in the case where $\theta\ll 1$.

A subtle, but important, point for measuring unwanted swapping in superconducting qubits is that one must pay attention to how phase tracking is implemented for single-qubit XY rotations (see~\cref{sec:single-qubit_gates}).
For two-qubit gates with $\theta = 0$ (or $\pi$) such as \CZ, in normal phase tracking one ``commutes" Z phases coming from qubit idle evolution through the CZ gate to update the phases of subsequent microwaves.
However, the small swap matrix element to be measured does not commute with these Z rotations.
As pointed in~\cite{wei_characterizing_2024}, this causes it to acquire an artificial phase during phase tracking that is linear in the cycle number $\nn$, preventing it from being amplified and measured.
To keep the swap phase constant, one must modify phase tracking such that the microwave phases do not follow the qubit idle evolution during the circuit.
The same problem also exists in characterizing the crosstalk discussed in~\cref{sec:crosstalk}, and it needs to be tackled with care.

In~\cref{fig:theta_snr_comparison}, we compare in simulation MEADD to previously employed techniques for measuring the swap angle: the phase method~\cite{neill_accurately_2021} and unitary tomography~\cite{foxen2020}.
The phase method differs from unitary tomography primarily in that it repeats the gate being characterized to amplify errors.
As seen in \cref{fig:theta_snr_comparison}, this coherent amplification gives the phase method higher signal to noise ratio (SNR) than unitary tomography, even when keeping the total number of measurement results constant across both characterization methods.
However, one sees that the advantage of coherent amplification is washed out once $\zeta$ is on the same order as $\theta$. In comparison, MEADD exhibits significantly higher SNR and maintains the advantage from coherent amplification even when $\zeta$ is significantly larger than $\theta$, as the dynamical-decoupling sequence cancels the swap-suppressing effects of $\zeta$.

\begin{figure*}[t]
    \centering
    \includegraphics[width=2\columnwidth]{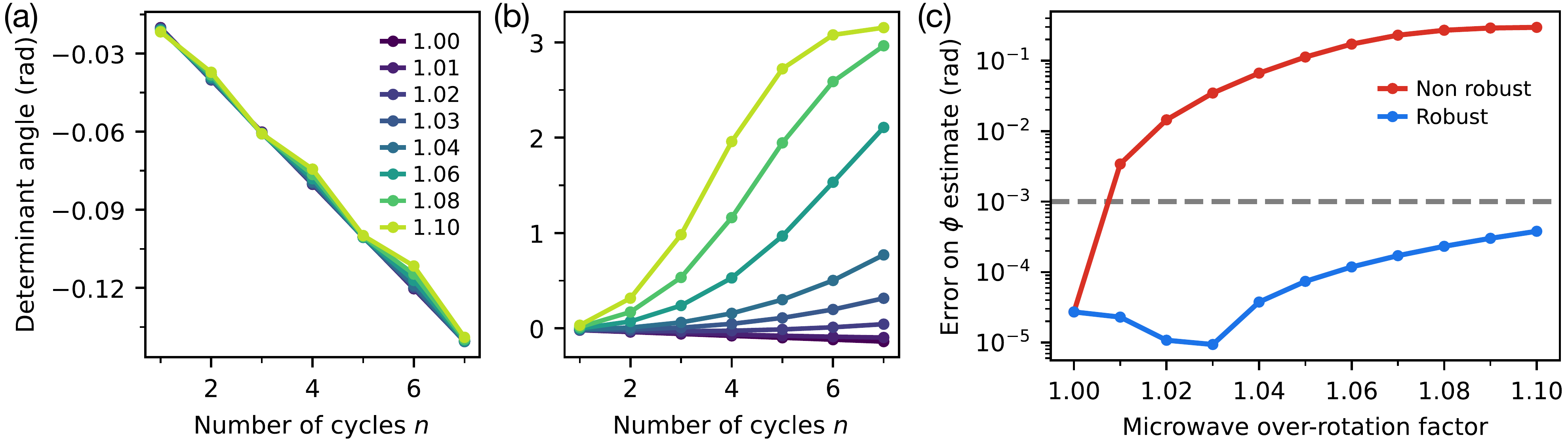}
    \caption{
    Numerical simulation results for characterizing the controlled phase of a $\CZ$ gate with $\phi=\pi-0.01$, using MEADD and imperfect DD gates.
    \textbf{(a)} Using the robust DD sequence described in~\cref{sec:characterize_two_qubit_gates}, the angle of the determinant in~\cref{eq:meadd-phi-determinant}, which corresponds to $-2n\phi$, builds up linearly even if we introduce microwave over-rotation errors on one of the qubits.
    The strength of the over-rotation, a factor multiplying the $\pi$ rotation around $X$, is varied between 0 and 10\% and illustrated with the different colors.
    \textbf{(b)} Similar simulations now using the XY4 DD sequence, which is \emph{not} robust to implementation errors in this protocol.
    \textbf{(c)} Resulting $\phi$-estimate error obtained from the fitting the slope of the MEADD data for the different over-rotation errors. Dashed line indicates millirad precision.
    }
    \label{fig:robustness_simulations}
\end{figure*}

\subsection{Robustness to implementation errors}\label{sec:robustness}
The dynamical decoupling sequences used by MEADD will unavoidably contain systematic errors, so we must design sequences that are robust to these imperfections.
It is well known that the so-called XY4 sequence $(XY)^{2\nn}$ is robust to any small systematic errors in $X$ and $Y$~\cite{maudsley_modified_1986,viola_dynamical_1999}, whereas the Carr-Purcell-Meiboom-Gill (CPMG) sequence $X^{2\nn}$ is susceptible to over-rotation errors in $X$~\cite{meiboom_modified_1958}.
Similar to XY4, we require our characterization sequence to be robust to any systematic errors in the single-qubit gates, though the exact robust sequence will depend on the two-qubit gate to be characterized.
These errors can lead to non-zero matrix elements between the even and odd-parity subspaces, which destroy the direct sum structure in Eq.~\eqref{eq:direct_sum_C} that we rely on.

In~\cref{fig:robustness_simulations}, we demonstrate the importance of using a robust DD sequence to characterize the entangling parameters of a $\CZ$ gate with MEADD.
We simulate the protocol to estimate the controlled phase detailed in~\cref{sec:characterize_two_qubit_gates}, both with the robust CPMG and the non-robust XY4 DD sequences; note how the robustness of the sequences is reversed when being interleaved with $\CZ$ gates.
Using the robust DD sequence in panel (a), we see that the phase builds up linearly even if we introduce microwave over rotation errors on one of the qubits.
Even for an over rotation of 10\% ($\SI{0.314}{rad}$), we still get linear phase accumulation to a good approximation that gives us an estimate of $\phi$ to sub milliradian precision, as shown in panel (c) with the blue curve.

On the other hand, using the \emph{non} robust DD sequence gives unusable data as we introduce microwave over rotation errors on one of the qubits, as shown in panel (b). For over rotation of just 2\% ($\SI{0.063}{rad}$), the phase accumulation is so distorted from being linear that the error in the $\phi$ estimate is larger than the value we are trying to measure ($\SI{0.01}{rad}$).
With this sequence, even 1\% over-rotation error introduces errors in the $\phi$ estimate larger than a milliradian, i.e. it is more than an order of magnitude more sensitive to implementation errors than the robust DD sequence.

In the following, we will first demonstrate how to design robust MEADD protocols for the case of a $\CZ$ gate, while general excitation-preserving two-qubit gates are addressed in Supplementary Information~\ref{sec:robustness_eptqg}.
We consider small arbitrary systematic errors of the X gate as
\begin{align}\label{eq:X_error}
    \widetilde X(\bm \epsilon)
    &=
    e^{-i(\epsilon_X X + \epsilon_Y Y + \epsilon_Z Z)} X
    \\
    &\approx
    (I-i\bm{\epsilon}\cdot\bm{\sigma})\xssp X
    \,,
\end{align}
where $\bm \epsilon = (\epsilonX,\,\epsilonY,\,\epsilonZ)$ and $\bm{\sigma} = (X,\, Y,\, Z)$.
Since the errors are assumed to be small, we will consider their effect to first order, which allows us to treat errors on the individual qubits separately in the following.

The perfect 2-cycle unitary for $\WW = \CZ$ reads
\begin{align}
\CC^2 &= (X\otimes X) \xssp\CZ\xssp (X\otimes X) \xssp\CZ\\
&= -Z\otimes Z\,.
\end{align}
Noticing that $C^4 = I\otimes I$, we have
\begin{align}\label{eq:c2_identity}
\bigl[(X\otimes X) \xssp\CZ\bigr]^3 &= (Y\otimes Y)\xssp \CZ
= \CZ\xssp (X\otimes X)\,.
\end{align}
Considering errors on the right qubit, the noisy 2-cycle unitary is
\begin{align}
    \widetilde \CC^2
    &=
    (X\otimes \widetilde X) \xssp\CZ\xssp (X\otimes \widetilde X) \xssp\CZ
    \\
    &\approx
    C^2 - i\pigl[
    (I\otimes\bm{\epsilon}\cdot\bm{\sigma})(X\otimes X) \xssp\CZ\xssp (X\otimes X) \xssp\CZ\xssp
    \nonumber\\
    &\qquad
    +(X\otimes X) \xssp\CZ\xssp (I\otimes\bm{\epsilon}\cdot\bm{\sigma})(X\otimes X) \xssp\CZ\xssp
    \pigr]
    \,,
\end{align}
With the identities in~\cref{eq:c2_identity}, we have
\begin{align}
\widetilde \CC^2 &= \VV\xssp \CC^2 \,,
\end{align}
where 
\begin{align}\label{eq:error_operator}
\VV = I -  i\pigl[
    (I\otimes\bm{\epsilon}\cdot\bm{\sigma})
    +\CZ\, (I\otimes Y\bm{\epsilon}\cdot\bm{\sigma}\, Y)\,\CZ \xssp
    \pigr]\,.
\end{align}
Since $X$ and $Y$ interact with the $\CZ$ gate in similar ways, we first consider $\epsilonX,\, \epsilonY\neq 0$ and $\epsilonZ = 0$, for which we have the anticommutation relations
\begin{gather}
\anticommute{(I\otimes\bm{\epsilon}\cdot\bm{\sigma})}{Z\otimes Z}= 0\,,\\[2pt]
\anticommute{\CZ\xssp (I\otimes Y\bm{\epsilon}\cdot\bm{\sigma} Y)\xssp\CZ}{Z\otimes Z}= 0\,.
\end{gather}
Both generator terms in the error unitary $\VV$ in~\cref{eq:error_operator} anticommute with ${\CC^2=-Z\otimes Z}$.
Therefore the noisy 4-cycle unitary $\widetilde \CC^4 = \VV\xssp\CC^2 \VV\xssp\CC^2$ is identical to the noiseless 4-cycle unitary $ \CC^4$ to first order,
and the sequence is robust to small errors in $X$ and $Y$ for a 4-cycle.

For $\epsilon_Z \neq 0$ and $\epsilonX = \epsilonY = 0$, we have
\begin{align}
\widetilde \CC^2
&= e^{-i\epsilon_Z (I \otimes Z - I \otimes Z)}\xssp \CC^2 = \CC^2\,,
\end{align}
and the error $\epsilon_Z$ cancels itself in a 2-cycle.
The complementary sequence $\overbar \CC$ from \cref{eq:complementary-cycle-unitary} is robust in exactly the same way, the only difference being the $Y$ operators in~\cref{eq:error_operator} may be replaced by $X$, which does not affect their anticommutation with $\overbar{\CC}^2 = -Z\otimes Z$.

One subtlety here is the gauge freedom related to the position assigned to Z errors around microwave gates.
It turns out that it can be totally ignored here because one can always commute half of $\epsilon_Z$ to the right of the X gate, leaving a pair of $\epsilon_Z/2$ and $-\epsilon_Z/2$ errors conjugating the $\CZ$ gate.
This gauge has thus no impact at all on the characterization of the controlled phase $\phi$ or swap angle $\theta$, but is indistinguishable from a change in the swapping phase $\chi$.
For the current case of $\CZ$ characterization this is acceptable, as for an ideal $\CZ$ gate there is no swapping between the qubits ($\theta=0$), at which point $\chi$ has no meaning.

We use MEADD to measure the crosstalk (stray coupling) by treating the Hamiltonian~\cref{eq:crosstalk_h} integrated over a given time period as a gate, and executing the dynamical decoupling sequence to measure the small swap angle, exactly as one does for $\CZ$ gates.
The main difference is that, because there is no $\CZ$ gate interleaving the microwaves, we use XY4 sequences on both qubits, which are robust to microwave over rotation errors.

\section{Supplementary Information}
Supplementary Information is available for this paper. 

\section{Data Availability}
The data that support the findings of this study are available upon request from the authors. 

\section{Code Availability}
The simulation code that support the results of this study is available upon request from the authors. 

\section{Competing Interests}
The authors declare no competing interests.

\section{Author contribution}
JG and ZJ conceived the original idea and developed the theory. JG, EG, and ZC carried out the experiments with help from DD and WM. JG and EG implemented the numerical simulations with help from YZ. ZJ supervised the project. All authors discussed the results and contributed to the final manuscript.

\section{Acknowledgements}
We are grateful to the Google Quantum AI team for building, operating, and maintaining the software and hardware infrastructure used in this work.
We thank Abraham Asfaw, Sergio Boixo, Yu Chen, Ben Chiaro, Michel Devoret, Vinicius S. Ferreira, Evan Jeffrey, Amir Karamlou, Julian Kelly, Mostafa Khezri, Paul V. Klimov, Alexander Korotkov, Kenny Lee, Will Livingston, Xiao Mi, Charles Neill, Murphy Yuezhen Niu, Will Oliver, Andre Petukhov, and Vadim N. Smelyanskiy for discussions and feedback on the draft, and Andreas Bengtsson, Jimmy Chen, Catherine Erickson, and Sabrina Hong for software support.

\putbib[characterization]
\end{bibunit}
\onecolumngrid
\newpage

\renewcommand\appendixname{}
\renewcommand\appendixpagename{\vspace*{\fill}\centering\Large{Supplementary Information for ``Characterizing Coherent Errors using Matrix-Element Amplification"}\vspace*{\fill}}
\makeatletter
\def\p@subsection{}
\makeatother
\renewcommand*{\thesubsection}{\Alph{section}\arabic{subsection}}
\renewcommand*{\thesubsubsection}{}
\renewcommand\thefigure{S\arabic{figure}}    
\setcounter{figure}{0}
\setcounter{equation}{0}
\makeatletter
\newcommand*{\newbibstartnumber}[1]{%
  \apptocmd{\thebibliography}{%
    \global\c@NAT@ctr #1\relax
    \addtocounter{NAT@ctr}{-1}%
  }{}{}%
}
\makeatother
\newbibstartnumber{1}

\begin{appendices}
%\appendixpage

\begin{bibunit}[apsrev4-1_with_title]
% \tableofcontents
% \title{Supplementary Information for}

\vspace{.5em}

\renewcommand{\author}[2]{#1$\!^\textrm{\scriptsize #2}$}
\renewcommand{\affiliation}[2]{$^\textrm{\scriptsize #1}$ #2 \\}
\newcommand{\Google}{1}
\newcommand{\Sherbrooke}{2}
\newcommand{\UMD}{3}
\newcommand{\xGoogle}{\affiliation{\Google}{Google Quantum AI, Venice, CA 90291, USA}}
\newcommand{\xSherbrooke}{\affiliation{\Sherbrooke}{Institut quantique \& D\'epartement de Physique, Universit\'e de Sherbrooke, Quebec J1K2R1, Canada}}
\newcommand{\xUMD}{\affiliation{\UMD}{Department of Physics \& Joint Quantum Institute, University of Maryland, MD 20742, USA}}

{\centering
\author{Jonathan A. Gross,}{\Google} 
\author{\'{E}lie Genois,}{\Google, \Sherbrooke}
\author{Dripto M. Debroy,}{\Google}
\author{Yaxing Zhang,}{\Google}\\[2pt]
\author{Wojciech Mruczkiewicz,}{\Google}
\author{Ze-Pei Cian,}{\Google, \UMD}
\author{and Zhang Jiang\hspace{2.5pt}}{\Google}

\vspace{.8em}

{\xGoogle}
{\xSherbrooke}
{\xUMD}}

\vspace{2em}

\twocolumngrid
\section{Floquet characterization for two-qubit gates}
\label{sec:floquet}

For complete characterization of the two-qubit gate, we need the single-qubit phases $\zeta$ and $\gamma$ in addition to the entangling parameters $\theta$, $\chi$, and $\phi$.
To accomplish this, we use a method very similar to Floquet characterization~\cite{arute_observation_2020}.
We are able to simplify the protocol by removing the variable Z rotations since we can use our precise estimate of $\theta$ to infer $\zeta$ from the energy splitting in the odd-parity sector.

We repeat the unitary for many different depths $\nn$, using preparation and measurement analogous to what is used to arrive at \cref{eq:cphase-trace-out-right,eq:cphase-trace-out-left} and shown in \cref{fig:meadd-circuits}, only this time we are directly measuring the matrix elements of the gate unitary $\uodd$ instead of $C_\mathrm{odd}$, which is interleaved with DD gates.
Computing the complex angle of the determinant at each depth $\nn$ allows us extract $\gamma$ from the slope exactly as we extracted $\phi$ from the slope in \cref{eq:meadd-phi-determinant}.

To extract $\zeta$, we use another technique for mitigating the effects of decoherence.
The singular value decomposition of the experimentally measured matrix $M$ (which in the ideal case is simply $\uodd$) give us
\begin{align}
    M
    &=
    U_M\Sigma_MV_M^\dagger
    \,.
\end{align}
Just as we took the complex angle of the determinant to mitigate decoherence effects, we will take the unitary part of $M$ to mitigate decoherence effects:
\begin{align}
    M^\prime
    &=
    U_MV_M^\dagger
\end{align}
In the case of interest where swapping is small, the eigenstates of $M^\prime$ can be unambiguously associated with $\ket{01}$ and $\ket{10}$, allowing us to consistently define the sign as well as the magnitude of the rotation angle of this unitary.
Fitting the rotation angle as a function of depth $\nn$ allows us to extract the rotation angle per gate $\Omega$.
Finally, we extract $\zeta$ using the relation $\cos\Omega=\cos\zeta\,\cos\theta$ and $\sgn\zeta=\sgn\Omega$:
\begin{align}
    \zeta
    &=
    \sgn\Omega\,\arccos(\cos\Omega/\cos\theta)
    \,.
\end{align}
There is a danger here that noise causes $\cos\Omega$ to exceed $\cos\theta$, in which case we set $\zeta=0$.

This method leverages the amplification and SPAM robustness of repeating the unitary to measure $\gamma$ and $\Omega$, while making use of our precise measurement of $\theta$ to ultimately extract $\zeta$ (something that had to be accomplished with additional experiments using physical Z rotations in Floquet Characterization~\cite{arute_observation_2020} and Phase-method~\cite{neill_accurately_2021} characterizations).

\section{Simulation parameters}
\label{sec:simulation-parameters}

In \cref{fig:theta_snr_comparison} of the main text, for each of the 100 executions at a particular parameter point, unitary tomography and phase method collected 160,000 measurement results.
Since the phase method uses four different repetition numbers (1, 2, 4, and 8), this meant only collecting 2,500 samples per circuit as opposed to 10,000 samples per circuit for unitary tomography.
Phase method uses two different Z rotations for each repetition number, without doing measurement flipping, while unitary tomography does not add any extra Z phases, but does do measurement flipping, so at each depth the number of circuits is the same.
We did not simulate realistic asymmetric readout noise, so the measurement flipping is essentially just giving twice as many samples at that point.
The MEADD circuits used in \cref{fig:theta_snr_comparison} only measures two of the five parameters ($\theta$ and $\chi$), so the number of measurement results were lowered to 64,000.

\section{Characterized CZ entangling parameters}\label{sec:cz-parameters}
\begin{figure*}
    \centering
    \includegraphics[width=.8\textwidth]{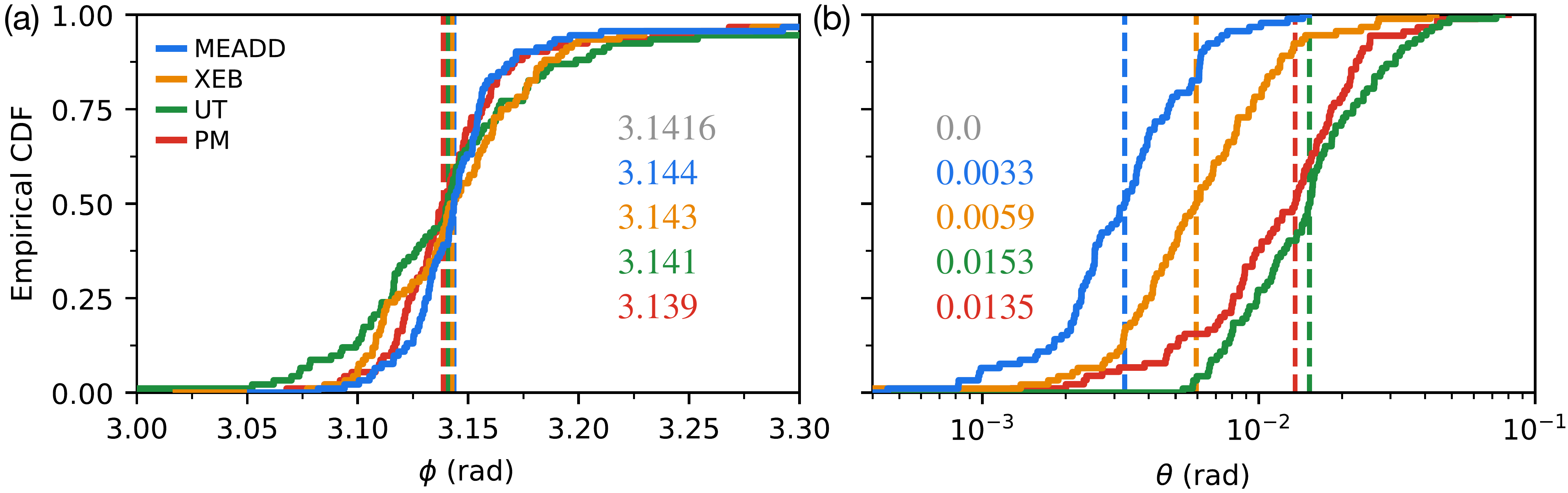}
    \caption{Characterized $\CZ$ entangling parameters. \textbf{(a)} Empirical CDFs of the conditional phase $\phi$ and \textbf{(b)} swap angle $\theta$ for the same 85 $\CZ$ gates as in~\cref{fig:experimental-results-cz} of the main text, when averaged over the 12 sequential repetitions.
    Results are shown for MEADD (blue), cross-entropy benchmarking (orange), unitary tomography (green) and the phase method (red).
    The median gate parameters of the different methods are reported in radian in the plots, with the targeted $\CZ$ parameters in gray.
    }
    \label{fig:app-cz-theta-phi}
\end{figure*}

We present the entangling $\CZ$ gate parameters characterized by the different methods in~\cref{fig:app-cz-theta-phi}.
We observe tight distributions around the targeted $\phi=\pi$ and $\theta=0$ parameters, thus demonstrating that the gates implemented here are close to the targeted $\CZ$ operations.
Notably, the significantly higher precision of MEADD allows the protocol to resolve the small residual swap angle in the experimental gates much better than existing approaches. 

\section{Robustness of MEADD for arbitrary excitation-preserving two-qubit gates}
\label{sec:robustness_eptqg}

Recall from \cref{fig:excite-preserve-kak} that the general excitation-preserving gate decomposes into the fundamental entangler $F(\theta,\phi)$ surrounded by Z phases.
Defining the Z rotations before and after the fundamental entangler to be $K_\text{pre}$ and $K_\text{post}$, respectively, we get
\begin{align}
    \WW
    &=
    K_\text{post}F(\theta,\phi)K_\text{pre}
    \,.
\end{align}
Combining the two-qubit gate with the dynamical-decoupling gate, which we notate as $D$, we represent the repeated cycle unitary as
\begin{align}
    (D\WW)^n
    &=
    \big[DK_\text{post}F(\theta,\phi)K_\text{pre}\big]^n
    \\
    &=
    K_\text{pre}^{-1}\big[K_\text{pre}DK_\text{post}F(\theta,\phi)\big]^nK_\text{pre}
    \\
    &=
    K_\text{pre}^{-1}\big[D'F\big]^nK_\text{pre}
    \,,
\end{align}
where we used the gauge freedom to put all the product unitaries after the entangler in the repeated block, and simplified the notation with $D'\equiv K_\text{pre}DK_\text{post}$ and $F\equiv F(\theta,\phi)$.
We can combine both phase fluctuations (which are errors in $K_\text{pre/post}$) and microwave imperfections (which are errors in $D$) into one expression 
\begin{align}
\widetilde{D}' \approx (I-i\xssp\epsilon E\xssp)\, D'\,,
\end{align}
where $\epsilon$ is a small parameter and $E$ an error matrix with $\lVert E\rVert = O(1)$. To the first order in $\epsilon$, the repeated cycle unitary with error reads
\begin{align}
    \big[\widetilde{D}' F\big]^n
    &\approx
    \Big(
    I-i\epsilon\sum_{j=0}^{n-1}
    \big[D'F\big]^j
    E
    \big[D'F\big]^{-j}
    \Big)
    \big[D'F\big]^n
    \,.
\end{align}
The condition that the protocol being first-order insensitive to an arbitrary error in $K_\text{pre/post}$ or $D$ is 
\begin{align}
\mathcal{T}(E) \equiv 
\sum_{j=0}^{n-1}
    \big[D'F\big]^j
    E
    \big[D'F\big]^{-j} = 0
    \,,
\end{align}
that is, the twirl of $E$ under the group generated by $D'F=K_\text{pre}DK_\text{post}F(\theta,\phi)$ equals zero. According to our robustness condition, this term must vanish for all $E$ corresponding to single-qubit errors:
\begin{align}
    E
    &\in
    \mathrm{span}\{I{\otimes}X,I{\otimes}Y,I{\otimes}Z,X{\otimes}I,Y{\otimes}I,Z{\otimes}I\}
    \,.
\end{align}

The action of conjugating $E$ by $D'F$, called the adjoint action, is an orthogonal map from the 15-dimensional Lie algebra of $\mathrm{SU}(4)$ to itself, which includes the subspace to which $E$ belongs.
We notate this map by
\begin{align}
    \big[\mathrm{Ad}(D'F)\big](A)
    &=
    (D'F)A(D'F)^{-1}
    \,.
\end{align}
As such, we can now analyze the robustness of a dynamical-decoupling sequence by inspecting the eigenvalues of $\mathrm{Ad}(D'F)$.
In this new notation,
\begin{align}
   \mathcal{T}(E) = \sum_{j=0}^{n-1}
    \big[\mathrm{Ad}(D'F)\big]^j(E)
    \,,
\end{align}
which is the action of the sum of $\big[\mathrm{Ad}(D'F)\big]^j$ on $E$.
Since $\mathrm{Ad}(D'F)$ is an orthogonal map, its eigenvalues are phases.
For non trivial phases, this sum will vanish, so we need to ensure that the eigenspaces overlapping with $E$ all have non trivial eigenvalues.
Note that small phases will require more cycles to cancel, in practice limiting the magnitude of errors that can be tolerated, as we see for $\sqrt{\iSWAP}$.

We simplify the required analysis by noting that $\mathrm{Ad}(D'F)$ acts trivially on $\mathrm{span}\{X{\otimes}X,Y{\otimes}Y,Z{\otimes}Z\}$.
It also respects the $\mathbf{Z}_2$ symmetry of permuting subsystems, so we can further block diagonalize the map into symmetric and antisymmetric subspaces, each 6-dimensional.
For representing matrices in these subspaces, we choose for our bases the symmetric/antisymmetric combinations of IX, YZ, IY, ZX, IZ, and XY, in that order.

Using $\mathrm{Ad}_\pm$ to denote the adjoint action restricted to these symmetric/antisymmetric combinations, consider the adjoint action of the fundamental entangler:
\begin{widetext}
\begin{align}
    \operatorname{Ad}_+\big(F(\theta,\phi)\big)
    &=
    \begin{pmatrix}
        \cos(\theta-\phi/2) & -\sin(\theta-\phi/2) & 0 & 0 & 0 & 0
        \\
        \sin(\theta-\phi/2) & \cos(\theta-\phi/2) & 0 & 0 & 0 & 0
        \\
        0 & 0 & \cos(\phi/2-\theta) & -\sin(\phi/2-\theta) & 0 & 0
        \\
        0 & 0 & \sin(\phi/2-\theta) & \cos(\phi/2-\theta) & 0 & 0
        \\
        0 & 0 & 0 & 0 & 1 & 0
        \\
        0 & 0 & 0 & 0 & 0 & 1
    \end{pmatrix}\,,
    \\[10pt]
    \operatorname{Ad}_-\big(F(\theta,\phi)\big)
    &=
    \begin{pmatrix}
        \cos(\theta+\phi/2) & -\sin(\theta+\phi/2) & 0 & 0 & 0 & 0
        \\
        \sin(\theta+\phi/2) & \cos(\theta+\phi/2) & 0 & 0 & 0 & 0
        \\
        0 & 0 & \cos(\theta+\phi/2) & -\sin(\theta+\phi/2) & 0 & 0
        \\
        0 & 0 & \sin(\theta+\phi/2) & \cos(\theta+\phi/2) & 0 & 0
        \\
        0 & 0 & 0 & 0 & \cos2\theta & -\sin2\theta
        \\
        0 & 0 & 0 & 0 & \sin2\theta & \cos2\theta
    \end{pmatrix}\,.
\end{align}
\end{widetext}
For the case where the dynamical decoupling layer is $D=X\otimes X$, the adjoint action in both subspaces is diagonal in this basis and respects the 2-dimensional subspaces we just identified:
\begin{align}
    \operatorname{Ad}_\pm\big(D)
    &=
    \begin{pmatrix}
        1 & 0 & 0 & 0 & 0 & 0
        \\
        0 & 1 & 0 & 0 & 0 & 0
        \\
        0 & 0 & -1 & 0 & 0 & 0
        \\
        0 & 0 & 0 & -1 & 0 & 0
        \\
        0 & 0 & 0 & 0 & -1 & 0
        \\
        0 & 0 & 0 & 0 & 0 & -1
    \end{pmatrix}
    \,.
\end{align}
The combination of $\CZ$ with the dynamical decoupling gives symmetric and antisymmetric actions
\begin{align}
    \operatorname{Ad}_+(D\,\mathrm{CZ})
    &=
    \begin{pmatrix}
        0 & 0 & 0 & 1 & 0 & 0
        \\
        0 & 0 & 1 & 0 & 0 & 0
        \\
        0 & -1 & 0 & 0 & 0 & 0
        \\
        -1 & 0 & 0 & 0 & 0 & 0
        \\
        0 & 0 & 0 & 0 & -1 & 0
        \\
        0 & 0 & 0 & 0 & 0 & 1
    \end{pmatrix}\,,
    \\[3pt]
    \operatorname{Ad}_-(D\,\mathrm{CZ})
    &=
    \begin{pmatrix}
        0 & 0 & 0 & 1 & 0 & 0
        \\
        0 & 0 & -1 & 0 & 0 & 0
        \\
        0 & 1 & 0 & 0 & 0 & 0
        \\
        -1 & 0 & 0 & 0 & 0 & 0
        \\
        0 & 0 & 0 & 0 & -1 & 0
        \\
        0 & 0 & 0 & 0 & 0 & -1
    \end{pmatrix}
    \,.
\end{align}
The eigenvalues are non trivial with values $-1$, $i$, and $-i$, except for the last diagonal entry in $\mathrm{Ad}_+$, which corresponds to the symmetric combination of XY.
Since this is a non-local, non excitation preserving error that is not expected to be present in our system, we do not concern ourselves with echoing this away.
Taking sums of powers of the other eigenvalues, $\sum_j\lambda^j$, we find that all vanish when including four terms in the sum, which agrees with our earlier analysis, where the 4-cycle unitary was identical with the noiseless 4-cycle unitary when considering all single-qubit errors.

We can carry this same analysis out for other fSim gates.
Consider $\sqrt{\mathrm{iSWAP}}=F(\pi/4,0)$, for instance:
\begin{align}
    \operatorname{Ad}_+&\big(D\sqrt{\mathrm{iSWAP}}\big)\nonumber \\
    &\qquad =
    \frac{1}{\sqrt{2}}
    \begin{pmatrix}
        1 & -1 & 0 & 0 & 0 & 0
        \\
        1 & 1 & 0 & 0 & 0 & 0
        \\
        0 & 0 & -1 & -1 & 0 & 0
        \\
        0 & 0 & 1 & -1 & 0 & 0
        \\
        0 & 0 & 0 & 0 & -\sqrt{2} & 0
        \\
        0 & 0 & 0 & 0 & 0 & -\sqrt{2}
    \end{pmatrix},
    \\[6pt]
    \operatorname{Ad}_-&\big(D\sqrt{\mathrm{iSWAP}}\big)\nonumber \\
    &\qquad =
    \frac{1}{\sqrt{2}}
    \begin{pmatrix}
        1 & -1 & 0 & 0 & 0 & 0
        \\
        1 & 1 & 0 & 0 & 0 & 0
        \\
        0 & 0 & -1 & 1 & 0 & 0
        \\
        0 & 0 & -1 & -1 & 0 & 0
        \\
        0 & 0 & 0 & 0 & 0 & \sqrt{2}
        \\
        0 & 0 & 0 & 0 & -\sqrt{2} & 0
    \end{pmatrix}
    \,.
\end{align}
None of these eigenvalues are trivial, so everything is echoed away.
Because the phases are smaller, however, it takes 8 cycles instead of 4 before complete cancellation occurs, which puts stricter limits on how noisy the microwaves can be before the errors start polluting the characterization.

Finally, consider the iSWAP gate itself, $F(\pi/2,0)$.
Looking at its adjoint action, we notice a problem:
\begin{align}
    \operatorname{Ad}_+\big(\mathrm{iSWAP}\big)
    &=
    \begin{pmatrix}
        0 & -1 & 0 & 0 & 0 & 0
        \\
        1 & 0 & 0 & 0 & 0 & 0
        \\
        0 & 0 & 0 & 1 & 0 & 0
        \\
        0 & 0 & -1 & 0 & 0 & 0
        \\
        0 & 0 & 0 & 0 & 1 & 0
        \\
        0 & 0 & 0 & 0 & 0 & 1
    \end{pmatrix}
    \\[3pt]
    \operatorname{Ad}_-\big(\mathrm{iSWAP}\big)
    &=
    \begin{pmatrix}
        0 & -1 & 0 & 0 & 0 & 0
        \\
        1 & 0 & 0 & 0 & 0 & 0
        \\
        0 & 0 & 0 & -1 & 0 & 0
        \\
        0 & 0 & 1 & 0 & 0 & 0
        \\
        0 & 0 & 0 & 0 & -1 & 0
        \\
        0 & 0 & 0 & 0 & 0 & -1
    \end{pmatrix}
    \,.
\end{align}
Since the symmetric (IZ, XY) subspace is degenerate with trivial eigenvalue $+1$, we need to use the dynamical-decoupling gates.
However, the antisymmetric (IZ, XY) subspace is also degenerate, with eigenvalue $-1$, using the dynamical-decoupling gate will flip that eigenvalue to $+1$, which will allow antisymmetric Z errors to be amplified.

One strategy for overcoming this problem is to only use the iSWAP gate in every other cycle.
In the cycles where we do not use the iSWAP gate, we will instead idle the qubits for the iSWAP gate time.
This ensures the Z phase errors are identical between cycles, and thus coherently cancel with one another.
Since adjacent ideal $X\otimes X$ gates cancel, the sum of adjoint actions we need to consider for this sequence is
\begin{multline}
    \sum_j\operatorname{Ad}_\pm(\mathrm{iSWAP}^j)
    +\operatorname{Ad}_\pm\big(D\,\mathrm{iSWAP}^j\big)
    \\
    =
    \big(1+\operatorname{Ad}_\pm(D)\big)\sum_j\operatorname{Ad}_\pm(\mathrm{iSWAP})^j
    \,.
\end{multline}
The prefactor $1+\operatorname{Ad}_\pm(D)$ annihilates all but the (IX, YZ) subspace (symmetric and antisymmetric), and we see that iSWAP effectively echos away those errors on its own, so this sequence echos away low-frequency Z noise and is robust to imperfections in the microwave gates.

\section{Derivative Removal by Adiabatic Gate (DRAG)}
\label{sec:drag}

Here, we present a simple derivation of the DRAG pulse used to mitigate gate leakage~\cite{motzoi_simple_2009}.
We model a superconducting qubit as a driven nonlinear oscillator similar to Eq.~\eqref{eq:int_hamiltonian} in the main text ($\hbar =1$),
\begin{align}\label{eq:RWA_Hamiltonian_boson_sideband_nonlinear}
H_I(t) 
&= \delta\omega_{01}(t)\, a^\dag a -\frac{\eta}{2}\, a^{\dag\, 2} a^2 + [\Omega(t)\xssp a^\dag + \hc]\,,
\end{align}
where $a^\dag$ ($a$) is the bosonic creation (annihilation) operators and $\eta$ the so-called anharmonicity. Setting the detuning $\delta\omega_{01}(t) = 0$ and truncating the Hamiltonian to the second level $\ket{2}$, we have
\begin{align}
H_I(t)  &= 
 \begin{pmatrix}
 0 & \Omega(t)^*  & 0\\[3pt]
\Omega(t)\,  & 0 &\sqrt2\, \Omega(t)^*\\[3pt]
 0 & \sqrt2\, \Omega(t) & -\eta
 \end{pmatrix}\,.
\end{align}
The time-dependent  Schr\"odinger equation under $H_I(t)$ reads
\begin{align}
i\frac{\partial}{\partial t }\ket{\psi_I(t)}= H_I(t) \ket{\psi_I(t)}\,.
\end{align}
Going to a second interaction picture $\ket{\psi_{I\! I}(t)} = e^{-i\eta t\, \ket{2}\bra{2}} \ket{\psi_I(t)}$, we have
\begin{align}\label{eq:schrodinger_psiL}
i \frac{\partial}{\partial t}\ket{\psi_{I\! I}(t)}= H_{I\! I}(t) \ket{\psi_{I\! I}(t)}\,,
\end{align}
where the off-diagonal Hamiltonian $H_{I\! I}$ reads
\begin{align}\label{eq:ladder_Hamiltonian}
H_{I\! I}(t)  &= 
 \begin{pmatrix}
 0 & \Omega(t)^*  & 0\\[3pt]
 \Omega(t)\,  & 0 &\sqrt2\, \Omega(t)^*\xssp e^{i\eta t}\\[3pt]
 0 & \sqrt2\, \Omega(t)\xssp e^{-i\eta t} & 0
 \end{pmatrix}\,. 
\end{align}
Using first-order time-dependent perturbation theory, we obtain the leakage amplitude at the end of the microwave pulse for $-T/2\leq t\leq  T/2$,
\begin{align}
 \braket{2}{\psi_{I\! I}(T/2)} &\approx -i\int_{-\frac{T}{2}}^{\frac{T}{2}}\bra{2} H_{I\! I}(t) \ket{1}\xssp \braket{1}{\psi_{I\! I}(t)}\,\dif t\\[2pt]
 &=\int_{-\frac{T}{2}}^{\frac{T}{2}} \LL(t)\xssp e^{-i\eta t}\,\dif t\\[3pt]
 &\equiv \tilde\LL(\eta)\,,
\end{align}
where $\LL(t)\equiv -i\sqrt2\, \Omega(t)\xssp \braket{1}{\psi_{I\! I}(t)}$ and $\tilde\LL(\eta)$ its Fourier transformation. For any differentiable $\LL(t)$, we have 
\begin{align}
\tilde\LL(\eta)
&= \frac{i}{\eta}\int_{-\frac{T}{2}}^{\frac{T}{2}} \LL(t)\xssp \,\dif e^{-i\eta t}\\[2pt]
&= \frac{i}{\eta}\, \LL(t)\xssp e^{-i\eta t}\Big\rvert_{-\frac{T}{2}}^{\frac{T}{2}} + \frac{i}{\eta}\int_{-\frac{T}{2}}^{\frac{T}{2}} \LL^{(1)}(t)\xssp e^{-i\eta t}\,\dif t \,,
\end{align}
where $\LL^{(\nn)}(t) =  \dif^{\xssp \nn}\nsp \LL(t)/\dif t^\nn$.
Using integration by parts again, we have 
\begin{align}
\tilde\LL(\eta)&\approx \frac{i}{\eta}\, \LL(t)\xssp e^{-i\eta t}\Big\rvert_{-\frac{T}{2}}^{\frac{T}{2}} - \frac{1}{\eta^2}\, \LL^{(1)}(t)\xssp e^{-i\eta t}\Big\rvert_{-\frac{T}{2}}^{\frac{T}{2}}\nonumber \\[3pt]
&\quad - \frac{i}{\eta^3}\LL^{(2)}(t)\xssp e^{-i\eta t}\Big\rvert_{-\frac{T}{2}}^{\frac{T}{2}} \,,\label{eq:2nd_expansion}
\end{align}
where higher order terms are truncated when $\norm{\LL^{(n+1)}(t)/\LL^{(n)}(t)} \ll \eta$. We choose a control pulse $\Omega(t)$ whose value and first derivative vanish at both boundaries,
\begin{gather}
\Omega(-T/2) = \Omega(T/2) = 0\,,\label{eq:0th_zero}\\
\Omega^{(1)}(-T/2) = \Omega^{(1)}(T/2) = 0\,.\label{eq:1st_zero}
\end{gather}
Putting Eq.~\eqref{eq:0th_zero} into the Schr\"odinger equation~\eqref{eq:schrodinger_psiL}, we have
\begin{align}
\braket{1}{\psi_{I\! I}^{(1)}(-T/2)} = \braket{1}{\psi_{I\! I}^{(1)}(T/2)} = 0\,.\label{eq:1st_wavefunction_zero}
\end{align}
With identities~\eqref{eq:0th_zero}-\eqref{eq:1st_wavefunction_zero}, the first two terms in Eq.~\eqref{eq:2nd_expansion} vanish, and we can express the leakage amplitude as
\begin{align}
 \tilde\LL(\eta) &\approx - \frac{i}{\eta^3}\LL^{(2)}(t)\, e^{-i\eta t}\Big\rvert_{-\frac{T}{2}}^{\frac{T}{2}} \\[3pt]
 &= - \frac{\sqrt2}{\eta^3}\,\Omega^{(2)}(t)\xssp \braket{1}{\psi_{I\! I}(t)}\, e^{-i\eta t}\Big\rvert_{-\frac{T}{2}}^{\frac{T}{2}}
 \,.\label{eq:leakage_amplitude}
\end{align}
To zero out the leakage, we add a derivative term to the original pulse $\Omega(t)$,
\begin{align}\label{eq:drag_pulse}
\mathpalette{\Omega_{\scalebox{0.68}{\,DRAG}}}\relax(t) =  \Omega(t) -\frac{i}{\eta}\, \Omega^{(1)}(t)\,,
\end{align}
and correspondingly, we have
\begin{align}
\mathpalette{\LL_{\scalebox{0.68}{\,DRAG}}}\relax(t) =  \LL(t) - \frac{\sqrt2}{\eta}\, \Omega^{(1)}(t)\xssp \braket{1}{\psi_{I\! I}(t)}\,.
\end{align}
Replacing $\LL(t)$ with $\mathpalette{\LL_{\scalebox{0.68}{\,DRAG}}}\relax(t)$ in Eq.~\eqref{eq:2nd_expansion} leads to the leakage amplitude for the DRAG pulse. 
The leading contribution from the added derivative term in $\mathpalette{\LL_{\scalebox{0.68}{\,DRAG}}}\relax(t)$ comes from the term with coefficient $1/\eta^2$ in Eq.~\eqref{eq:2nd_expansion}   
\begin{align}
 &-\frac{1}{\eta^2}\, \frac{\dif}{\dif t}\,\Bigl(-\frac{\sqrt2}{\eta}\, \Omega(t)^{(1)}\xssp \braket{1}{\psi_{I\! I}(t)}\Big)\, e^{-i\eta t}\xssp\Big\rvert_{-\frac{T}{2}}^{\frac{T}{2}}\nonumber\\[2pt]
 &\qquad\quad = \frac{\sqrt2}{\eta^3}\,\Omega^{(2)}(t)\xssp \braket{1}{\psi_{I\! I}(t)}\, e^{-i\eta t}\Big\rvert_{-\frac{T}{2}}^{\frac{T}{2}}\,,
\end{align}
which exactly cancels Eq.~\eqref{eq:leakage_amplitude} and eliminates gate leakages.

\section{Parameter estimation under decoherence}
\label{sec:parameter_estimation_decoherence}

Quantum parameter estimation protocols are sensitive to noise~\cite{shaji_qubit_2007}.
For example, depolarizing noise---no matter how small---ruins the possibility of sub-shot-noise performances of a quantum interferometer~\cite{ji_parameter_2008}. Here, we discuss the effects of decoherence on characterizing single-qubit Z-phase gates and two-qubit parameters. 

Consider a single-qubit system described by the master equation $\dot\rho=-\frac{i}{\hbar}[H,\,\rho] + \mathcal L(\rho)$, where the Lindblad operator $\mathcal L(\rho)$ takes the form
\begin{align}
\mathcal L(\rho) &=\frac{1}{T_1} \Big(\sigma_+ \rho\xssp  \sigma_- -\frac{1}{2}(\sigma_- \sigma_+\rho +\rho  \sigma_- \sigma_+)\Big)\nonumber\\
&\quad +
\frac{1}{2\xssp T_2} \left(\sigma^z \rho\xssp  \sigma^z - \rho \right)\,.
\end{align}
where $\sigma_+=\sigma_-^\dag = \ket{0}\!\bra{1}$.
The Hamiltonian evolution consists a sequence of $\nn$ phase gates $\RZ(\varphi)= e^{-i\varphi Z/2}$, where the angle $\varphi$ is to be determined.
The experiment involves the following steps:
\begin{enumerate}
\item Initialize the qubit to $\ket{+} = \frac{1}{\sqrt 2}\bigl(\ket{0} + \ket{1}\bigr)$,
\item Apply the phase gate $\RZ(\varphi)$ $\nn$ times,
\item Measure the qubit in the basis of $\RZ(-s) X \RZ(s)$.
\end{enumerate}
The probability that the state $\ket{1}$ does not decay to $\ket{0}$ before being measured is approximately $e^{-\nn \lambda_1}$, where $\lambda_1 = \text{gate time} / T_1$.  Therefore, the probability of getting the measurement outcome $+1$ is
\begin{align}
p_{\nn}(s) = e^{-\nn \lambda_1} q_{\nn}(s) + \frac{1-e^{-\nn \lambda_1}}{2}\,, \label{eq:p_varphi}
\end{align}
where $q_{\nn}(s)$ is the probability of getting the measurement outcome $+1$ with only $T_2$ error
\begin{align}
 q_{\nn}(s) = \frac{1 + e^{-\nn\lambda_2}\cos(\nn\xssp \varphi + s)}{2}\,, \label{eq:q_varphi}
\end{align}
where $\lambda_2 = \text{gate time} / T_2$.  An unbiased estimator of the probability $p_{\nn}$ is
\begin{align}
   \hat p_{\nn} =  \frac{\text{\# of outcome $+1$}}{\mm}\,,
\end{align}
where $\mm$ is the number of times that we repeat the experiment.
The variance of the estimator $\hat p_\nn$ is
\begin{align}
V(\hat p_{\nn}) &=  \frac{p_{\nn}(1-p_\nn)}{\mm}\\
&=  \frac{e^{-2\nn \lambda_1} q_{\nn}(1-q_\nn)}{\mm} + \frac{1-e^{-2\nn \lambda_1}}{4 \mm}\,.
\end{align}
The variance of the estimator of $\varphi$ can be determined using the chain rule
\begin{align}
V(\hat\varphi_\nn) &= \frac{V(\hat p_{\nn}) }{(\partial p_\nn/\partial q_\nn)^2\, (\partial q_\nn/\partial\varphi)^2}\\[3pt]
&= \frac{e^{2\nn \lambda_1}-e^{-2\nn\lambda_2}\cos(\nn\xssp \varphi+ s)^2}{\mm\xssp\nn^ 2\xssp  e^{-2\nn \lambda_2}\xssp \sin(\nn\xssp \varphi+ s)^2}\\
&\leq
\frac{e^{2\nn (\lambda_1+\lambda_2)}}{\mm\xssp\nn^ 2\xssp \sin(\nn\xssp \varphi+ s)^2}
\,.\label{eq:single_variance}
\end{align}
By running experiments with two values of $s$ that are $\pi/2$ apart from each other, e.g., $s = 0$ and $s = \pi/2$, the variance becomes
\begin{align}
V(\hat\varphi_\nn) \leq  \frac{e^{2\nn (\lambda_1+ \lambda_2)}}{\mm\xssp\nn^ 2} \,.
\end{align}
Therefore, the standard deviation of the estimator $\hat\varphi_\nn$ scales as $O(\nn^{-1})$ before it blows exponentially.  We have the optimal value of $\nn$ at $\partial V(\hat\varphi_\nn)/\partial n = 0$,
\begin{align}\label{eq:min_variance}
    \nn_\star = \frac{1}{\lambda_1 + \lambda_2}\,,\quad V(\hat\varphi_{\nn_\star}) \leq  \frac{e^{2}(\lambda_1 + \lambda_2)^2}{ \mm}\,.
\end{align}

In two-qubit gates, the impact of decoherence usually depends on the specific shapes of the control pulses. For simplicity, we consider the resonant case where the decoherence effects are pulse-shape independent. 
To estimate the swap angle $\theta$ of an excitation-preserving two-qubit gate~\eqref{eq:nc_2q_u}, we initialize the qubit in the state $\ket{01}$ and observe population oscillations between $\ket{01}$ and $\ket{10}$. We can postselect out the $T_1$ error, since the qubit state decays to $\ket{00}$ when an excitation is lost. Similar to~\cref{eq:q_varphi}, the measurement probability in the single-excitation subspace for estimating $\theta$ reads
\begin{align}
 p_{\nn} = \frac{1 - e^{-2\nn\lambda_2}\cos(2\nn\xssp \theta)}{2}\,, \label{eq:p_theta}
\end{align}
where the factor $2$ in the exponential comes from dephasing of the two qubits, assuming their have the same $T_1$ and $T_2$.  The variance of the estimator $\hat p_\nn$ is
\begin{align}
V(\hat p_{\nn}) =  \frac{p_{\nn}(1-p_\nn)}{\mm\xssp e^{-\nn \lambda_1}} \,,
\end{align}
where $\mm\xssp e^{-\nn \lambda_1}$ is the number of measurements that pass the $T_1$ post selection.
The variance of the estimator of $\theta$ takes the form
\begin{align}
V(\hat\theta_\nn) &= \frac{V(\hat p_{\nn}) }{(\partial p_\nn/\partial\theta)^2}\\[3pt]
&= \frac{1-e^{-4\nn\lambda_2}\cos(2\nn\xssp \theta)^2}{4\mm\xssp\nn^ 2\xssp  e^{-\nn(\lambda_1+ 4\lambda_2)}\xssp \sin(2\nn\xssp \theta)^2}\\
&\leq
\frac{e^{\nn (\lambda_1+4\lambda_2)}}{4\mm\xssp\nn^ 2\xssp \sin(2\nn\xssp \theta)^2}
\,.
\end{align}
Compared to the single-qubit case~\eqref{eq:single_variance}, the effect of $\lambda_1$ is reduced by a factor of 2 due to post selection while the effect of $\lambda_2$ is doubled due to dephasing from both qubits.  By setting $\partial V(\hat\theta_\nn)/\partial n = 0$, we have
\begin{align}
    \nn_\star = \frac{2}{\lambda_1+4\lambda_2}\,,\quad V(\hat\theta_{\nn_\star}) \leq  \frac{e^{2}\xssp(\lambda_1+4\lambda_2)^2}{4 \mm}\,.
\end{align}

\putbib[characterization]
\end{bibunit}
\end{appendices}

% \bibliography{characterization}

\begin{thebibliography}{33}%
\makeatletter
\providecommand \@ifxundefined [1]{%
 \@ifx{#1\undefined}
}%
\providecommand \@ifnum [1]{%
 \ifnum #1\expandafter \@firstoftwo
 \else \expandafter \@secondoftwo
 \fi
}%
\providecommand \@ifx [1]{%
 \ifx #1\expandafter \@firstoftwo
 \else \expandafter \@secondoftwo
 \fi
}%
\providecommand \natexlab [1]{#1}%
\providecommand \enquote  [1]{``#1''}%
\providecommand \bibnamefont  [1]{#1}%
\providecommand \bibfnamefont [1]{#1}%
\providecommand \citenamefont [1]{#1}%
\providecommand \href@noop [0]{\@secondoftwo}%
\providecommand \href [0]{\begingroup \@sanitize@url \@href}%
\providecommand \@href[1]{\@@startlink{#1}\@@href}%
\providecommand \@@href[1]{\endgroup#1\@@endlink}%
\providecommand \@sanitize@url [0]{\catcode `\\12\catcode `\$12\catcode
  `\&12\catcode `\#12\catcode `\^12\catcode `\_12\catcode `\%12\relax}%
\providecommand \@@startlink[1]{}%
\providecommand \@@endlink[0]{}%
\providecommand \url  [0]{\begingroup\@sanitize@url \@url }%
\providecommand \@url [1]{\endgroup\@href {#1}{\urlprefix }}%
\providecommand \urlprefix  [0]{URL }%
\providecommand \Eprint [0]{\href }%
\providecommand \doibase [0]{http://dx.doi.org/}%
\providecommand \selectlanguage [0]{\@gobble}%
\providecommand \bibinfo  [0]{\@secondoftwo}%
\providecommand \bibfield  [0]{\@secondoftwo}%
\providecommand \translation [1]{[#1]}%
\providecommand \BibitemOpen [0]{}%
\providecommand \bibitemStop [0]{}%
\providecommand \bibitemNoStop [0]{.\EOS\space}%
\providecommand \EOS [0]{\spacefactor3000\relax}%
\providecommand \BibitemShut  [1]{\csname bibitem#1\endcsname}%
\let\auto@bib@innerbib\@empty
%</preamble>
\bibitem [{\citenamefont {Kimmel}\ \emph {et~al.}(2015)\citenamefont {Kimmel},
  \citenamefont {Low},\ and\ \citenamefont {Yoder}}]{kimmel_robust_2015}%
  \BibitemOpen
  \bibfield  {author} {\bibinfo {author} {\bibfnamefont {S.}~\bibnamefont
  {Kimmel}}, \bibinfo {author} {\bibfnamefont {G.~H.}\ \bibnamefont {Low}}, \
  and\ \bibinfo {author} {\bibfnamefont {T.~J.}\ \bibnamefont {Yoder}},\
  }\bibfield  {title} {\enquote {\bibinfo {title} {Robust calibration of a
  universal single-qubit gate set via robust phase estimation},}\ }\href
  {\doibase 10.1103/PhysRevA.92.062315} {\bibfield  {journal} {\bibinfo
  {journal} {Physical Review A}\ }\textbf {\bibinfo {volume} {92}},\ \bibinfo
  {pages} {062315} (\bibinfo {year} {2015})},\ \bibinfo {note} {publisher:
  American Physical Society}\BibitemShut {NoStop}%
\bibitem [{\citenamefont {Rudinger}\ \emph {et~al.}(2017)\citenamefont
  {Rudinger}, \citenamefont {Kimmel}, \citenamefont {Lobser},\ and\
  \citenamefont {Maunz}}]{rudinger_experimental_2017}%
  \BibitemOpen
  \bibfield  {author} {\bibinfo {author} {\bibfnamefont {K.}~\bibnamefont
  {Rudinger}}, \bibinfo {author} {\bibfnamefont {S.}~\bibnamefont {Kimmel}},
  \bibinfo {author} {\bibfnamefont {D.}~\bibnamefont {Lobser}}, \ and\ \bibinfo
  {author} {\bibfnamefont {P.}~\bibnamefont {Maunz}},\ }\bibfield  {title}
  {\enquote {\bibinfo {title} {Experimental {Demonstration} of a {Cheap} and
  {Accurate} {Phase} {Estimation}},}\ }\href {\doibase
  10.1103/PhysRevLett.118.190502} {\bibfield  {journal} {\bibinfo  {journal}
  {Physical Review Letters}\ }\textbf {\bibinfo {volume} {118}},\ \bibinfo
  {pages} {190502} (\bibinfo {year} {2017})},\ \bibinfo {note} {publisher:
  American Physical Society}\BibitemShut {NoStop}%
\bibitem [{\citenamefont {Arute}\ \emph {et~al.}(2020)\citenamefont {Arute},
  \citenamefont {Arya}, \citenamefont {Babbush}, \citenamefont {Bacon},
  \citenamefont {Bardin}, \citenamefont {Barends}, \citenamefont {Bengtsson},
  \citenamefont {Boixo}, \citenamefont {Broughton}, \citenamefont {Buckley},
  \citenamefont {Buell}, \citenamefont {Burkett}, \citenamefont {Bushnell},
  \citenamefont {Chen}, \citenamefont {Chen} \emph
  {et~al.}}]{arute_observation_2020}%
  \BibitemOpen
  \bibfield  {author} {\bibinfo {author} {\bibfnamefont {F.}~\bibnamefont
  {Arute}}, \bibinfo {author} {\bibfnamefont {K.}~\bibnamefont {Arya}},
  \bibinfo {author} {\bibfnamefont {R.}~\bibnamefont {Babbush}}, \bibinfo
  {author} {\bibfnamefont {D.}~\bibnamefont {Bacon}}, \bibinfo {author}
  {\bibfnamefont {J.~C.}\ \bibnamefont {Bardin}}, \bibinfo {author}
  {\bibfnamefont {R.}~\bibnamefont {Barends}}, \bibinfo {author} {\bibfnamefont
  {A.}~\bibnamefont {Bengtsson}}, \bibinfo {author} {\bibfnamefont
  {S.}~\bibnamefont {Boixo}}, \bibinfo {author} {\bibfnamefont
  {M.}~\bibnamefont {Broughton}}, \bibinfo {author} {\bibfnamefont {B.~B.}\
  \bibnamefont {Buckley}}, \bibinfo {author} {\bibfnamefont {D.~A.}\
  \bibnamefont {Buell}}, \bibinfo {author} {\bibfnamefont {B.}~\bibnamefont
  {Burkett}}, \bibinfo {author} {\bibfnamefont {N.}~\bibnamefont {Bushnell}},
  \bibinfo {author} {\bibfnamefont {Y.}~\bibnamefont {Chen}}, \bibinfo {author}
  {\bibfnamefont {Z.}~\bibnamefont {Chen}},  \emph {et~al.},\ }\bibfield
  {title} {\enquote {\bibinfo {title} {Observation of separated dynamics of
  charge and spin in the {Fermi}-{Hubbard} model},}\ }\href
  {http://arxiv.org/abs/2010.07965} {\bibfield  {journal} {\bibinfo  {journal}
  {arXiv:2010.07965}\ } (\bibinfo {year} {2020})}\BibitemShut {NoStop}%
\bibitem [{\citenamefont {Neill}\ \emph {et~al.}(2021)\citenamefont {Neill},
  \citenamefont {McCourt}, \citenamefont {Mi}, \citenamefont {Jiang},
  \citenamefont {Niu}, \citenamefont {Mruczkiewicz}, \citenamefont {Aleiner},
  \citenamefont {Arute}, \citenamefont {Arya}, \citenamefont {Atalaya},
  \citenamefont {Babbush}, \citenamefont {Bardin}, \citenamefont {Barends},
  \citenamefont {Bengtsson}, \citenamefont {Bourassa}, \citenamefont
  {Broughton}, \citenamefont {Buckley}, \citenamefont {Buell}, \citenamefont
  {Burkett}, \citenamefont {Bushnell} \emph {et~al.}}]{neill_accurately_2021}%
  \BibitemOpen
  \bibfield  {author} {\bibinfo {author} {\bibfnamefont {C.}~\bibnamefont
  {Neill}}, \bibinfo {author} {\bibfnamefont {T.}~\bibnamefont {McCourt}},
  \bibinfo {author} {\bibfnamefont {X.}~\bibnamefont {Mi}}, \bibinfo {author}
  {\bibfnamefont {Z.}~\bibnamefont {Jiang}}, \bibinfo {author} {\bibfnamefont
  {M.~Y.}\ \bibnamefont {Niu}}, \bibinfo {author} {\bibfnamefont
  {W.}~\bibnamefont {Mruczkiewicz}}, \bibinfo {author} {\bibfnamefont
  {I.}~\bibnamefont {Aleiner}}, \bibinfo {author} {\bibfnamefont
  {F.}~\bibnamefont {Arute}}, \bibinfo {author} {\bibfnamefont
  {K.}~\bibnamefont {Arya}}, \bibinfo {author} {\bibfnamefont {J.}~\bibnamefont
  {Atalaya}}, \bibinfo {author} {\bibfnamefont {R.}~\bibnamefont {Babbush}},
  \bibinfo {author} {\bibfnamefont {J.~C.}\ \bibnamefont {Bardin}}, \bibinfo
  {author} {\bibfnamefont {R.}~\bibnamefont {Barends}}, \bibinfo {author}
  {\bibfnamefont {A.}~\bibnamefont {Bengtsson}}, \bibinfo {author}
  {\bibfnamefont {A.}~\bibnamefont {Bourassa}}, \bibinfo {author}
  {\bibfnamefont {M.}~\bibnamefont {Broughton}}, \bibinfo {author}
  {\bibfnamefont {B.~B.}\ \bibnamefont {Buckley}}, \bibinfo {author}
  {\bibfnamefont {D.~A.}\ \bibnamefont {Buell}}, \bibinfo {author}
  {\bibfnamefont {B.}~\bibnamefont {Burkett}}, \bibinfo {author} {\bibfnamefont
  {N.}~\bibnamefont {Bushnell}},  \emph {et~al.},\ }\bibfield  {title}
  {\enquote {\bibinfo {title} {Accurately computing the electronic properties
  of a quantum ring},}\ }\href {\doibase 10.1038/s41586-021-03576-2} {\bibfield
   {journal} {\bibinfo  {journal} {Nature}\ }\textbf {\bibinfo {volume}
  {594}},\ \bibinfo {pages} {508} (\bibinfo {year} {2021})}\BibitemShut
  {NoStop}%
\bibitem [{\citenamefont {Nielsen}\ \emph {et~al.}(2021)\citenamefont
  {Nielsen}, \citenamefont {Gamble}, \citenamefont {Rudinger}, \citenamefont
  {Scholten}, \citenamefont {Young},\ and\ \citenamefont
  {Blume-Kohout}}]{nielsen_gate_2021}%
  \BibitemOpen
  \bibfield  {author} {\bibinfo {author} {\bibfnamefont {E.}~\bibnamefont
  {Nielsen}}, \bibinfo {author} {\bibfnamefont {J.~K.}\ \bibnamefont {Gamble}},
  \bibinfo {author} {\bibfnamefont {K.}~\bibnamefont {Rudinger}}, \bibinfo
  {author} {\bibfnamefont {T.}~\bibnamefont {Scholten}}, \bibinfo {author}
  {\bibfnamefont {K.}~\bibnamefont {Young}}, \ and\ \bibinfo {author}
  {\bibfnamefont {R.}~\bibnamefont {Blume-Kohout}},\ }\bibfield  {title}
  {\enquote {\bibinfo {title} {Gate {Set} {Tomography}},}\ }\href {\doibase
  10.22331/q-2021-10-05-557} {\bibfield  {journal} {\bibinfo  {journal}
  {Quantum}\ }\textbf {\bibinfo {volume} {5}},\ \bibinfo {pages} {557}
  (\bibinfo {year} {2021})}\BibitemShut {NoStop}%
\bibitem [{\citenamefont {Chiorescu}\ \emph {et~al.}(2003)\citenamefont
  {Chiorescu}, \citenamefont {Nakamura}, \citenamefont {Harmans},\ and\
  \citenamefont {Mooij}}]{chiorescu_coherent_2003}%
  \BibitemOpen
  \bibfield  {author} {\bibinfo {author} {\bibfnamefont {I.}~\bibnamefont
  {Chiorescu}}, \bibinfo {author} {\bibfnamefont {Y.}~\bibnamefont {Nakamura}},
  \bibinfo {author} {\bibfnamefont {C.~J. P.~M.}\ \bibnamefont {Harmans}}, \
  and\ \bibinfo {author} {\bibfnamefont {J.~E.}\ \bibnamefont {Mooij}},\
  }\bibfield  {title} {\enquote {\bibinfo {title} {Coherent {Quantum}
  {Dynamics} of a {Superconducting} {Flux} {Qubit}},}\ }\href {\doibase
  10.1126/science.1081045} {\bibfield  {journal} {\bibinfo  {journal}
  {Science}\ }\textbf {\bibinfo {volume} {299}},\ \bibinfo {pages} {1869}
  (\bibinfo {year} {2003})}\BibitemShut {NoStop}%
\bibitem [{\citenamefont {Quintana}\ \emph {et~al.}(2017)\citenamefont
  {Quintana}, \citenamefont {Chen}, \citenamefont {Sank}, \citenamefont
  {Petukhov}, \citenamefont {White}, \citenamefont {Kafri}, \citenamefont
  {Chiaro}, \citenamefont {Megrant}, \citenamefont {Barends}, \citenamefont
  {Campbell}, \citenamefont {Chen}, \citenamefont {Dunsworth}, \citenamefont
  {Fowler}, \citenamefont {Graff}, \citenamefont {Jeffrey} \emph
  {et~al.}}]{quintana_observation_2017}%
  \BibitemOpen
  \bibfield  {author} {\bibinfo {author} {\bibfnamefont {C.~M.}\ \bibnamefont
  {Quintana}}, \bibinfo {author} {\bibfnamefont {Y.}~\bibnamefont {Chen}},
  \bibinfo {author} {\bibfnamefont {D.}~\bibnamefont {Sank}}, \bibinfo {author}
  {\bibfnamefont {A.~G.}\ \bibnamefont {Petukhov}}, \bibinfo {author}
  {\bibfnamefont {T.~C.}\ \bibnamefont {White}}, \bibinfo {author}
  {\bibfnamefont {D.}~\bibnamefont {Kafri}}, \bibinfo {author} {\bibfnamefont
  {B.}~\bibnamefont {Chiaro}}, \bibinfo {author} {\bibfnamefont
  {A.}~\bibnamefont {Megrant}}, \bibinfo {author} {\bibfnamefont
  {R.}~\bibnamefont {Barends}}, \bibinfo {author} {\bibfnamefont
  {B.}~\bibnamefont {Campbell}}, \bibinfo {author} {\bibfnamefont
  {Z.}~\bibnamefont {Chen}}, \bibinfo {author} {\bibfnamefont {A.}~\bibnamefont
  {Dunsworth}}, \bibinfo {author} {\bibfnamefont {A.~G.}\ \bibnamefont
  {Fowler}}, \bibinfo {author} {\bibfnamefont {R.}~\bibnamefont {Graff}},
  \bibinfo {author} {\bibfnamefont {E.}~\bibnamefont {Jeffrey}},  \emph
  {et~al.},\ }\bibfield  {title} {\enquote {\bibinfo {title} {Observation of
  classical-quantum crossover of $1/f$ flux noise and its paramagnetic
  temperature dependence},}\ }\href {\doibase 10.1103/PhysRevLett.118.057702}
  {\bibfield  {journal} {\bibinfo  {journal} {Phys. Rev. Lett.}\ }\textbf
  {\bibinfo {volume} {118}},\ \bibinfo {pages} {057702} (\bibinfo {year}
  {2017})}\BibitemShut {NoStop}%
\bibitem [{\citenamefont {Carr}\ and\ \citenamefont
  {Purcell}(1954)}]{carr_effects_1954}%
  \BibitemOpen
  \bibfield  {author} {\bibinfo {author} {\bibfnamefont {H.~Y.}\ \bibnamefont
  {Carr}}\ and\ \bibinfo {author} {\bibfnamefont {E.~M.}\ \bibnamefont
  {Purcell}},\ }\bibfield  {title} {\enquote {\bibinfo {title} {Effects of
  diffusion on free precession in nuclear magnetic resonance experiments},}\
  }\href {\doibase 10.1103/PhysRev.94.630} {\bibfield  {journal} {\bibinfo
  {journal} {Phys. Rev.}\ }\textbf {\bibinfo {volume} {94}},\ \bibinfo {pages}
  {630} (\bibinfo {year} {1954})}\BibitemShut {NoStop}%
\bibitem [{\citenamefont {Meiboom}\ and\ \citenamefont
  {Gill}(1958)}]{meiboom_modified_1958}%
  \BibitemOpen
  \bibfield  {author} {\bibinfo {author} {\bibfnamefont {S.}~\bibnamefont
  {Meiboom}}\ and\ \bibinfo {author} {\bibfnamefont {D.}~\bibnamefont {Gill}},\
  }\bibfield  {title} {\enquote {\bibinfo {title} {Modified {Spin}‐{Echo}
  {Method} for {Measuring} {Nuclear} {Relaxation} {Times}},}\ }\href {\doibase
  10.1063/1.1716296} {\bibfield  {journal} {\bibinfo  {journal} {Review of
  Scientific Instruments}\ }\textbf {\bibinfo {volume} {29}},\ \bibinfo {pages}
  {688} (\bibinfo {year} {1958})}\BibitemShut {NoStop}%
\bibitem [{\citenamefont {Viola}\ \emph {et~al.}(1999)\citenamefont {Viola},
  \citenamefont {Knill},\ and\ \citenamefont {Lloyd}}]{viola_dynamical_1999}%
  \BibitemOpen
  \bibfield  {author} {\bibinfo {author} {\bibfnamefont {L.}~\bibnamefont
  {Viola}}, \bibinfo {author} {\bibfnamefont {E.}~\bibnamefont {Knill}}, \ and\
  \bibinfo {author} {\bibfnamefont {S.}~\bibnamefont {Lloyd}},\ }\bibfield
  {title} {\enquote {\bibinfo {title} {Dynamical decoupling of open quantum
  systems},}\ }\href {\doibase 10.1103/PhysRevLett.82.2417} {\bibfield
  {journal} {\bibinfo  {journal} {Phys. Rev. Lett.}\ }\textbf {\bibinfo
  {volume} {82}},\ \bibinfo {pages} {2417} (\bibinfo {year}
  {1999})}\BibitemShut {NoStop}%
\bibitem [{\citenamefont {Khodjasteh}\ and\ \citenamefont
  {Lidar}(2005)}]{khodjasteh_fault_2005}%
  \BibitemOpen
  \bibfield  {author} {\bibinfo {author} {\bibfnamefont {K.}~\bibnamefont
  {Khodjasteh}}\ and\ \bibinfo {author} {\bibfnamefont {D.~A.}\ \bibnamefont
  {Lidar}},\ }\bibfield  {title} {\enquote {\bibinfo {title} {Fault-tolerant
  quantum dynamical decoupling},}\ }\href {\doibase
  10.1103/PhysRevLett.95.180501} {\bibfield  {journal} {\bibinfo  {journal}
  {Phys. Rev. Lett.}\ }\textbf {\bibinfo {volume} {95}},\ \bibinfo {pages}
  {180501} (\bibinfo {year} {2005})}\BibitemShut {NoStop}%
\bibitem [{\citenamefont {Uhrig}(2007)}]{uhrig_keeping_2007}%
  \BibitemOpen
  \bibfield  {author} {\bibinfo {author} {\bibfnamefont {G.~S.}\ \bibnamefont
  {Uhrig}},\ }\bibfield  {title} {\enquote {\bibinfo {title} {Keeping a quantum
  bit alive by optimized $\ensuremath{\pi}$-pulse sequences},}\ }\href
  {\doibase 10.1103/PhysRevLett.98.100504} {\bibfield  {journal} {\bibinfo
  {journal} {Phys. Rev. Lett.}\ }\textbf {\bibinfo {volume} {98}},\ \bibinfo
  {pages} {100504} (\bibinfo {year} {2007})}\BibitemShut {NoStop}%
\bibitem [{\citenamefont {Dong}\ \emph {et~al.}(2022)\citenamefont {Dong},
  \citenamefont {Gross},\ and\ \citenamefont {Niu}}]{dong_beyond_2022}%
  \BibitemOpen
  \bibfield  {author} {\bibinfo {author} {\bibfnamefont {Y.}~\bibnamefont
  {Dong}}, \bibinfo {author} {\bibfnamefont {J.}~\bibnamefont {Gross}}, \ and\
  \bibinfo {author} {\bibfnamefont {M.~Y.}\ \bibnamefont {Niu}},\ }\bibfield
  {title} {\enquote {\bibinfo {title} {Beyond {Heisenberg} limit quantum
  metrology through quantum signal processing},}\ }\href
  {http://arxiv.org/abs/2209.11207} {\bibfield  {journal} {\bibinfo  {journal}
  {arXiv.2209.11207}\ } (\bibinfo {year} {2022})}\BibitemShut {NoStop}%
\bibitem [{\citenamefont {Wang}\ \emph {et~al.}(2015)\citenamefont {Wang},
  \citenamefont {Deng},\ and\ \citenamefont {Duan}}]{wang_hamiltonian_2015}%
  \BibitemOpen
  \bibfield  {author} {\bibinfo {author} {\bibfnamefont {S.-T.}\ \bibnamefont
  {Wang}}, \bibinfo {author} {\bibfnamefont {D.-L.}\ \bibnamefont {Deng}}, \
  and\ \bibinfo {author} {\bibfnamefont {L.-M.}\ \bibnamefont {Duan}},\
  }\bibfield  {title} {\enquote {\bibinfo {title} {Hamiltonian tomography for
  quantum many-body systems with arbitrary couplings},}\ }\href {\doibase
  10.1088/1367-2630/17/9/093017} {\bibfield  {journal} {\bibinfo  {journal}
  {New Journal of Physics}\ }\textbf {\bibinfo {volume} {17}},\ \bibinfo
  {pages} {093017} (\bibinfo {year} {2015})}\BibitemShut {NoStop}%
\bibitem [{\citenamefont {Sundaresan}\ \emph {et~al.}(2020)\citenamefont
  {Sundaresan}, \citenamefont {Lauer}, \citenamefont {Pritchett}, \citenamefont
  {Magesan}, \citenamefont {Jurcevic},\ and\ \citenamefont
  {Gambetta}}]{sundaresan_reducing_2020}%
  \BibitemOpen
  \bibfield  {author} {\bibinfo {author} {\bibfnamefont {N.}~\bibnamefont
  {Sundaresan}}, \bibinfo {author} {\bibfnamefont {I.}~\bibnamefont {Lauer}},
  \bibinfo {author} {\bibfnamefont {E.}~\bibnamefont {Pritchett}}, \bibinfo
  {author} {\bibfnamefont {E.}~\bibnamefont {Magesan}}, \bibinfo {author}
  {\bibfnamefont {P.}~\bibnamefont {Jurcevic}}, \ and\ \bibinfo {author}
  {\bibfnamefont {J.~M.}\ \bibnamefont {Gambetta}},\ }\bibfield  {title}
  {\enquote {\bibinfo {title} {Reducing {Unitary} and {Spectator} {Errors} in
  {Cross} {Resonance} with {Optimized} {Rotary} {Echoes}},}\ }\href {\doibase
  10.1103/PRXQuantum.1.020318} {\bibfield  {journal} {\bibinfo  {journal} {PRX
  Quantum}\ }\textbf {\bibinfo {volume} {1}},\ \bibinfo {pages} {020318}
  (\bibinfo {year} {2020})}\BibitemShut {NoStop}%
\bibitem [{\citenamefont {Boixo}\ \emph {et~al.}(2018)\citenamefont {Boixo},
  \citenamefont {Isakov}, \citenamefont {Smelyanskiy}, \citenamefont {Babbush},
  \citenamefont {Ding}, \citenamefont {Jiang}, \citenamefont {Bremner},
  \citenamefont {Martinis},\ and\ \citenamefont
  {Neven}}]{boixo_characterizing_2018}%
  \BibitemOpen
  \bibfield  {author} {\bibinfo {author} {\bibfnamefont {S.}~\bibnamefont
  {Boixo}}, \bibinfo {author} {\bibfnamefont {S.~V.}\ \bibnamefont {Isakov}},
  \bibinfo {author} {\bibfnamefont {V.~N.}\ \bibnamefont {Smelyanskiy}},
  \bibinfo {author} {\bibfnamefont {R.}~\bibnamefont {Babbush}}, \bibinfo
  {author} {\bibfnamefont {N.}~\bibnamefont {Ding}}, \bibinfo {author}
  {\bibfnamefont {Z.}~\bibnamefont {Jiang}}, \bibinfo {author} {\bibfnamefont
  {M.~J.}\ \bibnamefont {Bremner}}, \bibinfo {author} {\bibfnamefont {J.~M.}\
  \bibnamefont {Martinis}}, \ and\ \bibinfo {author} {\bibfnamefont
  {H.}~\bibnamefont {Neven}},\ }\bibfield  {title} {\enquote {\bibinfo {title}
  {Characterizing quantum supremacy in near-term devices},}\ }\href {\doibase
  10.1038/s41567-018-0124-x} {\bibfield  {journal} {\bibinfo  {journal} {Nature
  Physics}\ }\textbf {\bibinfo {volume} {14}},\ \bibinfo {pages} {595}
  (\bibinfo {year} {2018})}\BibitemShut {NoStop}%
\bibitem [{\citenamefont {Arute}\ \emph {et~al.}(2019)\citenamefont {Arute},
  \citenamefont {Arya}, \citenamefont {Babbush}, \citenamefont {Bacon},
  \citenamefont {Bardin}, \citenamefont {Barends}, \citenamefont {Biswas},
  \citenamefont {Boixo}, \citenamefont {Brandao}, \citenamefont {Buell},
  \citenamefont {Burkett}, \citenamefont {Chen}, \citenamefont {Chen},
  \citenamefont {Chiaro}, \citenamefont {Collins} \emph
  {et~al.}}]{arute2019quantum}%
  \BibitemOpen
  \bibfield  {author} {\bibinfo {author} {\bibfnamefont {F.}~\bibnamefont
  {Arute}}, \bibinfo {author} {\bibfnamefont {K.}~\bibnamefont {Arya}},
  \bibinfo {author} {\bibfnamefont {R.}~\bibnamefont {Babbush}}, \bibinfo
  {author} {\bibfnamefont {D.}~\bibnamefont {Bacon}}, \bibinfo {author}
  {\bibfnamefont {J.~C.}\ \bibnamefont {Bardin}}, \bibinfo {author}
  {\bibfnamefont {R.}~\bibnamefont {Barends}}, \bibinfo {author} {\bibfnamefont
  {R.}~\bibnamefont {Biswas}}, \bibinfo {author} {\bibfnamefont
  {S.}~\bibnamefont {Boixo}}, \bibinfo {author} {\bibfnamefont {F.~G. S.~L.}\
  \bibnamefont {Brandao}}, \bibinfo {author} {\bibfnamefont {D.~A.}\
  \bibnamefont {Buell}}, \bibinfo {author} {\bibfnamefont {B.}~\bibnamefont
  {Burkett}}, \bibinfo {author} {\bibfnamefont {Y.}~\bibnamefont {Chen}},
  \bibinfo {author} {\bibfnamefont {Z.}~\bibnamefont {Chen}}, \bibinfo {author}
  {\bibfnamefont {B.}~\bibnamefont {Chiaro}}, \bibinfo {author} {\bibfnamefont
  {R.}~\bibnamefont {Collins}},  \emph {et~al.},\ }\bibfield  {title} {\enquote
  {\bibinfo {title} {Quantum supremacy using a programmable superconducting
  processor},}\ }\href {\doibase 10.1038/s41586-019-1666-5} {\bibfield
  {journal} {\bibinfo  {journal} {Nature}\ }\textbf {\bibinfo {volume} {574}},\
  \bibinfo {pages} {505–510} (\bibinfo {year} {2019})}\BibitemShut {NoStop}%
\bibitem [{\citenamefont {Foxen}\ \emph {et~al.}(2020)\citenamefont {Foxen},
  \citenamefont {Neill}, \citenamefont {Dunsworth}, \citenamefont {Roushan},
  \citenamefont {Chiaro}, \citenamefont {Megrant}, \citenamefont {Kelly},
  \citenamefont {Chen}, \citenamefont {Satzinger}, \citenamefont {Barends},
  \citenamefont {Arute}, \citenamefont {Arya}, \citenamefont {Babbush},
  \citenamefont {Bacon}, \citenamefont {Bardin} \emph {et~al.}}]{foxen2020}%
  \BibitemOpen
  \bibfield  {author} {\bibinfo {author} {\bibfnamefont {B.}~\bibnamefont
  {Foxen}}, \bibinfo {author} {\bibfnamefont {C.}~\bibnamefont {Neill}},
  \bibinfo {author} {\bibfnamefont {A.}~\bibnamefont {Dunsworth}}, \bibinfo
  {author} {\bibfnamefont {P.}~\bibnamefont {Roushan}}, \bibinfo {author}
  {\bibfnamefont {B.}~\bibnamefont {Chiaro}}, \bibinfo {author} {\bibfnamefont
  {A.}~\bibnamefont {Megrant}}, \bibinfo {author} {\bibfnamefont
  {J.}~\bibnamefont {Kelly}}, \bibinfo {author} {\bibfnamefont
  {Z.}~\bibnamefont {Chen}}, \bibinfo {author} {\bibfnamefont {K.}~\bibnamefont
  {Satzinger}}, \bibinfo {author} {\bibfnamefont {R.}~\bibnamefont {Barends}},
  \bibinfo {author} {\bibfnamefont {F.}~\bibnamefont {Arute}}, \bibinfo
  {author} {\bibfnamefont {K.}~\bibnamefont {Arya}}, \bibinfo {author}
  {\bibfnamefont {R.}~\bibnamefont {Babbush}}, \bibinfo {author} {\bibfnamefont
  {D.}~\bibnamefont {Bacon}}, \bibinfo {author} {\bibfnamefont {J.~C.}\
  \bibnamefont {Bardin}},  \emph {et~al.},\ }\bibfield  {title} {\enquote
  {\bibinfo {title} {Demonstrating a continuous set of two-qubit gates for
  near-term quantum algorithms},}\ }\href {\doibase
  10.1103/PhysRevLett.125.120504} {\bibfield  {journal} {\bibinfo  {journal}
  {Phys. Rev. Lett.}\ }\textbf {\bibinfo {volume} {125}},\ \bibinfo {pages}
  {120504} (\bibinfo {year} {2020})}\BibitemShut {NoStop}%
\bibitem [{\citenamefont {Debroy}\ \emph {et~al.}(2023)\citenamefont {Debroy},
  \citenamefont {Genois}, \citenamefont {Gross}, \citenamefont {Mruczkiewicz},
  \citenamefont {Lee}, \citenamefont {Hong}, \citenamefont {Chen},
  \citenamefont {Smelyanskiy},\ and\ \citenamefont {Jiang}}]{cafe_paper_2023}%
  \BibitemOpen
  \bibfield  {author} {\bibinfo {author} {\bibfnamefont {D.~M.}\ \bibnamefont
  {Debroy}}, \bibinfo {author} {\bibfnamefont {E.}~\bibnamefont {Genois}},
  \bibinfo {author} {\bibfnamefont {J.~A.}\ \bibnamefont {Gross}}, \bibinfo
  {author} {\bibfnamefont {W.}~\bibnamefont {Mruczkiewicz}}, \bibinfo {author}
  {\bibfnamefont {K.}~\bibnamefont {Lee}}, \bibinfo {author} {\bibfnamefont
  {S.}~\bibnamefont {Hong}}, \bibinfo {author} {\bibfnamefont {Z.}~\bibnamefont
  {Chen}}, \bibinfo {author} {\bibfnamefont {V.}~\bibnamefont {Smelyanskiy}}, \
  and\ \bibinfo {author} {\bibfnamefont {Z.}~\bibnamefont {Jiang}},\ }\bibfield
   {title} {\enquote {\bibinfo {title} {Context-aware fidelity estimation},}\
  }\href {\doibase 10.1103/PhysRevResearch.5.043202} {\bibfield  {journal}
  {\bibinfo  {journal} {Phys. Rev. Res.}\ }\textbf {\bibinfo {volume} {5}},\
  \bibinfo {pages} {043202} (\bibinfo {year} {2023})}\BibitemShut {NoStop}%
\bibitem [{\citenamefont {Flammia}\ and\ \citenamefont
  {Liu}(2011)}]{flammia_direct_2011}%
  \BibitemOpen
  \bibfield  {author} {\bibinfo {author} {\bibfnamefont {S.~T.}\ \bibnamefont
  {Flammia}}\ and\ \bibinfo {author} {\bibfnamefont {Y.-K.}\ \bibnamefont
  {Liu}},\ }\bibfield  {title} {\enquote {\bibinfo {title} {Direct fidelity
  estimation from few {Pauli} measurements},}\ }\href {\doibase
  10.1103/PhysRevLett.106.230501} {\bibfield  {journal} {\bibinfo  {journal}
  {Physical Review Letters}\ }\textbf {\bibinfo {volume} {106}},\ \bibinfo
  {pages} {230501} (\bibinfo {year} {2011})}\BibitemShut {NoStop}%
\bibitem [{\citenamefont {Knill}\ \emph {et~al.}(2008)\citenamefont {Knill},
  \citenamefont {Leibfried}, \citenamefont {Reichle}, \citenamefont {Britton},
  \citenamefont {Blakestad}, \citenamefont {Jost}, \citenamefont {Langer},
  \citenamefont {Ozeri}, \citenamefont {Seidelin},\ and\ \citenamefont
  {Wineland}}]{knill_randomized_2008}%
  \BibitemOpen
  \bibfield  {author} {\bibinfo {author} {\bibfnamefont {E.}~\bibnamefont
  {Knill}}, \bibinfo {author} {\bibfnamefont {D.}~\bibnamefont {Leibfried}},
  \bibinfo {author} {\bibfnamefont {R.}~\bibnamefont {Reichle}}, \bibinfo
  {author} {\bibfnamefont {J.}~\bibnamefont {Britton}}, \bibinfo {author}
  {\bibfnamefont {R.~B.}\ \bibnamefont {Blakestad}}, \bibinfo {author}
  {\bibfnamefont {J.~D.}\ \bibnamefont {Jost}}, \bibinfo {author}
  {\bibfnamefont {C.}~\bibnamefont {Langer}}, \bibinfo {author} {\bibfnamefont
  {R.}~\bibnamefont {Ozeri}}, \bibinfo {author} {\bibfnamefont
  {S.}~\bibnamefont {Seidelin}}, \ and\ \bibinfo {author} {\bibfnamefont
  {D.~J.}\ \bibnamefont {Wineland}},\ }\bibfield  {title} {\enquote {\bibinfo
  {title} {Randomized benchmarking of quantum gates},}\ }\href {\doibase
  10.1103/PhysRevA.77.012307} {\bibfield  {journal} {\bibinfo  {journal}
  {Physical Review A}\ }\textbf {\bibinfo {volume} {77}},\ \bibinfo {pages}
  {012307} (\bibinfo {year} {2008})}\BibitemShut {NoStop}%
\bibitem [{\citenamefont {Barends}\ \emph {et~al.}(2019)\citenamefont
  {Barends}, \citenamefont {Quintana}, \citenamefont {Petukhov}, \citenamefont
  {Chen}, \citenamefont {Kafri}, \citenamefont {Kechedzhi}, \citenamefont
  {Collins}, \citenamefont {Naaman}, \citenamefont {Boixo}, \citenamefont
  {Arute}, \citenamefont {Arya}, \citenamefont {Buell}, \citenamefont
  {Burkett}, \citenamefont {Chen}, \citenamefont {Chiaro}, \citenamefont
  {Dunsworth}, \citenamefont {Foxen}, \citenamefont {Fowler}, \citenamefont
  {Gidney}, \citenamefont {Giustina}, \citenamefont {Graff}, \citenamefont
  {Huang}, \citenamefont {Jeffrey}, \citenamefont {Kelly}, \citenamefont
  {Klimov}, \citenamefont {Kostritsa}, \citenamefont {Landhuis}, \citenamefont
  {Lucero}, \citenamefont {McEwen}, \citenamefont {Megrant}, \citenamefont
  {Mi}, \citenamefont {Mutus}, \citenamefont {Neeley}, \citenamefont {Neill},
  \citenamefont {Ostby}, \citenamefont {Roushan}, \citenamefont {Sank},
  \citenamefont {Satzinger}, \citenamefont {Vainsencher}, \citenamefont
  {White}, \citenamefont {Yao}, \citenamefont {Yeh}, \citenamefont {Zalcman},
  \citenamefont {Neven}, \citenamefont {Smelyanskiy},\ and\ \citenamefont
  {Martinis}}]{barends_diabatic_2019}%
  \BibitemOpen
  \bibfield  {author} {\bibinfo {author} {\bibfnamefont {R.}~\bibnamefont
  {Barends}}, \bibinfo {author} {\bibfnamefont {C.~M.}\ \bibnamefont
  {Quintana}}, \bibinfo {author} {\bibfnamefont {A.~G.}\ \bibnamefont
  {Petukhov}}, \bibinfo {author} {\bibfnamefont {Y.}~\bibnamefont {Chen}},
  \bibinfo {author} {\bibfnamefont {D.}~\bibnamefont {Kafri}}, \bibinfo
  {author} {\bibfnamefont {K.}~\bibnamefont {Kechedzhi}}, \bibinfo {author}
  {\bibfnamefont {R.}~\bibnamefont {Collins}}, \bibinfo {author} {\bibfnamefont
  {O.}~\bibnamefont {Naaman}}, \bibinfo {author} {\bibfnamefont
  {S.}~\bibnamefont {Boixo}}, \bibinfo {author} {\bibfnamefont
  {F.}~\bibnamefont {Arute}}, \bibinfo {author} {\bibfnamefont
  {K.}~\bibnamefont {Arya}}, \bibinfo {author} {\bibfnamefont {D.}~\bibnamefont
  {Buell}}, \bibinfo {author} {\bibfnamefont {B.}~\bibnamefont {Burkett}},
  \bibinfo {author} {\bibfnamefont {Z.}~\bibnamefont {Chen}}, \bibinfo {author}
  {\bibfnamefont {B.}~\bibnamefont {Chiaro}}, \bibinfo {author} {\bibfnamefont
  {A.}~\bibnamefont {Dunsworth}}, \bibinfo {author} {\bibfnamefont
  {B.}~\bibnamefont {Foxen}}, \bibinfo {author} {\bibfnamefont
  {A.}~\bibnamefont {Fowler}}, \bibinfo {author} {\bibfnamefont
  {C.}~\bibnamefont {Gidney}}, \bibinfo {author} {\bibfnamefont
  {M.}~\bibnamefont {Giustina}}, \bibinfo {author} {\bibfnamefont
  {R.}~\bibnamefont {Graff}}, \bibinfo {author} {\bibfnamefont
  {T.}~\bibnamefont {Huang}}, \bibinfo {author} {\bibfnamefont
  {E.}~\bibnamefont {Jeffrey}}, \bibinfo {author} {\bibfnamefont
  {J.}~\bibnamefont {Kelly}}, \bibinfo {author} {\bibfnamefont {P.~V.}\
  \bibnamefont {Klimov}}, \bibinfo {author} {\bibfnamefont {F.}~\bibnamefont
  {Kostritsa}}, \bibinfo {author} {\bibfnamefont {D.}~\bibnamefont {Landhuis}},
  \bibinfo {author} {\bibfnamefont {E.}~\bibnamefont {Lucero}}, \bibinfo
  {author} {\bibfnamefont {M.}~\bibnamefont {McEwen}}, \bibinfo {author}
  {\bibfnamefont {A.}~\bibnamefont {Megrant}}, \bibinfo {author} {\bibfnamefont
  {X.}~\bibnamefont {Mi}}, \bibinfo {author} {\bibfnamefont {J.}~\bibnamefont
  {Mutus}}, \bibinfo {author} {\bibfnamefont {M.}~\bibnamefont {Neeley}},
  \bibinfo {author} {\bibfnamefont {C.}~\bibnamefont {Neill}}, \bibinfo
  {author} {\bibfnamefont {E.}~\bibnamefont {Ostby}}, \bibinfo {author}
  {\bibfnamefont {P.}~\bibnamefont {Roushan}}, \bibinfo {author} {\bibfnamefont
  {D.}~\bibnamefont {Sank}}, \bibinfo {author} {\bibfnamefont {K.~J.}\
  \bibnamefont {Satzinger}}, \bibinfo {author} {\bibfnamefont {A.}~\bibnamefont
  {Vainsencher}}, \bibinfo {author} {\bibfnamefont {T.}~\bibnamefont {White}},
  \bibinfo {author} {\bibfnamefont {J.}~\bibnamefont {Yao}}, \bibinfo {author}
  {\bibfnamefont {P.}~\bibnamefont {Yeh}}, \bibinfo {author} {\bibfnamefont
  {A.}~\bibnamefont {Zalcman}}, \bibinfo {author} {\bibfnamefont
  {H.}~\bibnamefont {Neven}}, \bibinfo {author} {\bibfnamefont {V.~N.}\
  \bibnamefont {Smelyanskiy}}, \ and\ \bibinfo {author} {\bibfnamefont {J.~M.}\
  \bibnamefont {Martinis}},\ }\bibfield  {title} {\enquote {\bibinfo {title}
  {Diabatic gates for frequency-tunable superconducting qubits},}\ }\href
  {\doibase 10.1103/PhysRevLett.123.210501} {\bibfield  {journal} {\bibinfo
  {journal} {Physical Review Letters}\ }\textbf {\bibinfo {volume} {123}},\
  \bibinfo {pages} {210501} (\bibinfo {year} {2019})}\BibitemShut {NoStop}%
\bibitem [{\citenamefont {Drury}\ and\ \citenamefont
  {Love}(2008)}]{drury_constructive_2008}%
  \BibitemOpen
  \bibfield  {author} {\bibinfo {author} {\bibfnamefont {B.}~\bibnamefont
  {Drury}}\ and\ \bibinfo {author} {\bibfnamefont {P.~J.}\ \bibnamefont
  {Love}},\ }\bibfield  {title} {\enquote {\bibinfo {title} {Constructive
  quantum {Shannon} decomposition from {Cartan} involutions},}\ }\href
  {\doibase 10.1088/1751-8113/41/39/395305} {\bibfield  {journal} {\bibinfo
  {journal} {Journal of Physics A: Mathematical and Theoretical}\ }\textbf
  {\bibinfo {volume} {41}},\ \bibinfo {pages} {395305} (\bibinfo {year}
  {2008})}\BibitemShut {NoStop}%
\bibitem [{\citenamefont {{Google Quantum AI}}(2023)}]{google2023suppressing}%
  \BibitemOpen
  \bibfield  {author} {\bibinfo {author} {\bibnamefont {{Google Quantum AI}}},\
  }\bibfield  {title} {\enquote {\bibinfo {title} {Suppressing quantum errors
  by scaling a surface code logical qubit},}\ }\href {\doibase
  10.1038/s41586-022-05434-1} {\bibfield  {journal} {\bibinfo  {journal}
  {Nature}\ }\textbf {\bibinfo {volume} {614}},\ \bibinfo {pages} {676}
  (\bibinfo {year} {2023})}\BibitemShut {NoStop}%
\bibitem [{\citenamefont {Vool}\ and\ \citenamefont
  {Devoret}(2017)}]{vool_introduction_2017}%
  \BibitemOpen
  \bibfield  {author} {\bibinfo {author} {\bibfnamefont {U.}~\bibnamefont
  {Vool}}\ and\ \bibinfo {author} {\bibfnamefont {M.}~\bibnamefont {Devoret}},\
  }\bibfield  {title} {\enquote {\bibinfo {title} {Introduction to quantum
  electromagnetic circuits},}\ }\href {\doibase 10.1002/cta.2359} {\bibfield
  {journal} {\bibinfo  {journal} {International Journal of Circuit Theory and
  Applications}\ }\textbf {\bibinfo {volume} {45}},\ \bibinfo {pages} {897}
  (\bibinfo {year} {2017})}\BibitemShut {NoStop}%
\bibitem [{\citenamefont {Krantz}\ \emph {et~al.}(2019)\citenamefont {Krantz},
  \citenamefont {Kjaergaard}, \citenamefont {Yan}, \citenamefont {Orlando},
  \citenamefont {Gustavsson},\ and\ \citenamefont
  {Oliver}}]{krantz_quantum_2019}%
  \BibitemOpen
  \bibfield  {author} {\bibinfo {author} {\bibfnamefont {P.}~\bibnamefont
  {Krantz}}, \bibinfo {author} {\bibfnamefont {M.}~\bibnamefont {Kjaergaard}},
  \bibinfo {author} {\bibfnamefont {F.}~\bibnamefont {Yan}}, \bibinfo {author}
  {\bibfnamefont {T.~P.}\ \bibnamefont {Orlando}}, \bibinfo {author}
  {\bibfnamefont {S.}~\bibnamefont {Gustavsson}}, \ and\ \bibinfo {author}
  {\bibfnamefont {W.~D.}\ \bibnamefont {Oliver}},\ }\bibfield  {title}
  {\enquote {\bibinfo {title} {A quantum engineer's guide to superconducting
  qubits},}\ }\href {\doibase 10.1063/1.5089550} {\bibfield  {journal}
  {\bibinfo  {journal} {Applied Physics Reviews}\ }\textbf {\bibinfo {volume}
  {6}},\ \bibinfo {pages} {021318} (\bibinfo {year} {2019})}\BibitemShut
  {NoStop}%
\bibitem [{\citenamefont {Blais}\ \emph {et~al.}(2021)\citenamefont {Blais},
  \citenamefont {Grimsmo}, \citenamefont {Girvin},\ and\ \citenamefont
  {Wallraff}}]{blais_circuit_2021}%
  \BibitemOpen
  \bibfield  {author} {\bibinfo {author} {\bibfnamefont {A.}~\bibnamefont
  {Blais}}, \bibinfo {author} {\bibfnamefont {A.~L.}\ \bibnamefont {Grimsmo}},
  \bibinfo {author} {\bibfnamefont {S.~M.}\ \bibnamefont {Girvin}}, \ and\
  \bibinfo {author} {\bibfnamefont {A.}~\bibnamefont {Wallraff}},\ }\bibfield
  {title} {\enquote {\bibinfo {title} {Circuit {Quantum} {Electrodynamics}},}\
  }\href {\doibase 10.1103/RevModPhys.93.025005} {\bibfield  {journal}
  {\bibinfo  {journal} {Reviews of Modern Physics}\ }\textbf {\bibinfo {volume}
  {93}},\ \bibinfo {pages} {025005} (\bibinfo {year} {2021})}\BibitemShut
  {NoStop}%
\bibitem [{\citenamefont {Motzoi}\ \emph {et~al.}(2009)\citenamefont {Motzoi},
  \citenamefont {Gambetta}, \citenamefont {Rebentrost},\ and\ \citenamefont
  {Wilhelm}}]{motzoi_simple_2009}%
  \BibitemOpen
  \bibfield  {author} {\bibinfo {author} {\bibfnamefont {F.}~\bibnamefont
  {Motzoi}}, \bibinfo {author} {\bibfnamefont {J.~M.}\ \bibnamefont
  {Gambetta}}, \bibinfo {author} {\bibfnamefont {P.}~\bibnamefont
  {Rebentrost}}, \ and\ \bibinfo {author} {\bibfnamefont {F.~K.}\ \bibnamefont
  {Wilhelm}},\ }\bibfield  {title} {\enquote {\bibinfo {title} {Simple pulses
  for elimination of leakage in weakly nonlinear qubits},}\ }\href {\doibase
  10.1103/PhysRevLett.103.110501} {\bibfield  {journal} {\bibinfo  {journal}
  {Physical Review Letters}\ }\textbf {\bibinfo {volume} {103}},\ \bibinfo
  {pages} {110501} (\bibinfo {year} {2009})}\BibitemShut {NoStop}%
\bibitem [{\citenamefont {Rol}\ \emph {et~al.}(2019)\citenamefont {Rol},
  \citenamefont {Battistel}, \citenamefont {Malinowski}, \citenamefont
  {Bultink}, \citenamefont {Tarasinski}, \citenamefont {Vollmer}, \citenamefont
  {Haider}, \citenamefont {Muthusubramanian}, \citenamefont {Bruno},
  \citenamefont {Terhal},\ and\ \citenamefont {DiCarlo}}]{rol_fast_2019}%
  \BibitemOpen
  \bibfield  {author} {\bibinfo {author} {\bibfnamefont {M.~A.}\ \bibnamefont
  {Rol}}, \bibinfo {author} {\bibfnamefont {F.}~\bibnamefont {Battistel}},
  \bibinfo {author} {\bibfnamefont {F.~K.}\ \bibnamefont {Malinowski}},
  \bibinfo {author} {\bibfnamefont {C.~C.}\ \bibnamefont {Bultink}}, \bibinfo
  {author} {\bibfnamefont {B.~M.}\ \bibnamefont {Tarasinski}}, \bibinfo
  {author} {\bibfnamefont {R.}~\bibnamefont {Vollmer}}, \bibinfo {author}
  {\bibfnamefont {N.}~\bibnamefont {Haider}}, \bibinfo {author} {\bibfnamefont
  {N.}~\bibnamefont {Muthusubramanian}}, \bibinfo {author} {\bibfnamefont
  {A.}~\bibnamefont {Bruno}}, \bibinfo {author} {\bibfnamefont {B.~M.}\
  \bibnamefont {Terhal}}, \ and\ \bibinfo {author} {\bibfnamefont
  {L.}~\bibnamefont {DiCarlo}},\ }\bibfield  {title} {\enquote {\bibinfo
  {title} {Fast, high-fidelity conditional-phase gate exploiting leakage
  interference in weakly anharmonic superconducting qubits},}\ }\href {\doibase
  10.1103/PhysRevLett.123.120502} {\bibfield  {journal} {\bibinfo  {journal}
  {Phys. Rev. Lett.}\ }\textbf {\bibinfo {volume} {123}},\ \bibinfo {pages}
  {120502} (\bibinfo {year} {2019})}\BibitemShut {NoStop}%
\bibitem [{\citenamefont {Neg\^{\i}rneac}\ \emph {et~al.}(2021)\citenamefont
  {Neg\^{\i}rneac}, \citenamefont {Ali}, \citenamefont {Muthusubramanian},
  \citenamefont {Battistel}, \citenamefont {Sagastizabal}, \citenamefont
  {Moreira}, \citenamefont {Marques}, \citenamefont {Vlothuizen}, \citenamefont
  {Beekman}, \citenamefont {Zachariadis}, \citenamefont {Haider}, \citenamefont
  {Bruno},\ and\ \citenamefont {DiCarlo}}]{negirneac_high_2021}%
  \BibitemOpen
  \bibfield  {author} {\bibinfo {author} {\bibfnamefont {V.}~\bibnamefont
  {Neg\^{\i}rneac}}, \bibinfo {author} {\bibfnamefont {H.}~\bibnamefont {Ali}},
  \bibinfo {author} {\bibfnamefont {N.}~\bibnamefont {Muthusubramanian}},
  \bibinfo {author} {\bibfnamefont {F.}~\bibnamefont {Battistel}}, \bibinfo
  {author} {\bibfnamefont {R.}~\bibnamefont {Sagastizabal}}, \bibinfo {author}
  {\bibfnamefont {M.~S.}\ \bibnamefont {Moreira}}, \bibinfo {author}
  {\bibfnamefont {J.~F.}\ \bibnamefont {Marques}}, \bibinfo {author}
  {\bibfnamefont {W.~J.}\ \bibnamefont {Vlothuizen}}, \bibinfo {author}
  {\bibfnamefont {M.}~\bibnamefont {Beekman}}, \bibinfo {author} {\bibfnamefont
  {C.}~\bibnamefont {Zachariadis}}, \bibinfo {author} {\bibfnamefont
  {N.}~\bibnamefont {Haider}}, \bibinfo {author} {\bibfnamefont
  {A.}~\bibnamefont {Bruno}}, \ and\ \bibinfo {author} {\bibfnamefont
  {L.}~\bibnamefont {DiCarlo}},\ }\bibfield  {title} {\enquote {\bibinfo
  {title} {High-fidelity controlled-$z$ gate with maximal intermediate leakage
  operating at the speed limit in a superconducting quantum processor},}\
  }\href {\doibase 10.1103/PhysRevLett.126.220502} {\bibfield  {journal}
  {\bibinfo  {journal} {Phys. Rev. Lett.}\ }\textbf {\bibinfo {volume} {126}},\
  \bibinfo {pages} {220502} (\bibinfo {year} {2021})}\BibitemShut {NoStop}%
\bibitem [{\citenamefont {Johnson}(2011)}]{johnson_controlling_2011}%
  \BibitemOpen
  \bibfield  {author} {\bibinfo {author} {\bibfnamefont {B.~R.}\ \bibnamefont
  {Johnson}},\ }\emph {\bibinfo {title} {Controlling {Photons} in
  {Superconducting} {Electrical} {Circuits}}},\ \href
  {https://rsl.yale.edu/sites/default/files/2024-08/2011-RSL-Thesis-Blake-Johnson.pdf}
  {\bibinfo {type} {Doctoral dissertation}},\ \bibinfo  {school} {Yale}
  (\bibinfo {year} {2011})\BibitemShut {NoStop}%
\bibitem [{\citenamefont {Wei}\ \emph {et~al.}(2024)\citenamefont {Wei},
  \citenamefont {Pritchett}, \citenamefont {Zajac}, \citenamefont {McKay},\
  and\ \citenamefont {Merkel}}]{wei_characterizing_2024}%
  \BibitemOpen
  \bibfield  {author} {\bibinfo {author} {\bibfnamefont {K.~X.}\ \bibnamefont
  {Wei}}, \bibinfo {author} {\bibfnamefont {E.}~\bibnamefont {Pritchett}},
  \bibinfo {author} {\bibfnamefont {D.~M.}\ \bibnamefont {Zajac}}, \bibinfo
  {author} {\bibfnamefont {D.~C.}\ \bibnamefont {McKay}}, \ and\ \bibinfo
  {author} {\bibfnamefont {S.}~\bibnamefont {Merkel}},\ }\bibfield  {title}
  {\enquote {\bibinfo {title} {Characterizing non-{Markovian} off-resonant
  errors in quantum gates},}\ }\href {\doibase
  10.1103/PhysRevApplied.21.024018} {\bibfield  {journal} {\bibinfo  {journal}
  {Physical Review Applied}\ }\textbf {\bibinfo {volume} {21}},\ \bibinfo
  {pages} {024018} (\bibinfo {year} {2024})}\BibitemShut {NoStop}%
\bibitem [{\citenamefont {Maudsley}(1986)}]{maudsley_modified_1986}%
  \BibitemOpen
  \bibfield  {author} {\bibinfo {author} {\bibfnamefont {A.~A.}\ \bibnamefont
  {Maudsley}},\ }\bibfield  {title} {\enquote {\bibinfo {title} {Modified
  {Carr}-{Purcell}-{Meiboom}-{Gill} sequence for {NMR} fourier imaging
  applications},}\ }\href {\doibase
  https://doi.org/10.1016/0022-2364(86)90160-5} {\bibfield  {journal} {\bibinfo
   {journal} {Journal of Magnetic Resonance (1969)}\ }\textbf {\bibinfo
  {volume} {69}},\ \bibinfo {pages} {488} (\bibinfo {year} {1986})}\BibitemShut
  {NoStop}%
\end{thebibliography}%


\begin{thebibliography}{5}%
\makeatletter
\providecommand \@ifxundefined [1]{%
 \@ifx{#1\undefined}
}%
\providecommand \@ifnum [1]{%
 \ifnum #1\expandafter \@firstoftwo
 \else \expandafter \@secondoftwo
 \fi
}%
\providecommand \@ifx [1]{%
 \ifx #1\expandafter \@firstoftwo
 \else \expandafter \@secondoftwo
 \fi
}%
\providecommand \natexlab [1]{#1}%
\providecommand \enquote  [1]{``#1''}%
\providecommand \bibnamefont  [1]{#1}%
\providecommand \bibfnamefont [1]{#1}%
\providecommand \citenamefont [1]{#1}%
\providecommand \href@noop [0]{\@secondoftwo}%
\providecommand \href [0]{\begingroup \@sanitize@url \@href}%
\providecommand \@href[1]{\@@startlink{#1}\@@href}%
\providecommand \@@href[1]{\endgroup#1\@@endlink}%
\providecommand \@sanitize@url [0]{\catcode `\\12\catcode `\$12\catcode
  `\&12\catcode `\#12\catcode `\^12\catcode `\_12\catcode `\%12\relax}%
\providecommand \@@startlink[1]{}%
\providecommand \@@endlink[0]{}%
\providecommand \url  [0]{\begingroup\@sanitize@url \@url }%
\providecommand \@url [1]{\endgroup\@href {#1}{\urlprefix }}%
\providecommand \urlprefix  [0]{URL }%
\providecommand \Eprint [0]{\href }%
\providecommand \doibase [0]{http://dx.doi.org/}%
\providecommand \selectlanguage [0]{\@gobble}%
\providecommand \bibinfo  [0]{\@secondoftwo}%
\providecommand \bibfield  [0]{\@secondoftwo}%
\providecommand \translation [1]{[#1]}%
\providecommand \BibitemOpen [0]{}%
\providecommand \bibitemStop [0]{}%
\providecommand \bibitemNoStop [0]{.\EOS\space}%
\providecommand \EOS [0]{\spacefactor3000\relax}%
\providecommand \BibitemShut  [1]{\csname bibitem#1\endcsname}%
\let\auto@bib@innerbib\@empty
%</preamble>
\bibitem [{\citenamefont {Arute}\ \emph {et~al.}(2020)\citenamefont {Arute},
  \citenamefont {Arya}, \citenamefont {Babbush}, \citenamefont {Bacon},
  \citenamefont {Bardin}, \citenamefont {Barends}, \citenamefont {Bengtsson},
  \citenamefont {Boixo}, \citenamefont {Broughton}, \citenamefont {Buckley},
  \citenamefont {Buell}, \citenamefont {Burkett}, \citenamefont {Bushnell},
  \citenamefont {Chen}, \citenamefont {Chen} \emph
  {et~al.}}]{arute_observation_2020}%
  \BibitemOpen
  \bibfield  {author} {\bibinfo {author} {\bibfnamefont {F.}~\bibnamefont
  {Arute}}, \bibinfo {author} {\bibfnamefont {K.}~\bibnamefont {Arya}},
  \bibinfo {author} {\bibfnamefont {R.}~\bibnamefont {Babbush}}, \bibinfo
  {author} {\bibfnamefont {D.}~\bibnamefont {Bacon}}, \bibinfo {author}
  {\bibfnamefont {J.~C.}\ \bibnamefont {Bardin}}, \bibinfo {author}
  {\bibfnamefont {R.}~\bibnamefont {Barends}}, \bibinfo {author} {\bibfnamefont
  {A.}~\bibnamefont {Bengtsson}}, \bibinfo {author} {\bibfnamefont
  {S.}~\bibnamefont {Boixo}}, \bibinfo {author} {\bibfnamefont
  {M.}~\bibnamefont {Broughton}}, \bibinfo {author} {\bibfnamefont {B.~B.}\
  \bibnamefont {Buckley}}, \bibinfo {author} {\bibfnamefont {D.~A.}\
  \bibnamefont {Buell}}, \bibinfo {author} {\bibfnamefont {B.}~\bibnamefont
  {Burkett}}, \bibinfo {author} {\bibfnamefont {N.}~\bibnamefont {Bushnell}},
  \bibinfo {author} {\bibfnamefont {Y.}~\bibnamefont {Chen}}, \bibinfo {author}
  {\bibfnamefont {Z.}~\bibnamefont {Chen}},  \emph {et~al.},\ }\bibfield
  {title} {\enquote {\bibinfo {title} {Observation of separated dynamics of
  charge and spin in the {Fermi}-{Hubbard} model},}\ }\href
  {http://arxiv.org/abs/2010.07965} {\bibfield  {journal} {\bibinfo  {journal}
  {arXiv:2010.07965}\ } (\bibinfo {year} {2020})}\BibitemShut {NoStop}%
\bibitem [{\citenamefont {Neill}\ \emph {et~al.}(2021)\citenamefont {Neill},
  \citenamefont {McCourt}, \citenamefont {Mi}, \citenamefont {Jiang},
  \citenamefont {Niu}, \citenamefont {Mruczkiewicz}, \citenamefont {Aleiner},
  \citenamefont {Arute}, \citenamefont {Arya}, \citenamefont {Atalaya},
  \citenamefont {Babbush}, \citenamefont {Bardin}, \citenamefont {Barends},
  \citenamefont {Bengtsson}, \citenamefont {Bourassa}, \citenamefont
  {Broughton}, \citenamefont {Buckley}, \citenamefont {Buell}, \citenamefont
  {Burkett}, \citenamefont {Bushnell} \emph {et~al.}}]{neill_accurately_2021}%
  \BibitemOpen
  \bibfield  {author} {\bibinfo {author} {\bibfnamefont {C.}~\bibnamefont
  {Neill}}, \bibinfo {author} {\bibfnamefont {T.}~\bibnamefont {McCourt}},
  \bibinfo {author} {\bibfnamefont {X.}~\bibnamefont {Mi}}, \bibinfo {author}
  {\bibfnamefont {Z.}~\bibnamefont {Jiang}}, \bibinfo {author} {\bibfnamefont
  {M.~Y.}\ \bibnamefont {Niu}}, \bibinfo {author} {\bibfnamefont
  {W.}~\bibnamefont {Mruczkiewicz}}, \bibinfo {author} {\bibfnamefont
  {I.}~\bibnamefont {Aleiner}}, \bibinfo {author} {\bibfnamefont
  {F.}~\bibnamefont {Arute}}, \bibinfo {author} {\bibfnamefont
  {K.}~\bibnamefont {Arya}}, \bibinfo {author} {\bibfnamefont {J.}~\bibnamefont
  {Atalaya}}, \bibinfo {author} {\bibfnamefont {R.}~\bibnamefont {Babbush}},
  \bibinfo {author} {\bibfnamefont {J.~C.}\ \bibnamefont {Bardin}}, \bibinfo
  {author} {\bibfnamefont {R.}~\bibnamefont {Barends}}, \bibinfo {author}
  {\bibfnamefont {A.}~\bibnamefont {Bengtsson}}, \bibinfo {author}
  {\bibfnamefont {A.}~\bibnamefont {Bourassa}}, \bibinfo {author}
  {\bibfnamefont {M.}~\bibnamefont {Broughton}}, \bibinfo {author}
  {\bibfnamefont {B.~B.}\ \bibnamefont {Buckley}}, \bibinfo {author}
  {\bibfnamefont {D.~A.}\ \bibnamefont {Buell}}, \bibinfo {author}
  {\bibfnamefont {B.}~\bibnamefont {Burkett}}, \bibinfo {author} {\bibfnamefont
  {N.}~\bibnamefont {Bushnell}},  \emph {et~al.},\ }\bibfield  {title}
  {\enquote {\bibinfo {title} {Accurately computing the electronic properties
  of a quantum ring},}\ }\href {\doibase 10.1038/s41586-021-03576-2} {\bibfield
   {journal} {\bibinfo  {journal} {Nature}\ }\textbf {\bibinfo {volume}
  {594}},\ \bibinfo {pages} {508} (\bibinfo {year} {2021})}\BibitemShut
  {NoStop}%
\bibitem [{\citenamefont {Motzoi}\ \emph {et~al.}(2009)\citenamefont {Motzoi},
  \citenamefont {Gambetta}, \citenamefont {Rebentrost},\ and\ \citenamefont
  {Wilhelm}}]{motzoi_simple_2009}%
  \BibitemOpen
  \bibfield  {author} {\bibinfo {author} {\bibfnamefont {F.}~\bibnamefont
  {Motzoi}}, \bibinfo {author} {\bibfnamefont {J.~M.}\ \bibnamefont
  {Gambetta}}, \bibinfo {author} {\bibfnamefont {P.}~\bibnamefont
  {Rebentrost}}, \ and\ \bibinfo {author} {\bibfnamefont {F.~K.}\ \bibnamefont
  {Wilhelm}},\ }\bibfield  {title} {\enquote {\bibinfo {title} {Simple pulses
  for elimination of leakage in weakly nonlinear qubits},}\ }\href {\doibase
  10.1103/PhysRevLett.103.110501} {\bibfield  {journal} {\bibinfo  {journal}
  {Physical Review Letters}\ }\textbf {\bibinfo {volume} {103}},\ \bibinfo
  {pages} {110501} (\bibinfo {year} {2009})}\BibitemShut {NoStop}%
\bibitem [{\citenamefont {Shaji}\ and\ \citenamefont
  {Caves}(2007)}]{shaji_qubit_2007}%
  \BibitemOpen
  \bibfield  {author} {\bibinfo {author} {\bibfnamefont {A.}~\bibnamefont
  {Shaji}}\ and\ \bibinfo {author} {\bibfnamefont {C.~M.}\ \bibnamefont
  {Caves}},\ }\bibfield  {title} {\enquote {\bibinfo {title} {Qubit metrology
  and decoherence},}\ }\href {\doibase 10.1103/PhysRevA.76.032111} {\bibfield
  {journal} {\bibinfo  {journal} {Physical Review A}\ }\textbf {\bibinfo
  {volume} {76}},\ \bibinfo {pages} {032111} (\bibinfo {year}
  {2007})}\BibitemShut {NoStop}%
\bibitem [{\citenamefont {Ji}\ \emph {et~al.}(2008)\citenamefont {Ji},
  \citenamefont {Wang}, \citenamefont {Duan}, \citenamefont {Feng},\ and\
  \citenamefont {Ying}}]{ji_parameter_2008}%
  \BibitemOpen
  \bibfield  {author} {\bibinfo {author} {\bibfnamefont {Z.}~\bibnamefont
  {Ji}}, \bibinfo {author} {\bibfnamefont {G.}~\bibnamefont {Wang}}, \bibinfo
  {author} {\bibfnamefont {R.}~\bibnamefont {Duan}}, \bibinfo {author}
  {\bibfnamefont {Y.}~\bibnamefont {Feng}}, \ and\ \bibinfo {author}
  {\bibfnamefont {M.}~\bibnamefont {Ying}},\ }\bibfield  {title} {\enquote
  {\bibinfo {title} {Parameter {Estimation} of {Quantum} {Channels}},}\ }\href
  {\doibase 10.1109/TIT.2008.929940} {\bibfield  {journal} {\bibinfo  {journal}
  {IEEE Transactions on Information Theory}\ }\textbf {\bibinfo {volume}
  {54}},\ \bibinfo {pages} {5172} (\bibinfo {year} {2008})}\BibitemShut
  {NoStop}%
\end{thebibliography}%


\begin{thebibliography}{0}%
\makeatletter
\providecommand \@ifxundefined [1]{%
 \@ifx{#1\undefined}
}%
\providecommand \@ifnum [1]{%
 \ifnum #1\expandafter \@firstoftwo
 \else \expandafter \@secondoftwo
 \fi
}%
\providecommand \@ifx [1]{%
 \ifx #1\expandafter \@firstoftwo
 \else \expandafter \@secondoftwo
 \fi
}%
\providecommand \natexlab [1]{#1}%
\providecommand \enquote  [1]{``#1''}%
\providecommand \bibnamefont  [1]{#1}%
\providecommand \bibfnamefont [1]{#1}%
\providecommand \citenamefont [1]{#1}%
\providecommand \href@noop [0]{\@secondoftwo}%
\providecommand \href [0]{\begingroup \@sanitize@url \@href}%
\providecommand \@href[1]{\@@startlink{#1}\@@href}%
\providecommand \@@href[1]{\endgroup#1\@@endlink}%
\providecommand \@sanitize@url [0]{\catcode `\\12\catcode `\$12\catcode
  `\&12\catcode `\#12\catcode `\^12\catcode `\_12\catcode `\%12\relax}%
\providecommand \@@startlink[1]{}%
\providecommand \@@endlink[0]{}%
\providecommand \url  [0]{\begingroup\@sanitize@url \@url }%
\providecommand \@url [1]{\endgroup\@href {#1}{\urlprefix }}%
\providecommand \urlprefix  [0]{URL }%
\providecommand \Eprint [0]{\href }%
\providecommand \doibase [0]{https://doi.org/}%
\providecommand \selectlanguage [0]{\@gobble}%
\providecommand \bibinfo  [0]{\@secondoftwo}%
\providecommand \bibfield  [0]{\@secondoftwo}%
\providecommand \translation [1]{[#1]}%
\providecommand \BibitemOpen [0]{}%
\providecommand \bibitemStop [0]{}%
\providecommand \bibitemNoStop [0]{.\EOS\space}%
\providecommand \EOS [0]{\spacefactor3000\relax}%
\providecommand \BibitemShut  [1]{\csname bibitem#1\endcsname}%
\let\auto@bib@innerbib\@empty
%</preamble>
\end{thebibliography}%

\end{document}